\documentclass[10pt]{amsart}
\usepackage{graphicx}
\usepackage{amssymb}
\usepackage{bbm}
\usepackage{color}

\numberwithin{equation}{section}

\newcommand{\req}[1]{(\ref{#1})}

\def\cro{{\rm cr }}
\def\fdt{{\rm fdt}}

\def\b{{\beta}}

\def\Z{{\mathbb Z}}

\def\bJ{{\bf J}}
\def\R{{\mathbb R}}
\def\P{{\mathbb P}}
\def\E{{\mathbb E}}

\def\BB{{\bf B}}

\def\BB{{\bf B}}

\def\BJ{{\bf J}}

\def\BR{{\bf R}}
\def\BD{{\bf D}}
\def\BJ{{\bf J}}

\def\bmu{{\boldsymbol \mu}}

\def\bw{{\bf w}}
\def\bb{{\bf b}}
\def\ba{{\bf a}}
\def\Bx{{\bf x}}
\def\bx{{\bf x}}

\def\bD{{\bf \Gamma}}
\newcommand{\reals}{{\mathbb{R}}}

\def\bx{\Bx}

\def\CC{{\mathcal C}}
\def\CL{{\mathcal L}}

\def\CC{{\mathcal R}}
\def\CC{{\mathcal C}}
\def\CL{{\mathcal L}}

\def\Aa{{\mathcal A}}
\def\Ba{{\mathcal B}}
\def\Ca{{\mathcal C}}
\def\Da{{\mathcal D}}
\def\Ea{{\mathcal E}}
\def\Fa{{\mathcal F}}

\def\Ha{{\mathcal H}}

\def\Sa{{\mathcal S}}

\def\R{{\mathbb R}}

\def\E{{\mathbb E}}

\def\half{\frac{1}{2}}
\def\oneN{\frac{1}{N}}

\def\nn{\noindent}

\def\hh{\widehat}

\def\pm{{probability measure\ }}

\def\a{\alpha}
\def\b{\beta}
\def\d{\delta}
\def\e{\epsilon}
\def\g{\gamma}

\def\s{\sigma}

\def\NC{{\mbox{NC}}}
\def\D{\Delta}

\def\L{\Lambda}

\def\pars{\partial}

\def\bu{\bullet}

\def\ts{\times}

\def\ra{\rightarrow}

\def\tilde{\widetilde}
\def\hh{\widehat}

\cleardoublepage
\newtheorem{prop}{Proposition}[section]

\newtheorem{lem}[prop]{Lemma}

\newtheorem{theo}[prop]{Theorem}
\newtheorem{ass}[prop]{Hypothesis}

\begin{document}

\title[Spherical Spin Glasses]
{Limiting dynamics for spherical models of spin glasses\\
with magnetic field}

\author{Manuel Zamfir}
\address{Department of Mathematics\\
Stanford University\\ Stanford, CA 94305.}
\email{manuelzamfir@gmail.com}

%\newline
%{\bf AMS (2000) Subject Classification:}
%{82C44, 82C31, 60H10, 60F15, 60K35}
%\newline
%{\bf Keywords:} Interacting random processes, Disordered systems,
%Statistical mechanics, Langevin dynamics, Aging, $p$-spin models.}

\thanks{Research supported in part by NSF grants \#DMS-0406042 and \#DMS-0806211.}
\thanks{This work was carried out as a part of my PhD thesis at Stanford University and the author dearly expresses his gratitude to his advisor, Professor Amir Dembo for his helpful support and enlighten discussion.}

\begin{abstract}

We study the Langevin dynamics for the family of
spherical spin glass models of statistical physics, in the presence of a magnetic field.
We prove that in the limit of system size $N$ approaching infinity,
the {\em empirical state correlation}, the {\em response function},
the {\em overlap} and the {\em magnetization} for these
$N$-dimensional coupled diffusions converge to the non-random
unique strong solution of four explicit non-linear
integro-differential equations, that generalize the
system proposed by Cugliandolo and Kurchan in the presence of a magnetic field.

We then analyze the system and provide a rigorous derivation of
the FDT regime in a large area of the temperature-magnetization plane.
\end{abstract}

\maketitle

%%%%%%%%%%%%%%%%%%%%%%%%%%%%%%%%%%%%%%%%%%%%%%%%%%%%

%%%%%%%%%%%%%%%%%%%%%%%%%%%%%%%%%%%%%%%%%%%%%%%%%%%%
%%%%%%%%%%%%%%%%%%%%%%%%%%%%%%%%%%%%%%%%%%%%%%%%%%%%

\section{Introduction}

Many of the unique properties of magnetic systems with quenched random interactions,
namely spin glasses, are of dynamical nature (see \cite{FH}). Therefore, we would like to
understand not only the static properties, but also time dependent features
of the spin glass state. This is not an easy task, even for the
Sherrington and Kirkpatrick (SK) model.

The extended SK model can be described as follows. Let $\Gamma = \{-1,1\}$ be the space of spins. Fixing a positive integer $N$ (denoting the system size), define, for each configuration of the spins (i.e. for each $\Bx = (x^1,\dots,x^N) \in \Gamma^N$), a random Hamiltonian $H^N_\BJ(\Bx)$, as a function of the configuration $\Bx$ and of an exterior source of randomness $\BJ$ (i.e. a random variable defined on another probability space). For the extended SK model, the mean field random Hamiltonian is defined as:
$$H^N_{\BJ}(\Bx)= - \sum_{p=1}^m \frac{a_p}{p!}
 \sum_{1\le i_1,\ldots,
i_p \le N} J_{i_1\ldots i_p} x^{i_1}\ldots  x^{i_p}\,,$$
where $m \geq 2$, and the disorder parameters $J_{i_1 \ldots i_p}=J_{\{i_1,\ldots,i_p\}}$
are independent (modulo the permutation of the indices) centered
Gaussian variables. The variance of $J_{i_1\ldots i_p}$ is
$c(\{i_1,\ldots,i_p\}) N^{-p+1}$, where
\begin{equation}\label{eq:vardef}
c(\{i_1,\ldots,i_p\})=\prod_k l_k! \,,
\end{equation}
and $(l_1,l_2,\ldots)$ are
the multiplicities of the different elements of the set
$\{i_1,\ldots,i_p\}$ (for example, $c=1$ when $i_j \neq i_{j'}$ for
any $j \neq j'$, while $c=p!$ when all $i_j$ values are the same). Denoting by $F^N(\Bx)$ the {\em total magnetization} of the system:
\begin{equation}\label{eq:magnetization}
F^N (\Bx) = \sum_{i=1}^N x^i\,,
\end{equation}
the Gibbs measure for finitely many spins at inverse temperature $\beta = T^{-1}$ and intensity of the magnetic field $h > 0$ is defined as:
\begin{equation}
\label{eq:skmu}
\lambda^N_{\b,h,\BJ}(\Bx) = \frac{1}{Z_{\b, h, \BJ}^N} \exp\left(- \beta H^N_\BJ(\Bx) + h F^N(\Bx)\right)\mathbbm{1}_{\Bx \in \Gamma^N}\,.
\end{equation}
where $Z_{\b, h, \BJ}$ is a normalizing constant. The propagation of chaos for the dynamics is of much interest. It can be studied from the limit as $N \ra \infty$ of the empirical measure:
$$\mu_N = \frac{1}{N} \sum_{i=1}^N \delta_{x^i(t)}$$
Though the limit was established and characterized in \cite{BAG2} via an implicit non-Markovian stochastic differential equation for the continuous relaxation of the SK model with Langevin dynamics, the complexity of the latter equation prevents it from being amenable to a serious understanding.

{\em Spherical} models replace the product structure of the configuration space $\Gamma^N$ by the sphere $S^{N-1}(\sqrt{rN})$ in $\R^N$, for $r=1$, via imposing the hard constraint $\frac{1}{N}\sum_{i=1}^N x_i^2 = r$. The spherical Gibbs measure is then given by:
\begin{equation}
\label{eq:hardmu}
\mu^N_{\b,h,\BJ}(d\bx) = \frac{1}{Z^N_{\b, h, \BJ}} \exp\left( - 2\b H^N_{\BJ}(\Bx) + 2 h F^N(\Bx)\right)\nu_N (d\Bx)
\end{equation}
where the measure $\nu_N$ is the uniform measure on the sphere $S^{N-1}({\sqrt{rN}})$ (the presence of the extra factor of $2$ is just a matter of convenience and is equivalent to the rescaling $\beta \mapsto 2\beta$ and $h \mapsto 2h$). The Langevin dynamics for the normalized spherical mixed spin model (i.e. $r=1$) without magnetization (i.e. $h=0$), was rigurously studied in \cite{BDG2} and \cite{DGM}. The authors have shown that the dynamics of the system can be characterized via two functions, the so called {\em empirical correlation} and {\em empirical response} and they have derived the pair of coupled integro-differential equations that characterize them.

Here, we shall first extend their results to allow for a positive magnetic field (i.e. $h > 0$) and any radius of the underlying sphere. Due to the extra complexity introduced in the system via the presence of the magnetic field, that affects the symmetry of the spins, the dynamics will be characterized via a coupled system of four integro-differential equations. We rigurously analyze the behavior of the system in the high temperature regime and derive equations characterizing the phase transition curve. Along the way, we prove (see Theorem \ref{theo-sk}) that the system simplifies dramatically for large radii of the underlying sphere.

To work around the complexity induced by the Langevin dynamics on the sphere, we follow \cite{BDG2}, by a further relaxation of the {\em hard} spherical model, replacing the hard spherical constraint by a {\em soft} one. Namely, we first replace the uniform measure $\nu_N$  on the sphere $S^{N-1}({\sqrt{rN}})$ by a measure on $\R^N$,
$$\tilde \nu_N(d\Bx) = \frac{1}{Z_{N, f}}\exp\left(- N f\left(\frac{1}{N}\sum_{i=1}^N x_i^2\right)\right) d\Bx$$
where $f$ is a smooth function growing fast enough at infinity. The {\em soft spherical Gibbs measure} is then given by:
\begin{equation}
\label{eq:softmu}
d\tilde \mu^N_{\b,h,\BJ,f}(d\Bx) = \frac{1}{Z_{\b, h, \BJ, f}^N} \exp\left(-Nf\left(\frac{\parallel \Bx \parallel_2^2}{N}\right) - 2 \beta H^N_\BJ(\Bx) + 2 h F^N(\Bx)\right)\overset{N}{\underset{i=1}{\Pi}}dx^i\,.
\end{equation}
Thus, $\tilde \mu^N_{\b,h\BJ,f}$ is the invariant measure of the
randomly interacting particles described by the
(Langevin) stochastic differential system:
\begin{equation}\label{interaction}
dx^j_t=dB_t^j- f'(N^{-1} \|\Bx_t\|^2) x^j_t dt
+\beta G^j(\Bx_t) dt + hdt\,,
\end{equation}
where $\BB=(B^1,\ldots,B^N)$ is an N-dimensional standard
Brownian motion, independent of both the initial condition $\Bx_0$ and
the disorder $\BJ$, and $G^i(\Bx) := - \partial_{x^i}
\left( H^N_{\BJ} (\Bx) \right)$, for $i=1,\ldots,N$.
In Proposition \ref{thm-macro}, we characterize the long term behavior of the Langevin dynamics of this soft spherical model for a general class of functions $f$. We shall then choose an appropriate sequence of functions $f_n$, satisfying $\tilde \mu^N_{\b,h,\BJ,f_n} \ra \mu^N_{\b,h,\BJ}$, allowing us to derive, in Theorem  \ref{theo-sphere}, the limiting behavior of the hard spherical model.

We shall first prove that, fixing $f$, for a.e. disorder
$\BJ$, initial condition  $\Bx_0$ and Brownian path $\BB$,
there exists a unique strong solution of \req{interaction} for
all $t \geq 0$, whose law we denote by $\P^{N}_{\b,\Bx_0,{\BJ}}$.

We are interested in the time evolution for large $N$, of the {\em empirical covariance function}:
\begin{equation}\label{eq:covdef}
COV_N(s,t) = \oneN \sum_{i=1}^N \left[x_s^i x_t^i - \E_\BB [x_s^i]\E_\BB [x_t^i]\right]\,,
\end{equation}
where $\E_\BB[\cdot]$ represents the expectation with respect to the Brownian motion only (and not with respect to the Gaussian law of the couplings), under the quenched law $\P^{N}_{\b,\Bx_0,\BJ}$, as the
system size $N \to \infty$. In \cite{BDG2}, the authors have formally derived the limiting equations for the {\em empirical state correlation function}:
\begin{equation}\label{eq:cdef}
C_N(s,t):=\oneN \sum_{i=1}^N x^i_s x^i_t\,,
\end{equation}
in the absence of a magnetic field (i.e. $h=0$). The equations characterizing the limit as $N \ra \infty$ of $C_N(s,t)$ involve the analogous limit for the {\em empirical integrated response function}:
\begin{equation}\label{eq:chidef}
\chi_N(s,t):=\oneN \sum_{i=1}^N x^i_s B^i_t\,,
\end{equation}
and the limits are characterized as the unique solution of a system of two coupled integro-differential equations. The presence of the magnetic field requires us to consider also the {\em empirical averaged magnetization}:
\begin{equation}\label{eq:mdef}
M_N(s):=\oneN \sum_{i=1}^N x^{i}_s\,,
\end{equation}
the {\em averaged overlap}:
\begin{equation}\label{eq:udef}
L_N(s,t):=\oneN \sum_{i=1}^N \E_{\BB}\left[x^{i}_s\right]\E_{\BB} \left[x^{i}_t\right]\,,
\end{equation}
and the {\em empirical overlap}:
\begin{equation}\label{eq:qdef}
Q_N(s,t):=\oneN \sum_{i=1}^N x^{1,i}_s x^{2,i}_t\,,
\end{equation}
where $\left\{\Bx^{k}\right\}_s$, $k=1,2$ are two independent replicas, sharing the same couplings $\BJ$, with the noise given by two independent Brownian motions $\{\BB^k\}_s$. With these notations, our primary object of study, the empirical covariance can be written as:
$$COV_N(s,t) = C_N(s,t) - L_N(s,t)\,.$$

The empirical overlap defined in \eqref{eq:qdef} is the central quantity in the study of the static properties of the system (see \cite{Talagrand} for a comprehensive survey). Its dynamical properties were not rigurously analyzed until now. In the course of our proofs, we show that the limits as $N \ra \infty$ of $L_N$ (i.e. the averaged overlap - that we need to characterize in order to study the empirical covariance) and of $Q_N$ (i.e. the empirical overlap - that is interesting in its own right), coincide. Also, as opposed to the scenario analyzed in \cite{BDG2} (i.e. $h=0$), where the authors have characterized the dynamics via a coupled system of two integro-differential equations, the presence of the magnetic field will affect the symmetry of the spins and the dynamics of our system will be characterized via a coupled system of four integro-differential equations.

We shall analyze the solutions of the latter system in a non-perturbative high temperature region of the $(\beta,h)$-plane, rigorously establishing the existence of the so called {\em FDT} regime, where the Frequency Dissipation Theorem in statistical physics holds. We shall see that the phase plane diagram of the system in $(\beta, h)$ coordinates is the one shown in Figure \ref{figure-phase-plane} below. 

%\begin{figure}[ht]
%\includegraphics[scale=0.45]{phase_diagram.eps}
%\caption{The surfaces $\CL$.}
%\label{f1}
%\end{figure} 

\begin{figure}[h]
  \vskip 1.1in
  \noindent
  \begin{center}
  \begin{minipage}[t]{4.4in}
   \includegraphics[width=3.2in]{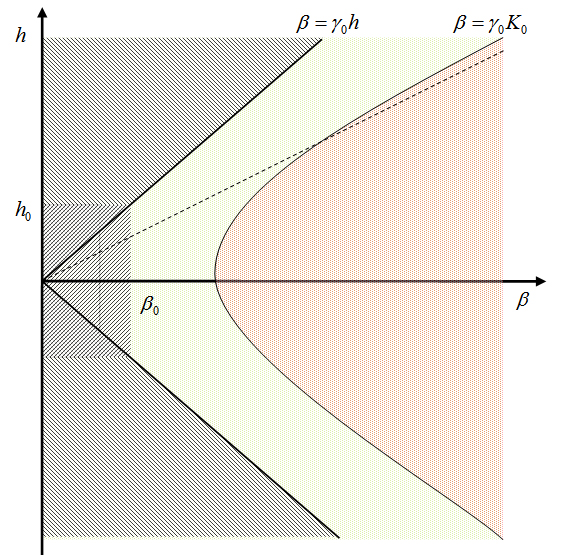}
   \label{figure-phase-plane}
   \caption{{\bf The Phase Plane Diagram:} The hashed region represents the area of applicability of Theorem \ref{FDT}, where we can rigorously prove the FDT regime, the light region represents the expected extend of the FDT regime and the red region, past the dynamical phase transition curve, represents the expected extent of the aging regime.}
  \end{minipage}
  \end{center}
\end{figure}

%%%%%%%%%%%%%%%%%%%%%%%%%%%%%%%%%%%%%%%%%%%%%%%%%%%%
%%%%%%%%%%%%%%%%%%%%%%%%%%%%%%%%%%%%%%%%%%%%%%%%%%%%

\section{Main Results}

We shall start by making the same assumptions on the initial conditions as in \cite{BDG2}. Namely, we assume that the initial condition $\Bx_0$ is
independent of the disorder $\BJ$, and the limits
\begin{equation}\label{eq:x0cond}
\lim_{N \to \infty} \E\left[C_N(0,0)\right] = C(0,0)\,,
\end{equation}
and
\begin{equation}\label{eq:x0cond1}
\lim_{N \to \infty} \E\left[M_N(0)\right] = M(0) \,,
\end{equation}
exists, and are finite. Further, we assume that
the tail probabilities $\P(|C_N(0,0)-C(0,0)|>x)$ and $\P(|M_N(0)-M(0)|>x)$ decay exponentially fast in $N$ (so the convergence $C_N(0,0) \to C(0,0)$ and $M_N(0) \to M(0)$
holds almost surely), and that
for each $k<\infty$, the sequence $N \mapsto \E [ C_N(0,0)^k ]$ and $N \mapsto \E [ M_N(0)^k ]$ is
uniformly bounded. Also, we will assume that each of the two replicas will have the same (random) initial conditions, hence $Q_N(0,0) = C_N(0,0)$.

Finally, consider the product probability space $\Ea_N
 = \R^N  \times \R^{d(N,m)}\times \Ca([0,T],\R^N) \times \Ca([0,T],\R^N)$
(here $T$ is a fixed time and $d(N,m)$
is the dimension of the space of the interactions $\bJ$),
equipped with the natural Euclidean norms for the finite dimensional parts,
i.e $(\Bx_0, \BJ)$, and the sup-norm for the Brownian motions $\BB^k$, $k=1,2$.
The space $\Ea_N$ is endowed with
the product probability measure $\P = \mu_N \otimes \gamma_N \otimes P_N \otimes P_N$,
where  $\mu_N$ denotes the distribution of $\Bx_0$, $\gamma_N$
is the (Gaussian) distribution of the coupling constants $\BJ$,
and $P_N$ is the distribution of the $N$-dimensional Brownian motion.
\begin{ass}\label{defconcentration}
For  $(\Bx_0, \BJ, \BB^1, \BB^2) \in \Ea_N$ we introduce the norms
$$\|(\Bx_0, \BJ, \BB^1, \BB^2)\|^2= \sum_{i=1}^N(x_0^i )^2
+ \sum_{p=1}^m \sum_{1 \leq i_1\ldots i_{p} \leq N}
(N^{\frac{p-1}{2}}J_{i_1\cdots i_{p}})^2
+ \sum_{k=1}^2\sup_{0 \leq t \leq T}\sum_{i=1}^N(B_t^{k,i})^2\,.$$
We shall assume that $\mu_N$ is such that
 the following concentration of measure property
holds
on $\Ea_N$;   there exists two finite positive
constants $C$ and $\alpha$, independent
on $N$, such that, if $V$
is a Lipschitz function on $\Ea_N$, with Lipschitz constant $K$, then
for all $\rho>0$,
$$\mu_N \otimes \gamma_N \otimes P_N \otimes P_N[ |V-\E[V]|\geq \rho ]
\leq C^{-1} \exp{\left(-C\left(\frac{\rho}{K}\right)^\alpha\right)}\,.$$
\end{ass}

Now, suppose that $f$ is a differentiable function on $\reals_+$ with
$f'$ locally Lipschitz, such that
\begin{equation}\label{eq:fcondu}
\sup_{\rho \geq 0} |f'(\rho)| (1+\rho)^{-r} < \infty
\end{equation}
for some $r<\infty$, and for some $A,\d>0$,
\begin{equation}\label{eq:fcond}
\inf_{\rho \geq 0} \{ f'(\rho) - A \rho^{m/2+\d-1} \} > -\infty
\end{equation}
(typically, $f(\rho)=\kappa (\rho-1)^r$ for some $r>m/2$
and $\kappa \gg 1$).
Then the normalization factor
$Z_{\b,h\BJ,f}= \int
e^{-\b H_{\BJ}^N (\Bx) - N f(N^{-1} \|\Bx\|^2) + h F^N(\Bx)} d\Bx$ is a.s.
finite
(by (\ref{eq:fcond})).

First, we shall show that, as $N \to \infty$
the functions $C_N(s,t)$, $\chi_N(s,t)$, $M_N(s)$, $Q_N(s,t)$ and $L_N(s,t)$ converge to
non-random continuous functions $C(s,t)$, $\chi(s,t)$, $M(s)$ and $Q(s,t) = L(s,t)$ that are
characterized as the solution of a system of coupled integro-differential equations. We denote by $\bD$ the upper half of the first quadrant, namely:
$$\bD:=\left\{(s,t)\in \R^2 : 0\leq t\leq s\right\}$$
Also, we denote by $\Ca_s^1$ the class of continuously differentiable symmetric functions of two variables and by $\Ca_s$ the class of continuous symmetric functions . These notations will be widely used and will appear through this work.
\begin{prop}\label{thm-macro}
Let $\psi(r)=\nu'(r)+r\nu''(r)$ and
\begin{equation}\label{eq:nudef}
\nu(r):=\sum_{p=1}^m \frac{a_p^2}{p!} r^p \,.
\end{equation}
Suppose $\mu_N$ satisfies hypothesis \ref{defconcentration} and $f$ satisfies \eqref{eq:fcondu} and \eqref{eq:fcond}. 
Fixing any $T<\infty$, as $N \to\infty$ the
random functions $M_N$, $\chi_N$, $C_N$, $Q_N$ and $L_N$ converge uniformly on $[0,T]^2$ (or $[0,T]$, whichever applies),
almost surely and in ${\bf L}^p$ with respect to $\Bx_0$, $\BJ$ and $\BB^k$, for $k = 1,2$,
to non-random functions $M(s)$, $\chi(s,t)=\int_0^t R(s,u) du$, $C(s,t)=C(t,s)$, $Q(s,t)=Q(t,s)$ and $L(s,t) = Q(s,t)$. Further, $R(s,t)=0$ for $t>s$, $R(s,s)=1$,
and for $s>t$ the absolutely continuous functions
$C$, $R$, $M$, $Q$, and $K(s)=C(s,s)$
are the unique solution in $\Ca^1(\R_+) \ts \Ca^1(\bD) \ts \Ca^1_s(\R_+^2) \ts \Ca^1_s(\R_+^2) \ts \Ca^1(\R_+)$
of the integro-differential equations:
{\allowdisplaybreaks
\begin{align}
\pars M(s) &\,=\,
 - f'(K(s)) M(s) + h + \b^2 \int_0^s M(u) R(s,u) \nu''(C(s,u)) du, && s\geq 0
 \label{eqM}\\
\pars_1 R(s,t) &\,=\,
 - f'(K(s)) R(s,t) + \b^2 \int_t^s
R(u,t) R(s,u) \nu''(C(s,u)) du, && s \geq t \geq 0\label{eqR}\\
\label{eqC}
\pars_1 C(s,t) &\,=\,
 - f'(K(s)) C(s,t) + \b^2 \int_0^s C(u,t) R(s,u) \nu''(C(s,u)) du && \\
&\qquad + \b^2 \int_0^t \nu'(C(s,u)) R(t,u) du + hM(t) + \mathbbm{1}_{s<t}R(t,s),&& s,t\geq 0
\nonumber \\
\pars_1 Q(s,t) &\,=\,
 - f'(K(s)) Q(s,t) +
\b^2 \int_0^s Q(u,t) R(s,u) \nu''(C(s,u)) du & &
\label{eqQ}\\
&\qquad + \b^2 \int_0^t \nu'(Q(s,u)) R(t,u) du + hM(t),&& s,t\geq 0
\nonumber\\
\pars K(s) &\,=\, -2 f'(K(s)) K(s) + 1 + 2\b^2
\int_0^s \psi(C(s,u)) R(s,u) du +2hM(s),&& s\geq 0
\label{eqZ}
\end{align}}
where the initial conditions
$K(0)=C(0,0)=Q(0,0)>0$ and $M(0)$ are determined by (\ref{eq:x0cond}) and (\ref{eq:x0cond1}), respectively. Moreover, $C(\cdot,\cdot)$ and $Q(\cdot,\cdot)$ are non-negative definite kernels, $K(s)\geq 0$, $|M(s)| \leq \sqrt{K(s)}$, for all $s\geq 0$ and
\begin{equation}\label{eq:rbd}
\left|\int_{t_1}^{t_2} R(s,u) du \right|^2 \le K(s)(t_2-t_1) \,,
\qquad 0 \leq t_1 \leq t_2  \leq s < \infty \,.
\end{equation}
\end{prop}

For every $r, L>0$, define the function:
\begin{align}
f(x)&:= f_{L,r}(x) = L(x-r)^2 + \frac{1}{4k} \left(\frac{x}{r}\right)^{2k} + \frac{\alpha h x}{r} \,, \quad k > m/4, \,
k \in \Z, \, L \geq 0 \,,
\label{eq:fdef}
\end{align}
that is easily seen to satisfy conditions \eqref{eq:fcondu} and \eqref{eq:fcond}. We will derive in Section
\ref{exactdynamics} the equations for the hard spherical constraint, by taking the limit $L \to \infty$. Notice that if there is no magnetic field (i.e. $h=0$), the equations for the correlation $C(\cdot,\cdot)$ and the response $R(\cdot,\cdot)$ will decouple from the magnetization, resulting with the system derived in \cite{DGM}.

\begin{theo}\label{theo-sphere}
For every $r>0$, let $(M_{L,r}, R_{L,r}, C_{L,r}, Q_{L,r}, K_{L,r})$ be the unique solution
of the system \eqref{eqM}-\eqref{eqZ} with potential $f_{L,r}(\cdot)$ as in \eqref{eq:fdef} and initial conditions $K_{L,r}(0) = Q_{L,r}(0,0) = r > 0$, $M_{L,r}(0) = \alpha\sqrt{r}$, $\alpha \in [0,1)$ and $R_{L,r}(t,t) = 1$ for every $t\geq 0$. Then, for any $T<\infty$, $(M_{L,r}, R_{L,r}, C_{L,r}, Q_{L,r}, K_{L,r})$ converges as $L \to \infty$, uniformly in $s,t \in [0,T]$, towards $(M, R, C, Q, K)$ that is the unique solution in $\Ca^1(\R_+) \ts \Ca^1(\bD) \ts \Ca^1_s(\R_+^2) \ts \Ca^1_s(\R_+^2) \ts \Ca^1(\R_+)$ of:
{\allowdisplaybreaks
\begin{align}
\pars M(s) &\,=\, - \mu(s) M(s) + h_r + \b^2 \int_0^s M(u) R(s,u) \nu''(C(s,u)) du, &s \geq 0&
\label{eqMs}\\
\pars_1 R(s,t) &\,=\, - \mu(s) R(s,t) + \b^2 \int_t^s R(u,t) R(s,u) \nu''(C(s,u)) du, &s\geq t \geq 0&
\label{eqRs} \\
\pars_1 C(s,t) &\,=\, -\mu(s) C(s,t) +
\b^2 \int_0^s C(u,t) R(s,u) \nu''(C(s,u)) du&&
\label{eqCs} \\
&\,+ \b^2 \int_0^t \nu'(C(s,u)) R(t,u) du + h_r M(t), &s\geq t \geq 0&
\nonumber\\
\pars_1 Q(s,t) &\,=\, -\mu(s) Q(s,t) +
\b^2 \int_0^s Q(u,t) R(s,u) \nu''(C(s,u)) du&&
\label{eqQs}\\
&\,+ \b^2 \int_0^t \nu'(Q(s,u)) R(t,u) du + h_r M(t), &s,t \geq 0&
\nonumber
\end{align}}
where $h_r = h$, $k = 1$ and
\begin{equation}\label{eqZs}
\mu(s) \,=\, \frac{1}{2r}
\left(k + 2\b^2
\int_0^s \psi(C(s,u)) R(s,u) du +2h_r M(s)\right)
\end{equation}
satisfying $M(0) = \alpha\sqrt{r}$, $C(t,t) = K(t) = r$, $R(t,t) = 1$, for all $t \geq 0$. Moreover, $C(\cdot,\cdot)$ and $Q(\cdot,\cdot)$ are non-negative definite kernels, with values in $[0,r]$,  $M(s)\in[0,\sqrt{r}]$, for all $s\geq 0$ , $R(s,t) \geq 0$ and
\begin{equation}\label{eq:rbd-new}
\left|\int_{t_1}^{t_2} R(s,u) du \right|^2 \le r(t_2-t_1) \,,
\qquad 0 \leq t_1 \leq t_2  \leq s < \infty \,.
\end{equation}
%----------------------------------
\end{theo}

The predicted structure of the solution is more complicated in the mixed spin case than in the pure spin one. However, we show in Section \ref{convergencepurespin} that as $r$ increases, only the highest level interactions will matter, effectively making the system behave like a pure spin one. (i.e. $\nu(x)$ is a monomial). Namely, we prove:

\begin{theo}\label{theo-sk}
For $\alpha\in(0,1)$ and $r > 0$, let $(M_r, R_r, C_r, Q_r)$ the unique solutions of \eqref{eqMs}-\eqref{eqZs} for $h_r=hr^{(m-1)/2}$, with initial conditions $M_r(0) = \alpha \sqrt{r}$, $C_r(t,t) = Q_r(0,0) = r > 0$, and $R_r(t,t) = 1$, for all $t \geq 0$. Then for any $T<\infty$, the appropriately scaled functions $\tilde M_r(s) = M_r(sr^{1-m/2})/\sqrt{r}$, $\tilde R_r(s,t) = R_r(sr^{1-m/2},tr^{1-m/2})$, $\tilde C_r(s,t) = C_r(sr^{1-m/2},tr^{1-m/2})/r$ and $\tilde Q_r(s,t) = Q_r(sr^{1-m/2},tr^{1-m/2})/r$, converge as $r \to \infty$, uniformly in $s,t \in [0,T]$, towards the solution of the corresponding pure spin system (i.e. towards the unique solution of \eqref{eqMs}-\eqref{eqZs} with $h_r = h$, $k=0$, $\tilde \nu(x)= a_m^2 (m!)^{-1} x^m$ and $\tilde \psi(x)= \tilde \nu'(x)+x \tilde \nu''(x)$, with initial conditions $M(0)=\alpha$, $C(t,t) =  Q(0,0) = 1$ and $ R(t,t)=1$ for all $t \geq 0$).

%----------------------------------
\end{theo}

In Section \ref{fdt_regime}, we will analyze the solutions of the system \eqref{eqMs}-\eqref{eqZs} in the high temperature region of the $(\beta,h)$-plane, formally establishing the existence of the FDT regime. The analysis is done in the absence of a random magnetic field (i.e. $\nu'(0)=0$). In this regime, the correlation, the response and the overlap are stationary for large $t$. Also, both the covariance and the response are decaying exponentially fast to $0$. The afore-mentioned region is $\{(\b,h)\,:\,\b\leq \b_0, h<h_0\}\,\bigcup\, \{(\b,h)\,:\,\b \leq \gamma_0 h\}$ for some non-trivial $\gamma_0$, $\b_0$ and $h_0$. The presence of the FDT regime for $\b$ small and $h$ small region comes as no surprise, in the light of the results proved in \cite{DGM}, where the authors have established similar results for $\b$ small and $h=0$. However, the occurrence of the same regime in the region bounded by $\frac{\beta}{h}<\gamma_0$ as well as the asymptotically linear relation between the critical inverse temperature and the intensity of the field is novel and represents an important contribution to the field.

\begin{theo}\label{FDT}
Suppose $\nu'(0)=0$. Let $(M, R, C, Q)$ be the unique solution of \eqref{eqMs}-\eqref{eqZs}, for $h_r=h$, $k=1/2$ and $r=1$, with initial conditions $R(t,t) = C(t,t) = Q(0,0) = 1$ and $M(0) = \alpha \in (0,1]$. Then there exist $\beta_0, h_0, \gamma_0 > 0$ such that if either $\gamma:=\frac{\beta}{h}<\gamma_0$ or $\beta<\beta_0$ and $h<h_0$, then for any $\tau\ge 0$,
\begin{align*}
\lim_{t\ra\infty} Q(t+\tau, t) &= Q^{\fdt}, & \lim_{t\ra\infty} M(t) &= M^{\fdt}=2h(1-Q^\fdt)\\
\lim_{t\ra\infty} C(t+\tau,t) &= C^{\fdt}(\tau), & \lim_{t\ra\infty} R(t+\tau, t) &= R^{\fdt}(\tau) = -2\pars C^\fdt(\tau)
\end{align*}
Furthermore, $M^\fdt, Q^\fdt, C^\fdt(\tau) \in [0,1]$, $R^\fdt(\tau) \geq 0$, $Q^\fdt$ is
only solution of the equation:
\begin{equation}
\label{FDTQ}
Q = 4(1-Q)^2[\beta^2 \nu'(Q) + h^2],\qquad Q\in [\left(1-(2h)^{-1}\right)\wedge 0,1]
\end{equation}
and $C^\fdt$ is the unique $[0,1]$-valued continuously differentiable solution of the equation:
\begin{equation}\label{FDTC}
 C'(s)=-\int_0^s\phi(C(v)) C'(s-v) dv - \frac{1}{2} ,\qquad C(0)=1 \,,
\end{equation}
for $\phi(x)=\frac{1}{2(1-Q^\fdt)} + 2\b^2 (\nu'(x)-\nu'(Q^\fdt))$. Moreover, $R^{\fdt}(\cdot)$ decays exponentially to zero at infinity and $C^{\fdt}(\cdot)$ converges exponentially fast to $Q^{\fdt}$
\end{theo}

Equation \eqref{FDTC} has been analyzed in detail in Proposition 1.4 of \cite{DGM}. The authors have shown that, for any choice of $\phi(\cdot)$ such that
\begin{equation}
\label{eq:existance-condition}
\sup_{x\in[0,1]}\{\phi(x)(1-x)\}\geq \frac{1}{2}
\end{equation}
the equation has an unique solution in $[0,1]$, that is decreasing, twice differentiable and converges as $s \ra \infty$ to $C^\infty = \sup\{x\in [0,1]\,|\, \phi(x)(1-x)\geq 1/2\}$. Furthermore, they show that the condition:
\begin{equation}\label{eq:exp-conv}
\phi(C^\infty)>\phi'(C^\infty) (1-C^\infty) \,,
\end{equation}
is necessary for the exponential convergence of $C'(s)$ to zero as $s \to \infty$ when $\phi(\cdot)$ is convex.

First, it is easy to check that for our $\phi(x)$ of Theorem \ref{FDT}, $\phi(Q^\fdt)(1-Q^\fdt) = 1/2$, hence \eqref{eq:existance-condition} is satisfied and furthermore, $C^\infty \geq Q^\fdt$. Setting $\b_c(h) \in (0,\infty)$ via
\begin{equation}\label{eqbetac}
\frac{1}{4\beta_c(h)^2}=\sup \left\{ \frac{(\nu'(x)-\nu'(Q^\fdt))(1-x)(1-Q^\fdt)}{x-Q^\fdt} : x \in (Q^\fdt,1] \right\} \,,
\end{equation}
it is easy to check that $C^\infty=Q^\fdt$ if
$\b<\b_c(h)$ whereas $C^\infty>Q^\fdt$ for $\b>\b_c(h)$.
Further, considering $x \to 0$ in \eqref{eqbetac}
we find that
\begin{equation}
\label{eqbetacbound}
\frac{1}{4\b_c(h)^2} \geq \nu''(Q^\fdt)(1-Q^\fdt)^2
\end{equation}
so, in particular, the condition \eqref{eq:exp-conv} then holds for any $\b<\b_c(h)$ (since in this case, as mentioned $C^\infty = Q^\fdt$). Furthermore, since $Q^\fdt$ is a solution of \eqref{FDTQ}, from \eqref{eqbetacbound} we get $\beta_c(h)^{-2}(\b_c(h)^2\nu'(Q^\fdt)+h^2) \geq Q^\fdt \nu''(Q^\fdt)$, so:
$$
\gamma_c(h)^2:=\left(\frac{\b_c(h)}{h}\right)^2 \leq \frac{1}{Q^\fdt\nu''(Q^\fdt)-\nu'(Q^\fdt)} \underset{h \ra \infty}{\longrightarrow} \frac{1}{\nu''(1)-\nu'(1)}
$$
This indicates that though the values of $\beta_0(h) \leq \gamma_0 h$ for which we have formally established the FDT regime in Theorem \ref{FDT} are quite small, they should match the predicted
dynamical phase transition point $\b_c(h)$ of our model. Furthermore, $0 < \underset{h \ra \infty}{\liminf}  \gamma_c(h) \leq \underset{h \ra \infty}{\limsup} \gamma_c(h) < \infty$, indicating that $\b_c(h)$ will be asymptotically linear in $h$. Figure \ref{figure-phase-plane} in the introduction summarizes all the information above.

%%%%%%%%%%%%%%%%%%%%%%%%%%%%%%%%%%%%%%%%%%%%%%%%%%%%
%%%%%%%%%%%%%%%%%%%%%%%%%%%%%%%%%%%%%%%%%%%%%%%%%%%%

\section{Limiting Soft Spherical Dynamics}

%%%%%%%%%%%%%%%%%%%%%%%%%%%%%%%%%%%%%%%%%%%%%%%%%%%%
%%%%%%%%%%%%%%%%%%%%%%%%%%%%%%%%%%%%%%%%%%%%%%%%%%%%

This section is dedicated to proving Proposition \ref{thm-macro}. The line of proof follows closely \cite{BDG2}, and references will be given, when appropriate. First, recall that:
\begin{equation}\label{eq:gdef}
G^i(\Bx) := - \partial_{x^i}
\Big( H^N_{\BJ} (\Bx) \Big) =\sum_{ p=1}^m \frac{a_p}{(p-1)!}
 \sum_{1\le i_1,\ldots, i_{p-1}
\le N} J_{i i_1\ldots i_{p-1}} x^{i_1}\ldots  x^{i_{p-1}} \,,
\end{equation}
We will start by introducing some notation. For $q_1, q_2 \in \{1,2\}$, define
\begin{align}
\nonumber
C_N^{q_1,q_2}(s,t):=\oneN \sum_{i=1}^N x^{q_1,i}_s x^{q_2,i}_t\,&,\qquad\qquad\qquad
K_N^{q_1,q_2}(s):=C_N^{q_1,q_2}(s,s)\,,\\
\nonumber
\chi_N^{q_1,q_2}(s,t):=\oneN \sum_{i=1}^N x^{q_1,i}_s B^{q_2,i}_t\,&,\qquad\qquad\qquad
M_N^{q_1}(s):=\oneN \sum_{i=1}^N x^{q_1,i}_s\,,\\
\nonumber
A_N^{q_1,q_2}(s,t):=\oneN \sum_{i=1}^N G^i(\Bx_s^{q_1}) x^{q_2,i}_t\,&,\qquad\qquad\qquad
F_N^{q_1,q_2}(s,t):=\oneN \sum_{i=1}^N G^i(\Bx_s^{q_1}) B^{q_2,i}_t\,,\\
\label{eq:hfdef}
R_N^{q_1}(s):=\oneN \sum_{i=1}^N G^i(\Bx_s^{q_1})\,&\qquad\qquad\qquad
W_N^{q_1}(s):=\oneN \sum_{i=1}^N B_s^{q_1,i}
\end{align}
and
\begin{align}
\nonumber
D_N^{q_1,q_2}(s,t)&:=-
f'(\E (K_N^{q_2,q_2}(t))) C_N^{q_1,q_2}(s,t) + A_N^{q_1,q_2}(t,s) \\
\nonumber
E_N^{q_1,q_2}(s,t)&:=-f'(\E(K_N^{q_1,q_1}(s))) \chi_N^{q_1,q_2}(s,t) + F_N^{q_1,q_2}(s,t)\\
\label{eq:dedef}
P_N^{q_1}(s)&:=-f'(\E (K_N^{q_1,q_1}(t))) M_N^{q_1}(s) + R_N^{q_1}(s)\,,
\end{align}
where for $q=1,2$, $\{\BB_s^{q}\}_{s\geq 0} = \{(B_s^{q,1},\dots,B_s^{q,N})\}_{s\geq 0}$, are two iid N-dimensional Brownian motions and
$\{\Bx_s^q\}_{s\geq 0} = \{(x_s^{q,1},\dots,x_s^{q,N})\}_{s\geq 0}$ are the two replicas sharing the same frustrations $\BJ$, with the noise given by the realization of the Brownian motions above. Also, when it is clear from the context that there is only one replica, for simplicity of the notation, the superscripts indicating the replica index will be omitted.

Also, in order to simplify the (already heavy) notations, we will
embed the constant $\beta$ into $\{a_p\}$ resulting with
$\beta G^j(\cdot) \mapsto G^j(\cdot)$ and then having $\beta=1$ in the stochastic differential
system \req{interaction}.
Adopting this convention, we will have from now on $\beta=1$.

%&&&&&&&&&&&&&&&&&&&&&&&&&&&&&&&&&&&&&&&&&&&&&&&&&
%&&&&&&&&&&&&&&&&&&&&&&&&&&&&&&&&&&&&&&&&&&&&&&&&&

\subsection{Strong Solutions and Self-Averaging}

%&&&&&&&&&&&&&&&&&&&&&&&&&&&&&&&&&&&&&&&&&&&&&&&&&
%&&&&&&&&&&&&&&&&&&&&&&&&&&&&&&&&&&&&&&&&&&&&&&&&&

First, by similar arguments to the ones employed in Proposition 2.1 of \cite{BDG2}, we will show that if $f'$ is locally Lipschitz, satisfying \eqref{eq:fcond}, then there exist an unique strong solution to \eqref{interaction}. Namely, we show:

\begin{prop}\label{existN}
Assume that $f'$ is locally Lipschitz, satisfying (\ref{eq:fcond}).
Then, for any $N\in \Z_+$,  almost any $\BJ$, initial condition $\Bx_0$
and Brownian path $\BB$,
there exists a unique strong solution to (\ref{interaction}).
This solution is also unique in law for almost any $\BJ$, and $\Bx_0$,
it is a \pm on $\Ca(\R_+,\R^N)$ which we denote $\P^N_{\Bx_0, \bJ}$.
Further, with
\begin{equation}\label{jnorm}
||\bJ||_\infty^{N}=\max_{1\le p\le m}\sup_{||{\bf u^i}||\leq
1, 1\le i\le p}\Big|
\sqrt{N}^{-1}\sum_{1\le i_k\le N, 1\le k\le p}N^{\frac{p-1}{2}}
 J_{i_1\cdots i_p}
u_{i_1}^1  \cdots u_{i_p}^p \Big| \,
\end{equation}
we have for $\d>0$ of (\ref{eq:fcond}), $q:=m/(2\d)+1$,
some $\kappa<\infty$, all $N$, $z>0$, $\bJ$, and $\Bx_0$, that
\begin{equation}\label{eq:conc}
\P^N_{{\Bx}_0,\bJ}\Big(\sup_{t\in\R^+}
K_N(t) \ge K_N(0) + \kappa (1 + \|\bJ\|_\infty^N + h)^q + z \Big)
\le  e^{-z N} \,.
\end{equation}
Consequently, for any $L>0$, there exists $z=z(L)<\infty$
such that
\begin{equation}\label{eq:ktail}
\P\Big(\sup_{t\in\R^+} K_N(t) \ge z \Big) \le  e^{-L N} \,.
\end{equation}
\end{prop}
\noindent{\bf Proof of Proposition \ref{existN}.} The proof follows the same lines as the proof of Proposition 2.1 of \cite{BDG2}. Namely, considering the truncated drift $b^M({\bf u})=(b^M_1(
{\bf u}),\ldots,b^M_N({\bf u}))$ given by
$b^M_i({\bf u})=G^i({\bf \phi_M(u)})-f'(N^{-1}|{\bf u}|^2 \wedge M)u^i + h$, where $\phi_M(\Bx) = \Bx$ when $\| \Bx \| \leq \sqrt {NM}$, we see that $\phi_M$ is globally Lipschitz, hence there exist an unique square-integrable strong solution ${\bf u}^{(M)}$ for the {\bf SDS}
$$
du^i_{t}= b^M_i ({\bf u}_t) dt  + dB^i_t
$$
(see, for example \cite[Theorems 5.2.5, 5.2.9]{Karatzas-shreve}).

Fixing $M$ and denoting ${\bf x}_t={\bf u}^{(M)}_{t \wedge \tau_M}$
and  $Z_s=2 N^{-1} \sum_{i=1}^N \int_0^{s \wedge \tau_M} x^i_t dB^i_t$, by
applying It\^o's formula for $C_N(t):=N^{-1}||{\bf x}_t||^2$ we see that
\begin{eqnarray}
\label{al2a}
C_N(s) \leq C_N(0) + 2
\sum_{p=1}^m \frac{a_p ||\bJ||_\infty^{N}}{(p-1)!}
\int_0^{s \wedge \tau_M} C_N(t)^{\frac{p}{2}}
 dt + Z_s
+ s \wedge \tau_M  \\
 -2 \int_0^{s \wedge \tau_M} f'(C_N(t)) C_N(t) dt + 2h\int_0^{s \wedge \tau_M} C_N(t)^{\frac{1}{2}}
 dt\,.
\nonumber
\end{eqnarray}
Since $x^{1 -{\frac{m}{2}}} f'(x) \to \infty$, it follows
from (\ref{al2a}) that there is an almost surely finite constant
$c(||\bJ||_\infty^N, h)$, independent of $M$, such that
\begin{equation}
C_N(s)
\leq C_N(0) + c(||\bJ||_\infty^N, h )s + Z_s
\label{al3a}
\end{equation}
As the quadratic variation of the martingale $Z_s$ is
$(4/N) \int_0^{s \wedge \tau_M} C_N(t) dt \leq 4 s N^{-1} M$,
applying Doob's inequality (c.f. \cite[Theorem 3.8, p. 13]{Karatzas-shreve})
for the exponential
martingale
% exp(mg-bracket)
$L_s^\lambda
=\exp(\lambda Z_s- 2 (\lambda^2/N) \int_0^{s \wedge \tau_M} C_N(t) dt)$
(with respect to the filtration $\{\Ha_t\}$ of $\BB_t$),
yields that
\begin{equation}\label{Doob}
\P\left( \sup_{s\le T} \{Z_s - 2 \int_0^s C_N(t) dt\}\ge z \right)
\le\P\left( \sup_{s\le T} L_s^N \geq e^{z N} \right) \le e^{-z N}
\,,
\end{equation}
for any $z>0$. Therefore, (\ref{al3a})
shows that with probability greater than $1- e^{-z N}$,
$$
C_N(s \wedge \tau_M) \le C_N(0) + c(||\bJ||_\infty^N, h) T + z
 +2 \int_0^{s \wedge \tau_M} C_N(t) dt \;,
$$
for all $s\le T$,
and by Gronwall's lemma then also
\begin{equation}
\sup_{t\le T}N^{-1} |{\bf u}^{(M)}_{t \wedge \tau_M} |^2
\le [C_N(0) + c(||\bJ||_\infty^N, h) T + z]e^{2 T} .
\label{new}
\end{equation}
Setting $z=M/3$,
for large enough $M$ (depending of $N$, $h$, $\bJ$, $\Bx_0$
and $T$ which are fixed here),
the right-side of (\ref{new}) is at most
$M/2$, resulting with
$$
\P(\tau_M\le T) \leq e^{-M N/3},
$$
where $\tau_M= \inf \{ t : ||{\bf u}^{(M)}_t|| \geq \sqrt{N M} \}$.
and hence that
\begin{equation}\label{al1a}
\sum_{M=1}^\infty \P\left(\tau_M\le T\right)<\infty.
\end{equation}
so establishing the existence of the solution after an application of the Borel-Cantelli lemma.

We also have  weak uniqueness
of our solutions for almost
all ${\bf J}$ since the restriction of
any weak solution to the stopped $\sigma$-field $\Ha_{\tau_M}$
for the filtration $\Ha_t$ of $\BB_t$ is unique. We denote
this unique  weak solution of (\ref{interaction})
by $\P^{N}_{\Bx_0,\BJ}$.

Turning to the proof of (\ref{eq:conc}),
by (\ref{eq:fcond}), for any $c>0$
there exists $\kappa<\infty$ such that for all $r,x \geq 0$,
$$2\left[ f'(x)x - r \sum_{p=1}^m \frac{a_p x^{\frac{p}{2}}}{(p-1)!} - h x ^{\frac{1}{2}}\right] -1
\ge c x- \kappa (1 + r + h)^q.$$
Taking $r= \|\bJ\|_\infty^{N}$,
we see that by (\ref{al2a}), for all $N$ and $s \geq 0$,
$$
C_N(s\wedge\tau_M) \leq C_N(0) -
\int_0^{s \wedge \tau_M}  \left[ c C_N(t) - \kappa (1+ \|\bJ\|_\infty^{N} + h)^q \right] dt
 + Z_s \,,
$$
where $(Z_s)_{s\ge 0}$ is a martingale with bracket
$( 4 N^{-1}\int_0^{s\wedge \tau_M} C_N(t) dt,s\ge 0)$.

By Doob's inequality (\ref{Doob}), with probability
at least $1-e^{- z N}$,
$$\sup_{u\le s \wedge \tau_M} Z_u \le 2 \int_0^{s \wedge \tau_M}
C_N(t)dt+z,$$
for all $s \geq 0$. Setting $c=3$ we then have that
\begin{eqnarray}
C_N(s\wedge\tau_M) &\leq& C_N(0) +z
-\int_0^{s \wedge \tau_M}  C_N(t) dt  + \kappa  (1+ \|\bJ\|_\infty^{N} + h)^q (s \wedge \tau_M) \,,
\label{al2bb}
\end{eqnarray}
so that by Gronwall's lemma,
$$C_N(s\wedge\tau_M)\le e^{-s \wedge \tau_M} (C_N(0)+z)
+ \kappa  (1+ \|\bJ\|_\infty^{N} + h)^q
\int_0^{s \wedge \tau_M} e^{-t} dt$$
from which the conclusion (\ref{eq:conc}) is obtained by
considering $M \to \infty$.

In view of the assumed
exponential in $N$ decay of the tail probabilities for $K_N(0)$
and the bound (B.7) of \cite{BDG2} on the corresponding probabilities for
$\|\bJ\|_\infty^N$ we thus get also the bounds of \req{eq:ktail}.
\hfill
$\Box$

The next is to extend the arguments in Propositions 2.2 - 2.8 of \cite{BDG2}, in order to show that any of the functions $A_N^{q_1,q_2}, F_N^{q_1,q_2}, \chi_N^{q_1,q_2}, C_N^{q_1,q_2}, W_N^{q_1}, R_N^{q_1}, M_N^{q_1}$ and $L_N$ self-averages for $N$ large. More precisely, we show that:

\begin{prop}\label{prop-self}
Suppose that
$\Psi: \R^\ell \to \R$ is locally Lipschitz with
$|\Psi(z)| \leq M \|z \|_k^k$ for some $M,\ell,k<\infty$,
and $Z_N \in \R^\ell$
is a random vector, where for $j=1,\ldots,\ell$,
the $j$-th coordinate of $Z_N$ is one of
the functions $A_N^{q_1,q_2}, F_N^{q_1,q_2}, \chi_N^{q_1,q_2}, C_N^{q_1,q_2}$, $W_N^{q_1}, M_N^{q_1}$ or $L_N$,
evaluated at some $(s_j,t_j) \in [0,T]^2$ (or at $s_j \in [0,T]$, whichever applies). Then,
$$
\lim_{N \to \infty}
\sup_{s_j,t_j} \left|\E [\Psi(Z_N)] - \Psi(\E [Z_N])\right| = 0 \;.
$$
\end{prop}

\noindent{\bf Proof of Proposition \ref{prop-self}.} The proof is structured as follows: first we show that $\E \big[ \sup_{s,t \leq T} |U_N(s,t)|^k \big]$ and $\E \big[ \sup_{s \leq T} |V_N(s)|^k \big]$ are bounded uniformly in $N$ and also that for any fixed $T < \infty$, the sequences $U_N(s,t)$ and $V_N(s)$ are pre-compact almost surely and in expectation with respect to the uniform
topology on $[0,T]^2$, respectively $[0,T]$. Here $U$ is any of the functions $C^{q_1,q_2}$, $F^{q_1,q_2}$, $\chi^{q_1,q_2}, A^{q_1,q_2}$ or $L$ and $V$ is one of the functions $M^{q_1}$ or $W^{q_1}$. The next step is to establish, similarly to Proposition 2.4 of \cite{BDG2}, that all the functions $U$ and $V$ above {\em self-averages}, namely:
\begin{align}
\nonumber
&\sum_N \P \left[ \sup_{s,t \le T} |U_N(s,t)- \E[U_N(s,t)]| \geq \rho\right] < \infty\\
\label{eq:asself}
&\sum_N \P \left[ \sup_{s\le T} |V_N(s)- \E[V_N(s)]| \geq \rho\right] < \infty
\intertext{implying by the uniform moment bounds on $\|U_N\|_\infty$ and $\|V_N\|_\infty$ that we have just established, that:}
\nonumber
&\lim_{N \rightarrow \infty} \sup_{s,t \leq T} \E \left[ \left|U_N(s,t)-\E [U_N(s,t)]\right|^{2} \right]=0\\
\label{eq:l2self}
&\lim_{N \rightarrow \infty} \sup_{s \leq T} \E \left[ \left|V_N(s)-\E [V_N(s)]\right|^{2} \right]=0
\end{align}
The final step is to establish the claim of the proposition, by using \eqref{eq:l2self} and the uniform bounds on the moments that we have just established.

By our hypothesis, the mapping $N \mapsto \E [ K_N(0)^k ]$ is bounded. Since both replicas have the same starting point $K_N^{q,q}(0) = K_N(0)$, for $q\in\{1,2\}$. Also,
by the estimate (B.6) of Appendix B, of \cite{BDG2},
\begin{equation}\label{eq:ger2}
\sup_N \E \left[ (||\bJ||_\infty^{N})^k \right] < \infty \,,
\end{equation}
for any $k<\infty$, for the norm $||\bJ||_\infty^{N}$ of \eqref{jnorm}, the bound \eqref{eq:conc} immediately implies that for each $k<\infty$, and any $q\in\{1,2\}$ also
\begin{equation}\label{eq:conc2}
\sup_{N} \E \left[ \sup_{t\in\R^+} [K_N^{q,q}(t)]^{k} \right] <\infty\,.
\end{equation}
Define $\|V_N\|_\infty:=\sup\{V_N(t): 0\leq t \leq T\}$ and $\|U_N\|_\infty:=\sup\{U_N(s,t): 0\leq s,t \leq T\}$. Also let $B_N^q(t):=\oneN \sum_{i=1}^N (B_t^{q,i})^2$, $G_N^q(t):=\oneN \sum_{i=1}^N (G^i(\Bx^q_t))^2$ and $L_N(t):=\oneN \sum_{i=1}^N (\E_{\BB}[x^i_t])^2$. A key result is the bound:
\begin{eqnarray}
\label{ger:moments}
&&\sup_N \E \left[ (||\bJ||_\infty^{N})^k \right] + \sup_N \E[\|L_N\|_\infty^k] +
\sup_N\E[\|K_N\|_\infty^k]+\sup_N\E[\|B_N\|_\infty^k] + \sup_N\E[\|G_N\|_\infty^k]<\infty \,,
\end{eqnarray}
for every fixed $k$, where we have dropped the replica index (since we are taking the expected value anyway). Indeed,
the bounds on $\|\bJ\|_\infty^N$ and
$\|K_N^{q,q}\|_\infty$ are already obtained in \req{eq:ger2} and
\req{eq:conc2},
and by Lemma 2.2 of \cite{BDG2} we have that
\begin{equation}\label{eq:gubd}
(G_N^q(t))^{\frac{1}{2}}\leq c ||\bJ||_\infty^N [1+ K_N^{q,q}(t)^{\frac{m-1}{2}}]\,,
\end{equation}
yielding by \req{eq:ger2} and \req{eq:conc2}
the uniform moment bound on $\|G_N^q\|_\infty$. Also, by Jensen's inequality, $\E [\|L_N\|^k_\infty] \leq \E [\|K_N\|_\infty^k]$ and finally, the exponential tails of $B_N^q$ (c.f. \cite[(2.16)]{BDG2}),
will provide an uniform bound for each moment of $\|B_N^q\|_\infty$, thus concluding the derivation of \req{ger:moments}.

Similarly, by \req{eq:ktail}, \req{eq:gubd}, the exponential tails of $B_N^q$ mentioned above and the exponential tails of $\|\bJ\|^N_\infty$ (c.f \cite[(B.7)]{BDG2}), we have for each $L>0$ the bound:
\begin{equation}\label{ger:moments2}
\P \left(\|\bJ\|^N_\infty + \|L_N\|_\infty + \sum_{q=1}^2\left[\|K_N^{q,q}\|_\infty
+ \|B_N^q\|_\infty + \|G_N^q\|_\infty\right]\ge M\right) \le e^{-LN} \,.
\end{equation}
will hold for some $M=M(L) < \infty$ and for all $N$. Applying Cauchy-Schwartz inequality to
$U_N$ and $V_N$ and using the estimates \eqref{ger:moments} and \eqref{ger:moments2}, we see that $\E \big[ \sup_{s,t \leq T} |U_N(s,t)|^k \big]$ and $\E \big[ \sup_{s \leq T} |V_N(s)|^k \big]$ are bounded uniformly in $N$. The argument is similar to the one employed in Proposition 2.3 of the cited paper.

With the previous controls
on  $ \| U_N\|_\infty$ and $\| V_N\|_\infty$ already
established, by the Arzela-Ascoli theorem, the
pre-compactness of $U_N$, respectively $V_N$ follows by showing that they are
equi-continuous sequences. We notice that such
$U_N(s,t)$ and $V_N(s)$ are all of the form $\oneN \sum_{i=1}^N a^i_s b^i_t$
hence,
\begin{eqnarray}
\nonumber
|  U_N(s,t) -  U_N(s',t') | &\leq&
  \oneN \sum_{i=1}^N  |a^i_s-a^i_{s'}| |b^i_t|
+  \oneN \sum_{i=1}^N |a^i_{s'}| |b^i_t-b^i_{t'}| \\
\label{cauchyu}
&\leq&
\left[\oneN \sum_{i=1}^N  |a^i_s-a^i_{s'}|^2 \right]^{1/2}
\left[\oneN \sum_{i=1}^N  |b^i_t|^2\right]^{1/2}  +
\left[\oneN \sum_{i=1}^N  |b^i_t-b^i_{t'}|^2\right]^{1/2}
\left[\oneN \sum_{i=1}^N  |a^i_{s'}|^2\right]^{1/2} \,.
\end{eqnarray}
and the same is true also for $|V_N(s) -  V_N(s')|$, where the functions $\ba_s$ and $\bb_s$ are either $\Bx_s^{q}$, $\BB_s^{q}$, $G(\Bx_s^q)$, for some $q\in \{1,2\}$, $\E_{\BB}[\Bx_s]$ or $1$. So, in view of \req{ger:moments}
and \req{ger:moments2}, it suffices to show that for
any $\e>0$, some function $L(\d,\e)$ going
to infinity as $\d$ goes to zero and all $N$,
\begin{eqnarray}
\nonumber
&&\P\left(\sup_{|t-t'|<\delta}  \left[ \oneN \sum_{i=1}^N |b^i_t-b^i_{t'}|^2
\right] >\e\right)\le e^{-L(\d,\e)N}\\
\label{tailbds}
&&\sup_{|t-t'|<\delta}  \E\left[ \oneN \sum_{i=1}^N |b^i_t-b^i_{t'}|^2
\right]\le L(\d,\e)^{-1}\,,
\end{eqnarray}
for $\bb=\Bx^q$, $\BB^q$, $G(\Bx^q)$ and $\E_\BB[\Bx]$. Obviously, this holds for
$\bb=\BB^q$. Also, since by \req{interaction}
$$|x^{q,i}_t-x^{q,i}_{t'}|\le
 |B^{q,i}_t-B^{q,i}_{t'}| + \|f'(K^{q,q}_N)\|_\infty
\int_t^{t'}|x^{q,i}_u|du
+\int_t^{t'}|G^i (\Bx^q_u)|du + h(t'-t)\,.
$$
we get, by \req{eq:fcondu}, for some universal constant $\rho_1<\infty$, all $t,t'$ and $N$,
\begin{eqnarray*}
\oneN \sum_{i=1}^N |x^{q,i}_t-x^{q,i}_{t'}|^2 &\leq&
\frac{4}{N} \sum_{i=1}^N |B^{q,i}_t-B^{q,i}_{t'}|^2 \\
&& + 4 |t-t'|^2 \Big[ \rho_1 (1+\|K_N^{q}\|_\infty)^{2r}\|K_N^{q}\|_\infty
+\|G_N^{q}\|_\infty + h^2\Big]
\end{eqnarray*}
hence by the bounds established on $\|G_N^{q}\|_\infty$ and $\|K_N^{q}\|_\infty$, we establish \eqref{tailbds} for $\bb=\Bx^q$. An application of Jensen's inequality will imply the same result for $\bb = \E_{\BB}[\Bx]$. Using the results in Lemma 2.2 of \cite{BDG2}, we can now establish \eqref{tailbds} for $\bb=G(\Bx^q)$, thus concluding the equi-continuity of $U_N$ and $V_N$, hence the fist step of the proof. Note that we have actually shown a stronger result that we will use later, namely that for all $\e>0$ there exists $\widetilde{L}(\d,\e) \to \infty$ for $\d \to 0$, such that for all $N$,
\begin{eqnarray}
\nonumber
\P\left(\sup_{|s-s'|+|t-t'|<\d} |U_N(s,t)-U_N(s',t')|>\e\right) \leq e^{-
\widetilde{L}(\d,\e) N}\\
\label{tailbdsgen}
\P\left(\sup_{|s-s'|<\d} |V_N(s)-V_N(s')|>\e\right) \leq e^{-
\widetilde{L}(\d,\e) N}
\end{eqnarray}
and also
\begin{eqnarray}
\nonumber
\sup_{|s-s'|+|t-t'|<\d} \left|\E [U_N(s,t)]- \E [U_N(s',t')]\right| \leq
\widetilde{L}(\d,\e)^{-1}\\
\label{tailbdsgen1}
\sup_{|s-s'|<\d} |\E[V_N(s)]- \E [V_N(s')]| \leq
\widetilde{L}(\d,\e)^{-1} \,.
\end{eqnarray}

The next step, as mentioned earlier is to establish \eqref{eq:asself} and \eqref{eq:l2self}. We will use the same approach as in the proof of Proposition 2.4 of \cite{BDG2}, by applying the estimate in Lemma 2.5 to $U_N(s,t)$ and $V_N(s)$, respectively,   for any fixed pair of times $s,t$. For every $M<\infty$ and any $N$, define the subset:
\begin{align*}
\CL_{N,M} &= \Big\{
(\Bx_0, \BJ, \BB^1, \BB^2)\in \Ea_N \,:\, ||\BJ||^{N}_\infty + \|L_N\|_\infty + \sum_{q=1}^2\left[\|B_N^q\|_\infty + \| K_N^{q,q}\|_\infty + \| G_N^q \|_\infty\right]
\le M \;\Big\}
\end{align*}
of $\Ea_N$.
For $M$ sufficiently large, the probability of the complement set
$\CL_{N,M}^c$ decays exponentially in $N$ by \req{ger:moments2}. Since the uniform moment bounds for the functions $U_N(s,t)$ and $U_N(s)$ has been established, as well as the stated pointwise bound in $\CL_{N,M}$, the only other ingredient that we need to be able to apply the bound in Lemma 2.5 in the cited paper is the Lipschitz constant
of $U_N$ and $V_N$ on $\CL_{N,M}$.

To this end, let $\bx^q,\widetilde{\bx^q}$ be the two strong solutions
of (\ref{interaction}) constructed from $(\Bx_0,\BJ,\BB^1,\BB^2)$
and $(\widetilde{\Bx}_0,\widetilde{\BJ}, \widetilde{\BB}^1, \widetilde{\BB}^2)$, respectively. If $(\Bx_0,\BJ,\BB^1,\BB^2)$
and $(\widetilde{\Bx}_0,\widetilde{\BJ},\widetilde{\BB}^1, \widetilde{\BB}^2)$
are both in $\CL_{N,M}$, then
\begin{align}
\label{xislip}
\sup_{t\le T}\frac{1}{N}\sum_{1\le i\le N}
|x^{q,i}_t-\widetilde{x}^{q,i}_t|^2 &\le \frac{D_o(M,T)}{N}\|(\Bx_0, \BJ, \BB^q)
- (\widetilde{\Bx}_0,\widetilde{\BJ},\widetilde{\BB^q})\|^2 \\
\nonumber
&\le \frac{D_o(M,T)}{N}\|(\Bx_0, \BJ, \BB^1, \BB^2)
- (\widetilde{\Bx}_0,\widetilde{\BJ},\widetilde{\BB^1}, \widetilde{\BB^2})\|^2\,,
\end{align}
for some $D_o(M,T)$ independent of $N$, where the first inequality is due to Lemma 2.6 of \cite{BDG2}. Now, equipped with \eqref{xislip}, we can easily show the desired Lipschitz estimate for all of the functions of interest $U_N(s,t)$ and $V_N(s)$, namely:
\begin{equation}\label{eq:lippru}
\sup_{s,t \le T}|U_N(s,t)- \widetilde{U_N}(s,t)|
\le \frac{D(M,T)}{\sqrt{N}}
\|(\Bx_0, \BJ, \BB^1, \BB^2) - (\widetilde{\Bx}_0,\widetilde{\BJ},\widetilde{\BB}^1,\widetilde{\BB}^2)\|\,,
\end{equation}
and
\begin{equation}\label{eq:lipprv}
\sup_{s\le T}|V_N(s)- \widetilde{V_N}(s)|
\le \frac{D(M,T)}{\sqrt{N}}
\|(\Bx_0, \BJ, \BB^1, \BB^2) - (\widetilde{\Bx}_0,\widetilde{\BJ},\widetilde{\BB}^1,\widetilde{\BB}^2)\|\,,
\end{equation}
where the constant $D(M,T)$ depends only on $M$ and $T$ and not on $N$. Indeed, since every $U_N(s,t)$ and every $V_N(s)$ is of the form $\oneN \sum_{i=1}^N a^i_s b^i_t$, then \eqref{cauchyu} will hold, with the functions $\ba_t$ and $\bb_t$ being one of $\Bx_t^{q}$, $\BB_t^q$, $G(\Bx_t^q)$, $\E_{\BB}[\Bx]$ or $1$. By the same proof as the one employed in Lemma 2.7 of \cite{BDG2}, we see that:
\begin{align*}
\left[\oneN \sum_{i=1}^N  |G^i(\Bx^q_s)-\tilde{G}^i(\tilde{\Bx}^q_s)|^2 \right]^{1/2} &\le \frac{C(M,T)}{\sqrt{N}} \|(\Bx_0, \BJ, \BB^1, \BB^2) - (\widetilde{\Bx}_0,\widetilde{\BJ},\widetilde{\BB}^1, \widetilde{\BB}^2)\|\,.
\intertext{and}
\left[\oneN \sum_{i=1}^N  |G^i(\Bx^q_t)|^2\right]^{1/2} &\le
c ||\bJ||_\infty^N (1+M^{m-1}) \le C(M) \,.
\intertext{Also, Jensen's inequality applied to \eqref{xislip} shows:}
\sup_{t\le T}\frac{1}{N}\sum_{1\le i\le N}
|\E_{\BB}[x^{i}_t]-\E_{\BB}[\widetilde{x}^{i}_t]|^2 &\le \frac{D_o(M,T)}{N}\|(\Bx_0, \BJ, \BB^1, \BB^2)
- (\widetilde{\Bx}_0,\widetilde{\BJ},\widetilde{\BB^1}, \widetilde{\BB^2})\|^2\,,
\end{align*}

The last three bounds, together with the \eqref{xislip} plugged into equation \eqref{cauchyu} and is's analogue for $V$, will show the Lipschitz bounds \eqref{eq:lippru} and \eqref{eq:lipprv}, whenever $(\Bx_0,\BJ,\BB^1,\BB^2)$
and $(\widetilde{\Bx}_0,\widetilde{\BJ},\widetilde{\BB}^1, \widetilde{\BB}^2)$
are both in $\CL_{N,M}$.

As noticed before, we have all the ingredients for applying Lemma 2.5 of \cite{BDG2} to $V_N:=U_N(s,t)$ and $V_N:=V_N(s)$, for any fixed $s,t \leq T$, yielding:
\begin{eqnarray}
\label{concentration}
\P[ |V_N-\E[V_N]|\geq \rho ] &\leq& C^{-1}
\exp{\left(- C\left({\frac{\rho}{2 D(M(L))}}\right)^\alpha
N^{\frac{\alpha}{2}}\right)}\\
\nonumber
&& + 4 (K+M(L)) \rho^{-1} e^{-LN/2} + e^{-NL} \;.
\end{eqnarray}
for constants $K$ and $D=D(M(L),T)$ independent of $s,t$, $\rho$ and $N$.
Consequently, by the union bound, for any finite subset $\Aa$ of $[0,T]^2$ and $\Ba$ of $[0,T]$ and any $\rho>0$, the sequences
$N \mapsto \P[ \sup_{(s,t) \in \Aa} |U_N(s,t)- \E[U_N(s,t)]| \geq \rho/3]$
and $N \mapsto \P[ \sup_{s \in \Ba} |V_N(s)- \E[V_N(s)]| \geq \rho/3]$ are
summable. Recalling \eqref{tailbdsgen} and \eqref{tailbdsgen1}, we choose $\d>0$ small enough so that
$\widetilde{L}(2\d,\rho/3) > 3/\rho>0$,
we thus get \req{eq:asself} by considering the finite subsets
$\Aa=\{(i\d,j\d): i,j=0,1,\ldots,T/\d\}$ and respectively $\Ba=\{i\d: i=0,1,\ldots,T/\d\}$.

Now, we have all the ingredients needed for finalizing the proof. For each $r \geq R$
let $c_r$ denote the finite Lipschitz constant of
$\Psi(\cdot)$ (with respect to $\|\cdot\|_2$),
on the compact set $\Gamma_r := \{ z : \|z\|_k \leq r \}$.
Then,
\begin{eqnarray*}
\left|\E [\Psi(Z_N)] - \Psi(\E [Z_N])\right|&\leq&
\E |\Psi(Z_N) - \Psi(\E [Z_N])| \mathbbm{1}_{Z_N \in \Gamma_r} \\
&& + \E |\Psi(Z_N)| \mathbbm{1}_{Z_N \notin \Gamma_r}
+|\Psi(\E [Z_N])| \P [Z_N \notin \Gamma_r]  \\
&\leq& c_r \E[\|Z_N - \E [Z_N] \|_2] + 2 \ell M r^{-k} \E \|Z_N \|_{k}^{2k} \,.
\end{eqnarray*}
We have by \req{eq:l2self} and
the uniform moment bounds of $U_N(s,t)$ and $V_N(s)$ that
$\sup_{s_j,t_j} \E[\|Z_N - \E Z_N \|_2] \to 0$
as $N \to \infty$, while
$c' = \sup_{s_j,t_j,N} \E \|Z_N \|_{k}^{2k} < \infty$,
implying that:
$$
\lim_{N \to \infty}
\sup_{s_j,t_j} |\E [\Psi(Z_N)] - \Psi(\E [Z_N])| \leq 2 c' \ell M r^{-k}
\,,
$$
which we make arbitrarily small by taking $r \to \infty$.
\hfill
$\Box$

Notice that, since $L_N(s,t)$ and $Q_N(s,t)$ have the same first moment, for every $s$ and $t$, the above proposition implies that any limit point of $L_N(s,t)$ is also a limit point of $Q_N(s,t)$.

%&&&&&&&&&&&&&&&&&&&&&&&&&&&&&&&&&&&&&&&&&&&&&&&&&
%&&&&&&&&&&&&&&&&&&&&&&&&&&&&&&&&&&&&&&&&&&&&&&&&&

\subsection{Getting the Limiting Equations}

%&&&&&&&&&&&&&&&&&&&&&&&&&&&&&&&&&&&&&&&&&&&&&&&&&
%&&&&&&&&&&&&&&&&&&&&&&&&&&&&&&&&&&&&&&&&&&&&&&&&&

The key step of the proof of Proposition \ref{thm-macro} is summarized by
\begin{prop}\label{prop-macro}
Fixing any $T<\infty$, any limit point of the
sequences
$\E [M_N]$, $\E [\chi_N]$, $\E [C_N]$ and $\E [Q_N] = \E [C^{1,2}_N]$ with respect to
uniform convergence on $[0,T]^2$, satisfies the integral
equations
{\allowdisplaybreaks
\begin{align}
M(s) \,=&\, M(0) + hs + \int_0^s P(u)du,\label{eqM1}\\
\chi(s,t) \,=&\, s\wedge t+ \int_0^s E(u,t)du,\label{eq:chi}\\
C(s,t) \,=&\, C(s,0)+\chi(s,t)+\int_0^t D(s,u)du + htM(s),\label{eqC1}\\
Q(s,t) \,=&\, Q(s,0)+\int_0^t H(s,u)du + htM(s),\label{eqQ1}\\
P(t) \,=&\, -f'(C(t,t))M(t) + \nu'(C({t},t)) M(t) - \nu'(C(0,t)) M(0)
\label{eqP}\\
&\, - \int_0^{t} \nu'(C (t,u)) P (u) du - \int_0^{t} M(u) \nu''(C(t,u)) D (u,t) du
\nonumber \\
&\, - h\left[M(t)\int_0^{t} M(u)\nu''(C(t,u))du + \int_0^{t} \nu'(C(t,u))du\right]
\nonumber \\
E(s,t)  \,=&\, -f'(C(s,s))\chi(s,t) + \chi(s,t) \nu'(C(s,s)) - h Q(s)\int_0^s \nu''(C(s,u))\chi(u,t) du
\label{eqE}\\
&\, - \int_0^s\chi(u,t)\nu''(C(s, u))D(s,u) du - \int_0^{t \wedge s} \nu'(C(s,u)) du - \int_0^s  \nu'(C(s, u)) E(u,t) du,
 \nonumber\\
D(s,t) \,=&\, C(s,t \vee s)\nu'(C(t \vee s,t)) - C(s,0)\nu'(C(0,t)) - f'(C(t,t))C(t,s)
\label{eqD}\\
&\, - \int_0^{t \vee s} \nu'(C(t, u))D(s,u) du
- \int_0^{t \vee s} C(s,u)\nu''(C(t,u)) D(t,u) du
\nonumber\\
&\, - h\left[M(t)\int_0^{t \vee s} C(s,u)\nu''(C(t,u))du + M(s)\int_0^{t \vee s} \nu'(C(t,u))du\right]
\nonumber\\
H(s,u) \,=&\, -f'(C(t,t))Q(t,s) + X(s,u) + Y(s,u)\label{eqDQ}\\
X(s,t) \,=&\, Q(s,t \vee s)\nu'(C(t \vee s,t)) - Q(s,0)\nu'(C(0,t))
\nonumber\\
&\, - \int_0^{t \vee s} \nu'(C(t, u))H(s,u) du - \int_0^{t \vee s} Q(s,u)\nu''(C(t,u)) D(t,u) du
\label{eqFrep}\\
&\, - h\left[M(t)\int_0^{s \vee t} Q(s,u)\nu''(C(t,u))du + M(s)\int_0^{s \vee t} \nu'(C(t,u))du\right]
\nonumber
\end{align}}
and $Y(s,y)$ is defined similarly to $X(s,t)$, with the roles of $C$ and $Q$ and respectively $D$ and $H$ reversed, in the space of bounded continuous functions on $[0,T]^2$, subject to the symmetry conditions $C(s,t)=C(t,s)$ and $Q(s,t)=Q(t,s)$ and the
boundary conditions $E(s,0)=0$ for all $s$, and
$E(s,t)=E(s,s)$ for all $t \geq s$.
\end{prop}

We will then show in Lemma \ref{lem-diff} that every solution of \eqref{eqM1}-\eqref{eqFrep} is necessarily a solution of \eqref{eqM}-\eqref{eqZ}, thus allowing us to conclude the proof of Proposition \ref{thm-macro}, upon showing, in Proposition \ref{uniqueness}, the uniqueness of the solution of \eqref{eqM}-\eqref{eqZ}.

\begin{lem}\label{lem-diff}
Fixing $T<\infty$, suppose $(M, \chi, C, Q, D, E, P, H)$ is a solution
of the integral equations \req{eqM1}--\req{eqFrep}
in the space of continuous functions on $[0,T]^2$
subject to the symmetry conditions $C(s,t)=C(t,s)$ and $Q(s,t)=Q(t,s)$ and the
boundary conditions $E(s,0)=0$ for all $s$, and
$E(s,t)=E(s,s)$ for all $t \geq s$.
Then, $\chi(s,t)=\int_0^t R(s,u)du$ where
$R(s,t)=0$ for $t>s$, $R(s,s)=1$ and
for $T \geq s>t$, the bounded and
absolutely continuous functions
$M, C, R, Q$ and $K(s)=C(s,s)$ necessarily satisfy
the integro-differential equations \req{eqM}--\req{eqZ}.
\end{lem}

\begin{prop}\label{uniqueness}
Let $T\ge 0$. There exists at most one solution $(M,R,C,Q,K)$ in
$\Ca^1(\R_+) \ts \Ca^1(\bD) \ts \Ca^1_s(\R_+^2) \ts \Ca_s^1(\R_+^2) \ts \Ca^1(\R_+)$ to \eqref{eqM}-\eqref{eqZ} with $R(s,s) = 1$, $C(s,s) = K(s)$, $\forall s\ge 0$,
$C(0,0) = Q(0,0) = K(0)$ and $M(0)$ known.
\end{prop}

We will now change the notations in \cite{BDG2}, denoting in short $\hh U_N^{q_1,q_2}:=\E[U_N^{q_1,q_2}]$, whenever $U$ is one of the functions of interest $A, C, F, K, \chi, D, E$ and respectively, $\hh V_N^{q}:=\E[V_N^{q}]$, whenever $V$ is one of the functions $M, P$ or $R$. As before, when there is only one replica present, we will drop the index superscript (for example $\hh C = \hh C^{1,1}$).

Recall the integrated form of the equation \eqref{interaction}, for $q = 1,2$ and $i=1,\dots,N$:
\begin{equation}
\label{integratedSDE}
x^{q,i}_s=x^{q,i}_0+B^{q,i}_s -\int_0^s f'(K_N^{q,q}(u))x^{q,i}_u du +\int_0^s
G^i(\bx^{q}_u) du + hs
\end{equation}
From now on, we will write $X \equiv Y$ whenever the random variables $X$ and $Y$ have the same law and $a_N\simeq b_N$ when
$a_N(\cdot,\cdot)-b_N(\cdot,\cdot) \to 0$ (or $a_N(\cdot)-b_N(\cdot) \to 0$) as $N \to \infty$,
uniformly on $[0,T]^2$ (or $[0,T]$, whichever applies). Let us denote by $\hh Q_N(s,t) := \hh C_N^{1,2}(s,t) = \hh C_N^{2,1}(s,t)$ (since $C_N^{1,2}(s,t) \equiv C_N^{2,1}(s,t)$). Applying Proposition \ref{prop-self}
(for $\Psi(z)=z_1 f'(z_2)$ whose polynomial growth is guaranteed by
our assumption \req{eq:fcondu}), we deduce that:
$$\E\left[ f'(K_N^{q_1,q_2}(u)) U_N^{q_3,q_4}(u,t)\right]\quad\simeq\quad f'(\hh K_N^{q_1,q_2}(u)) \hh U_N^{q_3,q_4}(u,t)$$
and
$$\E\left[ f'(K_N(u)) M_N(u)\right] \quad\simeq\quad f'(\hh K_N(u)) \hh M_N(u)$$
whenever $U$ is one of the functions $C$ or $\chi$. Hence, upon multiplying \eqref{integratedSDE} with $x^{q,i}_t$, $B^{q,i}_t$, $x^{3-q,i}_t$ and $1$, respectively,  followed by averaging over $i$ and taking the expected value, we get that for any $s,t\in\R^+$,
{\allowdisplaybreaks
\begin{eqnarray}
\hh M_N(s)&\simeq& \hh M_N(0)+ hs
-\int_0^s f'(\hh K_N(u)) \hh M_N(u)du
+\int_0^s \hh R_N (u) du\label{eqMa}\\
\hh \chi_N(s,t)&\simeq& \hh \chi_N(0,t) +
t\wedge s-\int_0^s f'(\hh K_N(u)) \hh \chi_N(u,t)du + \int_0^s \hh F_N (u,t) du \label{eqchi} \\
\hh C_N(s,t)&\simeq& \hh C_N(0,t)+\hh \chi_N(t,s)
-\int_0^s f'(\hh K_N(u)) \hh C_N(u,t)du + \int_0^s \hh A_N (u,t) du + hs\hh M_N(t)\label{eqCa}\\
\hh Q_N(s,t) &\simeq& \hh Q_N(0,t)-\int_0^s f'(\hh K_N(u))\hh Q_N(u,t)du + \int_0^s \hh A_N^{1,2}(u,t)du + hs \hh M_N(t) \,.
\label{eqQa}
\end{eqnarray}}
In the following proposition, we will approximate the terms $\hh R_N$, $\hh F_N$, $\hh R_N$ and $\hh A_N^{1,2}$, in order
to compute the limits
of \req{eqMa}-\req{eqQa} as $N \to \infty$.

\begin{prop}\label{comp1}
We have that
{\allowdisplaybreaks
\begin{align}
\hh A_N(t,s) \quad\simeq&\quad
\nu'(\hh C_N(t,t \vee s)) \hh C_N(s,t \vee s) -
\nu'(\hh C_N(t,0))\hh C_N(s,0) \label{eq:apxa} \\
-&\quad \int_0^{s\vee t} \nu''(\hh C_N(t,u))\hh C_N(s,u) \hh D_N (t,u) du
- \int_0^{s\vee t} \nu'(\hh C_N(t,u)) \hh D_N(s,u) du \nonumber \\
-&\quad h\left[\hh M_N(t)\int_0^{s \vee t} \hh C_N(s,u)\nu''(\hh C_N(t,u))du + \hh M_N(s)\int_0^{s \vee t} \nu'(\hh C_N(t,u))du\right],
\nonumber\\
\hh A_N^{1,2}(t,s) \quad\simeq&\quad \sum_{r=1}^2 \left[\nu'(\hh C_N^{r,1}(t, t \vee s)) \hh C_N^{2,r}(s, t \vee s) - \nu'(\hh C_N^{r,1}(t,0)) \hh C_N^{2,r}(s,0)\right]
\label{eq:apxa1} \\
-&\quad \sum_{r=1}^2 \left[\int_0^{s \vee t} \nu'(\hh C_N^{r,1}(t,u)) \hh D_N^{r,2} (s,u) du + \int_0^{s \vee t}
\nu''(\hh C_N^{r,1}(t,u)) \hh C_N^{2,r}(s,u) \hh D_N^{r,1} (t,u) du\right]  \nonumber \\
-&\quad h\sum_{r=1}^2 \left[\hh M_N(t) \int_0^{s \vee t} \hh C_N^{2,r}(s,u) \nu''(\hh C_N^{2,r}(t,u))du + \hh M_N(s) \int_0^{s \vee t} \nu'(C_N^{2,r}(t,u))du\right], \nonumber\\
\hh F_N (s,t) \quad\simeq&\quad \hh \chi_N(s,t \wedge s) \nu'(\hh C_N(s,s)) -
\int_0^s \nu'(\hh C_N(s,u)) \hh E_N (u,t \wedge u) du \label{eq:apxf} \\
-&\quad \int_0^{t \wedge s} \nu'(\hh C_N(s,u)) du - \int_0^s \hh \chi_N (u,t \wedge u) \nu''(\hh C_N(s,u)) \hh D_N(s,u) du\nonumber\\
-&\quad h \hh M_N(s)\int_0^s \nu''(\hh C_N(s,u))\hh \chi_N(u,t \wedge u) du,
\nonumber
\intertext{and}
\label{eq:apxr}
\hh R_N(t) \quad\simeq&\quad \nu'(\hh C_N({t},t))\hh M_N(t) - \nu'(\hh C_N(0,t)) \hh M_N(0)\\
\nonumber
-&\quad \int_0^{t}
 \hh M_N(u) \nu''(\hh C_N(t,u)) \hh D_N (u,t) du - \int_0^{t} \nu'(\hh C_N (t,u)) \hh P_N (u) du \\
-&\quad h\left[\hh M_N(t)\int_0^{t} \hh M_N(u)\nu''(\hh C_N(t,u))du + \int_0^{t} \nu'(\hh C_N(t,u))\right]du
\nonumber\,.
\end{align}}
\end{prop}
It is clear that using the results in Proposition \ref{comp1} in formulas \eqref{eqMa}-\eqref{eqQa}, we have proved Proposition \ref{prop-macro}. We shall start by developing the tools needed to conclude the proof of Proposition \ref{comp1}. To begin, we first prove a slightly more general version of Lemma 3.2 of \cite{BDG2}. The proof is essentially the same, replacing $x_t^j$ by $x_t^{q_1,j}$ and $x_s^i$ by $x_s^{q_2,i}$, respectively and will not be repeated.
\begin{lem}\label{kval}
Let $\E_{\BJ}$ denotes the expectation with respect to the
Gaussian law $\P_{\BJ}$ of the disorder $\BJ$. Then,
for the continuous paths $\Bx^{q} \in \Ca(\R_+,\R^N)$, $q\in \{q_1,q_2\}$, and all
$s,t \in [0,T]$ and $i,j \in \{1,\ldots,N\}$,
\begin{equation}\label{eq:kval}
k^{q_1,q_2,ij}_{ts}(\Bx) = \frac{x_t^{q_1,j} x_s^{q_2,i}}{N}
\nu''( C_N^{q_2,q_1} (s,t) ) + {\mathbbm{1}}_{i=j} \nu' (C_N^{q_2,q_1}(s,t)) \;.
\end{equation}
where $k^{q_1,q_2,ij}_{ts}(\Bx) :=
\E_{\BJ} [G^i(\Bx_t^{q_1})G^j(\Bx_s^{q_2})]$.
\end{lem}

Fixing continuous paths $\Bx^q$, let $k_t^{q_1,q_2}$ denote the operator
on ${\bf L}^2(\{1,\cdots N\}\ts[0,t])$
with the kernel $k=k^{q_1, q_2}(\Bx)$ of \req{eq:kval}.
That is, for $f\in {\bf L}^2(\{1,\cdots N\}\ts[0,t])$,
$u\le t$, $i\in\{1,\cdots,N\}$
\begin{equation}\label{eq:kfdef}
[k_t^{q_1, q_2} f]_u^i =\sum_{j=1}^N \int_0^t k_{uv}^{q_1,q_2,ij} f^j_v dv,
\end{equation}
which is clearly also in ${\bf L}^2(\{1,\cdots N\}\ts[0,t])$.
We next extend the definition \req{eq:kfdef} to the
stochastic integrals of the form
$$
[k_t^{q_1, q_2} \circ dZ]_u^i=\sum_{j=1}^N \int_0^t k_{uv}^{q_1,q_2,ij} dZ^j_v,
$$
where $Z^j_v$ is a continuous semi-martingale with respect to
the filtration $\Fa_t=\sigma(\Bx_u^{q_1}, \Bx_u^{q_2} : 0 \leq u \leq t )$ and is
composed for each $j$, of a squared-integrable continuous martingale
and a continuous, adapted, squared-integrable finite variation
part. In doing so, recall that by \req{eq:kval},
each $k_{uv}^{ij}(\Bx)$ is the finite sum of terms such as
$x^{q_1,i_1}_{u}\cdots x^{q_1,i_a}_{u}  x^{q_2,j_1}_v\cdots x^{q_2,j_b}_v$,
where in each term $a$, $b$ and $i_1,\ldots,i_a,j_1,\ldots,j_b$ are
some non-random integers. Keeping for simplicity
the implicit notation $\int_0^t k_{uv}^{q_1,q_2,ij} dZ^j_v$ we thus adopt
hereafter the convention of accordingly decomposing such integral
to a finite sum, taking for
each of its terms the variable $x^{q_1,i_1}_{u}\cdots x^{q_1,i_a}_{u}$
outside the integral, resulting with the usual It\^o
adapted stochastic integrals. The latter are well defined, with
$[k_t^{q_1,q_2} \circ dZ]_u^i$ being in ${\bf L}^2(\{1,\cdots N\}\ts[0,t])$.

Our next step is to generalize Proposition C.1 of \cite{BDG2}:
\begin{prop}\label{prop-alice}
Let $m\in \Z_+$ and suppose under the law $\P$ we have a
finite collection $\BJ = \{ J_\a \}_\a$ of
non-degenerate, independent, centered Gaussian random variables,
and $G_s^{q,i}=\sum_\a J_\a L_s^{q,i}(\a)$, for $q=1,\dots,m$, for
all $s \in [0,\tau]$ and $i \leq N$, where for each $\a$
the coefficients $L_s^{q,i}$ which are independent of $\BJ$ and also of each other, for different $q$'s,
are in ${\bf L}^2(\{1,\ldots,N\}\ts [0,\tau])$. Suppose further that
$U_s^{q,i}$ are continuous semi-martingales, independent of $\BJ$
and such that for each $\a$ and $q$, the stochastic integral
$$
\mu_{\a}^{q} := \sum_{i=1}^N \int_0^\tau L_u^{q,i}(\a) dU^{q,i}_u \,,
$$
is well defined and almost surely finite. Let
$\P^*$ denote the law of $\BJ$ such that
$\P^* = \prod_{q=1}^m\L_{\tau}^{q}/\E\left(\prod_{q=1}^m\L_{\tau}^{q}\right)\P$, where
\begin{equation}\label{eq:rn}
\L_{\tau}^{q} =\exp\left\{ \sum_{i=1}^N \int_0^\tau G^{q,i}_s dU^{q,i}_s
-\half\sum_{i=1}^N \int_0^\tau (G^{q,i}_s)^2 ds\right\}  \,.
\end{equation}
Let $V_{s}^{q,i}=\E^*(G_s^{q,i})$, $k_{ts}^{q_1,q_2,ij}=\E(G_t^{q_1, i} G_s^{q_2, j})$ and
$\Gamma_{ts}^{q_1,q_2,ij}=\E^*[(G_t^{q_1,i}-V_t^{q_1,i})(G_s^{q_2,j}-V_s^{q_2,j})]$. Then,
for any $s \leq \tau$, $i \leq N$ and $q\in\{1,\dots,m\}$,
\begin{equation}\label{eq:mean}
V^{q,i}_s + \sum_{r=1}^m[k_\tau^{q,r} V^{r}]^i_s = \sum_{r=1}^m [k_\tau^{q,r} \circ dU^{r}]^i_s\,,
\end{equation}
and for any $s,t \leq \tau$, $i,l \leq N$ and $q_1,q_2\in\{1,\dots,m\}$
\begin{equation}\label{eq:cov}
\sum_{r=1}^N \sum_{j=1}^N \int_0^\tau k_{su}^{q_1,r,ij} \Gamma_{ut}^{r,q_2,jl}
+ \Gamma_{st}^{q_1,q_2,il} = k_{st}^{q_1,q_2,il} \;.
\end{equation}
\end{prop}

\noindent{\bf Proof of Proposition \ref{prop-alice}.} Let $v_\a=\E(J_\a^2)>0$ denote the variance of $J_\a$ and
\begin{equation}\label{eq:chmes}
R_{\a\g}^{q} := \sum_{i=1}^N \int_0^\tau L_u^{q,i}(\a) L_u^{q,i}(\g) du \,,
\end{equation}
observing that
$$
\L_\tau^{q} =\exp\Big\{ \sum_{\a} J_\a \mu_\a^{q} -
\half\sum_{\a,\g} J_\a J_\g R_{\a\g}^{q} \Big\} \,.
$$
With $\BD= {\rm diag} (v_\a)$ a positive definite matrix and
$\BR := \sum_{q=1}^m \BR^{q} = \{\sum_{q=1}^m R_{\a\g}^{q}\}$ positive semi-definite, it
follows from this representation of $\L_\tau^{q}$,
that under $\P^*$ the random vector $\BJ$ has a Gaussian law
with covariance matrix $(\BD^{-1}+\sum_{q=1}^m\BR^{q})^{-1}$ and mean vector
$\bw=\{w_\a\}=(\BD^{-1}+\sum_{q=1}^m\BR^{q})^{-1} (\sum_{q=1}^m\bmu^{q})$. Hence, for any $\a$,
\begin{equation}\label{eqq}
w_\a + v_\a \sum_\g \left(\sum_{q=1}^m R_{\a\g}^{q}\right) w_\g = v_\a \sum_{q=1}^m \mu_\a^{q} \,.
\end{equation}
As $k_{su}^{q_1,q_2,ij}=\sum_\a L_s^{q_1,i} (\a) v_\a L_u^{q_2,j} (\a)$, it is not
hard to check that
\begin{eqnarray*}
[k_\tau^{q_1,q_2} \circ dU^{q_2}]^i_s &:=& \sum_{j=1}^N \int_0^\tau k_{su}^{q_1,q_2,ij} dU^{q_2,j}_u
= \sum_\a L_s^{q_1,i} (\a) v_\a \sum_{j=1}^N \int_0^\tau  L_u^{q_2,j} (\a) dU^{q_2,j}_u\\
 &=& \sum_\a L_s^{q_1,i} (\a) v_\a \mu_\a^{q_2} \;.
\end{eqnarray*}
Obviously,
$$V^{q,i}_s=\sum_\a L_s^{q,i}(\a) w_\a$$
and also,
\begin{eqnarray*}
[k_\tau^{q_1,q_2} V^{q_2}]^i_s &:=& \sum_{j=1}^N \int_0^\tau k_{su}^{q_1,q_2,ij} V^{q_2,j}_u du
= \sum_{\a,\g} L_{s}^{q_1,i} (\a) v_\a w_\g \sum_{j=1}^N
\int_0^\tau L_u^{q_2,j} (\a) L_u^{q_2,j}(\g) du\\
&=& \sum_{\a,\g} L_s^{q_1,i}(\a) v_\a R_{\a\g}^{q_2} w_\g \,,
\end{eqnarray*}
so we get \req{eq:mean}
out of \req{eqq}, with the last identity due to \req{eq:chmes}.
Turning to prove \req{eq:cov}, since $\Gamma_{ut}^{q_1,q_2,jl}$ is the
covariance of $G_u^{q_1,j}$ and $G_t^{q_2,l}$ under the tilted law $\P^*$,
we have that
$$
\Gamma_{ut}^{q_1,q_2,jl} = \sum_{\a,\g} L_u^{q_1,j} (\a)\left[(\BD^{-1}+\sum_{q=1}^m \BR^{q})^{-1}\right]_{\a\g}
 L_t^{q_2,l}(\g) \,,
$$
and hence by \req{eq:chmes} we see that
$$
\sum_{j=1}^N \int_0^\tau k_{su}^{q_1,r,ij} \Gamma_{ut}^{r,q_2,jl} du =
 \sum_{j=1}^N \int_0^\tau k_{su}^{q_1,r,ij} \sum_{\a,\g} L_u^{r,j} (\a)\left[(\BD^{-1}+\sum_{q=1}^m\BR^{q})^{-1}\right]_{\a\g}
 L_t^{q_2,l}(\g) du$$
$$=\sum_{j=1}^N \int_0^\tau \sum_{\sigma} L_s^{q_1,i} (\sigma) v_\sigma L_u^{r,j} (\sigma) \sum_{\a,\g} L_u^{r,j} (\a)\left[(\BD^{-1}+\sum_{q=1}^m\BR^{q})^{-1}\right]_{\a\g}
 L_t^{q_2,l}(\g) du
$$
$$=\sum_{\sigma,\a,\g} L_s^{q_1,i} (\sigma) v_\sigma R_{\sigma \alpha}^{r} \left[(\BD^{-1}+\sum_{q=1}^m\BR^{q})^{-1}\right]_{\a\g}
 L_t^{q_2,l}(\g) du
$$
$$=\sum_{\sigma,\g} L_s^{q_1,i} (\sigma) v_\sigma [R^{r} \left[(\BD^{-1}+\sum_{q=1}^m\BR^{q})^{-1}\right]_{\sigma \g}
 L_t^{q_2,l}(\g) du
$$
With $\BD={\rm diag} (v_\a)$ we easily
get \req{eq:cov} out of the
matrix identity:
$$\left({\bf I}+\BD \left(\sum_{q=1}^m\BR^{q}\right)\right)\left(\BD^{-1}+\sum_{q=1}^m\BR^{q}\right)^{-1} = \BD.$$
\hfill
$\Box$

Now, the same proof as in Lemma 3.2 of \cite{BDG2}, with $\L^N_\tau$ replaced by:
$$\L^N_\tau
=\exp\left\{ \sum_{q=1}^m\left[\sum_{i=1}^N \int_0^\tau G^{i}(\Bx_s^q) dU^{q, i}_s(\Bx)
-\half\sum_{i=1}^N \int_0^\tau (G^{i}(\Bx_s^q))^2 ds\right]\right\}$$
and using Proposition \ref{prop-alice} above instead of Proposition C.1 of \cite{BDG2}, will show:
\begin{lem}\label{condexp}
Let $m\in \Z_+$ and consider $m$ replicas $\{\Bx^q\}_s$, for $q = 1,\dots,m$, sharing the same couplings $\BJ$, with the noise given by $m$ independent N-dimensional Brownian motions $\{\BB^q\}_s$. Fixing $\tau \in \R_+$ and denoting $\Bx=(\Bx^1,\dots,\Bx^m)$, let $V_s^{q,i}(\Bx)=\E[G^i(\Bx_s^{q})|\Fa_\tau]$  and $Z_s^{q,i}(\Bx) =\E[B^{q,i}_s|\Fa_\tau]$ for $s \in [0,\tau]$. Then,
under $\P_\BJ\otimes \P^N_{\bx_0,\BJ}$ we can choose a version of
these conditional expectations such that the stochastic processes
\begin{eqnarray}
U^{q,i}_s(\Bx)&:=& x^{q,i}_s-x^{q,i}_0+\int_0^s f'(K_N^{q,q}(u)) x^{q,i}_u du - hs\label{eq:Udef} \\
Z^{q,i}_s(\Bx)&:=& U^{q,i}_s(\Bx) - \int_0^s V^{q,i}_u (\Bx^{q}) du \;,
\label{eq:UBdef}
\end{eqnarray}
are both continuous semi-martingales with respect to
the filtration $\Fa_t=\sigma(\Bx_u^k : 0 \leq u \leq t, 1\leq k \leq m )$, composed of squared-integrable continuous
martingales and finite variation parts.
Moreover, such choice satisfies for any $i,q$ and $s \in [0,\tau]$,
\begin{equation}\label{eq:Vid}
V^{q,i}_s + \sum_{r=1}^m[k_\tau^{q,r} V^{r}]^i_s  = \sum_{r=1}^m[k_\tau^{q,r} \circ dU^{r}]^i_s\,,
\end{equation}
and $V^{q,i}_s = \sum_{r=1}^m[k_\tau^{r,q} \circ dZ^{r}]_s^i$ for any $i,q$ and all $s \leq \tau$. Further, for any $u,v \in [0,\tau]$ and $i,j \leq N$, let
\begin{equation}\label{eq:gammadef}
\Gamma_{uv}^{q_1,q_2,ij}(\Bx) := \E \Big[(G^{i}(\Bx_u^{q_1}) - V_u^{q_1,i}(\Bx))(G^{j}(\Bx_v^{q_2}) - V_v^{q_2,j}(\Bx))|\Fa_\tau\Big]
\end{equation}
Further, we can choose a version of $\Gamma_{uv}^{q_1,q_2,il}$
such that for any $s,v \leq \tau$, any $q_1, q_2\in\{1,\dots,m\}$ and all $i,l \leq N$,
\begin{equation}\label{eq:gaid}
\sum_{r=1}^m \sum_{j=1}^N \int_0^\tau k_{su}^{q_1,r,ij} \Gamma_{ut}^{r,q_2,jl}
+ \Gamma_{st}^{q_1,q_2,il} = k_{st}^{q_1,q_2,il} \;.
\end{equation}
\end{lem}

\noindent{\bf Proof of Proposition \ref{comp1}.} We first apply \req{eq:Vid} to derive \req{eq:apxa1}. Fix $s,t \in [0,T]^2$, let $\tau=t \vee s$ and define:
$$
a_N^{q_1,q_2}(t,s)=\oneN \sum_{i=1}^N V^{q_1,i}_t(\Bx) x^{q_2,i}_s \,,
$$
Since $x^{q,i}_s$ is measurable on $\Fa_\tau$, $q=1,2$, we see that:
$$\hh A_N^{q_1,q_2} (t,s) = \E \left[ \oneN \sum_{i=1}^N \E [G^i(\Bx_t^{q_1}) x^{q_2,i}_s|\Fa_\tau] \right]
= \E [ a_N^{q_1,q_2}(t,s) ] = \hh a_N^{q_1,q_2} (t,s) \;.$$
Hence, with $t \leq \tau$, combining \req{eq:Vid} and \req{eq:Udef}, and suppressing in the notation the dependence of $k_{tu}^{q_1,q_2,ij}$ and $V_u^{q_1,j}$ of $\Bx$, we get:
\begin{align}
\label{eqA12temp}
&a_N^{1,2}(t,s) + \sum_{r=1}^2\left[\oneN \sum_{i,j=1}^N \int_0^\tau
x_s^{2,i} k_{tu}^{1,r,ij} V_u^{r,j} du + h \oneN \sum_{i,j=1}^N \int_0^\tau
x_s^{2,i} k_{tu}^{1,r,ij}du \right] \\
&= \sum_{r=1}^2 \left[\oneN \sum_{i,j=1}^N \int_0^\tau f'(K_N^{r,r}(u)) x_s^{2,i} k_{tu}^{1,r,ij} x_u^{r,j} du
+ \oneN \sum_{i,j=1}^N \int_0^\tau x_s^{2,i} k_{tu}^{1,r,ij} dx_u^{r,j} \right]
\nonumber
\end{align}
Using the explicit expression of $k^{q_1,q_2,ij}_{tu}$ from Lemma \ref{kval}, and collecting terms while
changing the order of summation and integration, we arrive at the identity:
{\allowdisplaybreaks
\begin{align}
 a_N^{1,2}(t,s) \,=&\, -\sum_{r=1}^2 \left[\int_0^\tau
C_N^{2,r}(s,u) \nu''(C_N^{r,1}(t,u)) a_N^{r,1} (u,t) du + \int_0^\tau \nu'(C_N^{r,1}(t,u)) a_N^{r,2} (u,s) du \right]\label{eq:Ahatid}\\
&\, - h\sum_{r=1}^2 \left[\int_0^\tau M_N^{1}(t)C_N^{2,r}(s,u)\nu''(C_N^{r,1}(t,u))du +
\int_0^\tau M_N^{2}(s)\nu'(C_N^{r,1}(t,u))du\right]
\nonumber \\
&\, +
\sum_{r=1}^2 \left[\int_0^\tau f'(K_N^{r,r}(u)) C_N^{2,r}(s,u) \nu''(C_N^{r,1}(t,u)) C_N^{1,r} (t,u) du \right]\nonumber \\
&\, + \sum_{r=1}^2 \left[\int_0^\tau f'(K_N^{r,r}(u)) \nu'(C_N^{r,1}(t,u)) C_N^{r,2}(u,s) du \right]
\nonumber \\
&\, + \sum_{r=1}^2 \left[\int_0^\tau C_N^{2,r}(s,u) \nu''(C_N^{r,1}(t,u)) d_u C_N^{r,1}(u,t) + \int_0^\tau \nu'(C_N^{r,1}(t,u)) d_u C_N^{r,2} (u,s)\right]
\nonumber\,.
\end{align}}
Applying Lemma A.1 of \cite{BDG2} for the semi-martingales $x = w = \Bx^r$, $y = \Bx^1$, $z = \Bx^2$ and polynomials $P(x)=x$ and $Q(x)=\nu'(x)$, the stochastic integrals in the last line of \req{eq:Ahatid} can be replaced with:
\begin{align}
&\quad \sum_{r=1}^2 \left[\nu'(C_N^{r,1}(\tau,t)) C_N^{2,r}(\tau,s) - \nu'(C_N^{r,1}(0,t)) C_N^{2,r}(0,s)\right] \label{eq:nostoch} \\
&- \sum_{r=1}^2 \left[\frac{1}{2N} C_N^{1,1}(t,t) \int_0^\tau \nu''(C_N^{r,1}(u,t))C_N^{2,r}(u,s) du
+\frac{1}{N} C_N^{1,r}(s,t) \int_0^\tau \nu'(C_N^{r,1}(u,t)) du \right]\;.
\nonumber
\end{align}
Now, it is easy to see that since $\E [\sup_{s,t\leq T} |A_N^{q_1,q_2}(s,t)|]$ is uniformly bounded in $N$ (see the discussion prior to Proposition \ref{prop-self}), then the same is true for $a_N^{q_1,q_2}(t,s)$, hence the terms in the second line \eqref{eq:nostoch} above will converge almost surely to $0$, as $N \ra \infty$. Furthermore, $a_N^{q_1,q_2}(t,s) = \E [ A_N^{q_1,q_2}(t,s) | \Fa_\tau ]$
inherits the self-averaging property from $A_N^{q_1,q_2}$, hence, we can apply Corollary \ref{prop-self}
with possibly $a_N^{q_1,q_2}$ as one of the arguments of the
locally Lipschitz function $\Psi(z)$ of at most polynomial growth
at infinity. Doing so for the functions $z_1 z_2 \nu''(z_3)$,
and $z_1 \nu'(z_2)$ and applying Proposition \ref{prop-self}
also for $f'(z_1) z_2 \nu''(z_3) z_3$ and
$f'(z_1) \nu'(z_2) z_3$, we deduce
from \req{eq:Ahatid} and \req{eq:nostoch} that
{\allowdisplaybreaks
\begin{align*}
 \hh A_N^{1,2}(t,s) \,\simeq&\, -\sum_{r=1}^2 \left[\int_0^\tau
\hh C_N^{2,r}(s,u) \nu''(\hh C_N^{r,1}(t,u)) \hh A_N^{r,1} (u,t) du
+ \int_0^\tau \nu'(\hh C_N^{r,1}(t,u)) \hh A_N^{r,2} (u,s) du\right]
\nonumber \\
&\, - h\sum_{r=1}^2 \left[\int_0^\tau \hh M_N^{1}(t) \hh C_N^{2,r}(s,u)\nu''(\hh C_N^{r,1}(t,u))du +
\int_0^\tau \hh M_N^{2}(s)\nu'(\hh C_N^{r,1}(t,u))du\right]
\nonumber \\
&\, +
\sum_{r=1}^2 \int_0^\tau f'(\hh K_N^{r,r}(u)) \hh C_N^{2,r}(s,u) \nu''(\hh C_N^{r,1}(t,u)) \hh C_N^{1,r} (t,u) du \nonumber \\
&\, + \sum_{r=1}^2 \int_0^\tau f'(\hh K_N^{r,r}(u)) \nu'(\hh C_N^{r,1}(t,u)) \hh C_N^{r,2}(u,s) du
\nonumber \\
&\, + \sum_{r=1}^2 \left[\nu'(\hh C_N^{r,1}(\tau,t)) \hh C_N^{2,r}(\tau,s) - \nu'(\hh C_N^{r,1}(0,t)) \hh C_N^{2,r}(0,s)\right]
\,.
\end{align*}}
Finally, recalling that
\begin{equation*}
\hh A_N^{q_1,q_2}(t,s) = \hh D_N^{q_1,q_2}(s,t) + f'(\hh K_N(t)) \hh C_N^{q_1,q_2}(s,t)\,,
\end{equation*}
and noting that $\hh K_N^{r,r}(t) = \hh K_N(t)$, for all $t$ and $r$, setting $\tau=t \vee s$, we indeed arrive at:
{\allowdisplaybreaks
\begin{align*}
\hh A_N^{1,2}(t,s) \,\simeq&\, -\sum_{r=1}^2 \left[\int_0^{t \vee s}
\hh C_N^{2,r}(s,u) \nu''(\hh C_N^{r,1}(t,u)) \hh D_N^{r,1} (t,u) du
+ \int_0^{t \vee s} \nu'(\hh C_N^{r,1}(t,u)) \hh A_N^{r,2} (s,u) du\right]
\nonumber \\
&\, - h\sum_{r=1}^2 \left[\int_0^{t \vee s} \hh M_N(t)\hh C_N^{2,r}(s,u)\nu''(\hh C_N^{r,1}(t,u))du
+ \int_0^{t \vee s} \hh M_N(s)\nu'(\hh C_N^{r,1}(t,u))du\right]
\nonumber \\
&\, + \sum_{r=1}^2 \left[\nu'(\hh C_N^{r,1}(t \vee s,t)) \hh C_N^{2,r}(t \vee s,s) - \nu'(\hh C_N^{r,1}(0,t)) \hh C_N^{2,r}(0,s)\right]
\,.
\end{align*}}
that is \eqref{eq:apxa1}.

For deriving \eqref{eq:apxa} next, the single-replica equivalent of \eqref{eq:apxa1}, we can apply the same strategy as above. Namely, defining:
$$
a_N(t,s)=\oneN \sum_{i=1}^N V^{i}_t(\Bx) x^{i}_s \,,
$$
we see that $a_N(t,s)$ has the same first moment with $A_N(t,s)$. Furthermore, since $\Fa_\tau$ is generated only by the realization of one replica up to time $\tau$, \req{eq:Vid} will imply that:
\begin{align*}
&a_N(t,s) + \oneN \sum_{i,j=1}^N \int_0^\tau
x_s k_{tu}^{ij} V_u^{j} du + h \oneN \sum_{i,j=1}^N \int_0^\tau
x_s^{i} k_{tu}^{ij}du\\
&= \oneN \sum_{i,j=1}^N \int_0^\tau f'(K_N(u)) x_s^{i} k_{tu}^{ij} x_u^{j} du
+ \oneN \sum_{i,j=1}^N \int_0^\tau x_s^{i} k_{tu}^{ij} dx_u^{j}
\end{align*}
Note that the above equation is indeed the one-dimensional version of \eqref{eqA12temp} (without the sums and the replica indices), so we would expect the results to be similar. Indeed, using the explicit expression of $k^{ij}_{tu}$ from Lemma \ref{kval}, we arrive at the identity:
{\allowdisplaybreaks
\begin{eqnarray}
\label{eq:Ahatid1}
 a_N(t,s) &=& -\int_0^\tau
C_N(s,u) \nu''(C_N(t,u)) a_N (u,t) du - \int_0^\tau \nu'(C_N(t,u)) a_N (u,s) du \\
&& - h\left[\int_0^\tau M_N(t)C_N(s,u)\nu''(C_N(t,u))du +
\int_0^\tau M_N(s)\nu'(C_N(t,u))du\right]
\nonumber \\
&& + \int_0^\tau f'(K_N(u)) C_N(s,u) \nu''(C_N(t,u)) C_N (t,u) du \nonumber \\
&& + \int_0^\tau f'(K_N(u)) \nu'(C_N(t,u)) C_N(u,s) du  \nonumber\\
&& + \int_0^\tau C_N(s,u) \nu''(C_N(t,u)) d_u C_N(u,t)
+ \int_0^\tau \nu'(C_N(t,u)) d_u C_N (u,s)
\,.
\nonumber
\end{eqnarray}}
Applying again Lemma A.1 of \cite{BDG2}, this time for the semi-martingales $x = y = z = w = \Bx$ and polynomials $P(x)=x$ and $Q(x)=\nu'(x)$, the stochastic integrals in the last line of \req{eq:Ahatid1} can be replaced with:
\begin{align*}
&\quad \nu'(C_N(\tau,t)) C_N(\tau,s) - \nu'(C_N(0,t)) C_N(0,s)  \\
&- \left[\frac{1}{2N} C_N(t,t) \int_0^\tau C_N(u,s) \nu''(C_N(u,t)) du
+\frac{1}{N} C_N(s,t) \int_0^\tau \nu'(C_N(u,t)) du \right]\;.
\end{align*}
As before, the terms in the second line above will converge to $0$ as $N \ra \infty$, and, $a_N(t,s) = \E [ A_N(t,s) | \Fa_\tau ]$ inherits the self-averaging property from $A_N$. Hence applying Corollary \ref{prop-self} with possibly $a_N$ as one of the arguments of the locally Lipschitz function $\Psi(z)$, setting $\tau=t \vee s$ and recalling that
$\hh A_N(t,s) = \hh D_N(s,t) + f'(\hh K_N(t)) \hh C_N(s,t)$, we arrive at \eqref{eq:apxa1}.

Now, for \eqref{eq:apxr}, denoting $r_N(s)=\oneN \sum_{i=1}^N V_t^i(\Bx)$ and we easily see that $\hh r_N (s) =\hh R_N (s)$, so by \req{eq:Vid} and \req{eq:Udef} we
get that:
\begin{align*}
&r_N(t) + \oneN \sum_{i,j=1}^N \int_0^\tau k_{tu}^{ij} V_u^j du = \oneN \sum_{i,j=1}^N \int_0^\tau f'(K_N(u)) k_{tu}^{ij} x_u^j du
+ \oneN \sum_{i,j=1}^N \int_0^\tau k_{tu}^{ij} dx_u^j
- h\oneN \sum_{i,j=1}^N \int_0^\tau k_{tu}^{ij}du
\end{align*}
So, as before, using the explicit expression of $k^{ij}_{tu}$, we get to:
{\allowdisplaybreaks
\begin{eqnarray}
\label{eq:Rhatid}
r_N(t) &=& - \int_0^\tau
M_N(u) \nu''(C_N(t,u)) a_N (u,t) du
- \int_0^\tau \nu'(C_N(t,u)) r_N (u) du
\\
&& - h\left[\int_0^\tau M_N(t)M_N(u)\nu''(C_N(t,u))du +
\int_0^\tau \nu'(C_N(t,u))du\right]
\nonumber \\
&& + \int_0^\tau f'(K_N(u)) M_N(u) \nu''(C_N(t,u)) C_N (u,t) du \nonumber \\
&& + \int_0^\tau M_N(u) \nu''(C_N(t,u)) d_u C_N(u,t)
+ \int_0^\tau \nu'(C_N(t,u)) d_u M_N (u)
\nonumber \\
&& + \int_0^\tau f'(K_N(u)) \nu'(C_N(t,u)) M_N (u) du
\nonumber \,.
\end{eqnarray}}
Once again, Lemma A.1 of \cite{BDG2}, helps, this time for the semi-martingales $x = z = w = \Bx$, $y = 1$ and polynomials $P(x)=x$ and $Q(x)=\nu'(x)$, hence we replace the stochastic integrals above with:
\begin{align*}
&\quad M_N(\tau) \nu'(C_N(\tau,t)) - M_N(0) \nu'(C_N(0,t))\\
&- \frac{1}{2N} C_N(t,t) \int_0^\tau M_N(u) \nu''(C_N(u,t)) du
-\frac{1}{N} M_N(t) \int_0^\tau \nu'(C_N(u,t)) du
\end{align*}
As before, the terms in the second line above will converge to $0$ as $N \ra \infty$, and, $r_N(t) = \E [ R_N(t) | \Fa_\tau ]$ inherits the self-averaging property from $R_N$. Hence applying Corollary \ref{prop-self} with possibly $a_N$ and $r_N$ as some of the arguments of $\Psi(z)$ and recalling that
$\hh P_N(t) = \hh R_N(t) + f'(\hh K_N(t)) \hh M_N(t)$, we arrive at \eqref{eq:apxr}.

Now the derivation of \req{eq:apxf} is similar to the derivation of its analogue in the proof of Proposition 3.1 in \cite{BDG2}. Namely, since:
\begin{equation}\label{eq2}
\E[G^i_s B^i_t ] +\E [ \int_0^t \Gamma_{sv}^{ii} dv ]
= \E \left[ [k_s \circ dZ]_s^i Z^i_t \right] \,.
\end{equation}
the equation \eqref{eq:hfdef} implies:
\begin{eqnarray*}
\hh F_N (s,t) &=&
\E \left[ \oneN \sum_{i=1}^N \E\left[ [k_s \circ d\BB]_s^i |\Fa_s \right] B^i_t \right] - \E \left[ \oneN \sum_{i=1}^N \int_0^t \Gamma_{sv}^{ii} dv \right]
\end{eqnarray*}
hence by \eqref{integratedSDE} and \eqref{eq:kval}:
\begin{eqnarray}
&&\hh F_N (s,t) + h\oneN \sum_{i=1}^N \E \left[\E \left[ [k_s]_s^i|\Fa_s\right] B^i_t\right] \label{eqfamir} \\
&=&\oneN \sum_{i=1}^N \left( \E \left[ \E\left[[k_s \circ d\Bx]_s^i + [k_s f'(K_N) \Bx]_s^i
- [k_s G]_s^i|\Fa_s\right] B^i_t \right] - \E \left[ \int_0^t \Gamma_{sv}^{ii} dv \right]\right)
\,,
\nonumber
\end{eqnarray}
The right hand side of \eqref{eqfamir} was evaluated in the proof of Proposition 3.1 of \cite{BDG2}. Using their result into \eqref{eqfamir}, we get, for $s\geq t$:
\begin{align*}
\hh F_N (s,t) + h\oneN \sum_{i=1}^N \E \left[ \E \left[ [k_s]_s^i|\Fa_s\right] B^i_t\right] &\,\simeq\,
 \hh \chi_N(s,t) \nu'(\hh C_N(s,s)) - \int_0^{t \wedge s} \nu'(\hh C_N(s,u)) du\\
&\quad - \int_0^s \nu'(\hh C_N(s,u)) \hh E_N (u,t \wedge u) du
- \int_0^{t \wedge s} \nu'(\hh C_N(s,u)) du \\
&\quad- \int_0^s \hh \chi_N (u,t \wedge u) \nu''(\hh C_N(s,u)) \hh D_N(s,u) du
\end{align*}

Now, using the explicit formula for $k_s$ to compute the remaining term:
\begin{align*}
h \oneN \sum_{i=1}^N \E \left[ \E \left[ [k_s]_s^i|\Fa_s\right] B^i_t\right] &\,=\, h \oneN \sum_{i=1}^N \E \left[ \left(M_N(s)\int_0^s \nu''(C_N(s,u))x_u^i du + \int_0^s \nu'(C_N(s,u))du\right) B^i_t\right]\\
&\,=\, h \E \left[ \int_0^s M_N(s)\nu''(C_N(s,u))\chi_N(u,t) du + \int_0^s \nu'(C_N(s,u))W_N(t)du \right]\\
&\,\simeq\, h \hh M_N(s)\int_0^s \nu''(\hh C_N(s,u))\hh \chi_N(u,t) du
\end{align*}
where the last line is obtained by two applications of Proposition \ref{prop-self} (eventually with the zero mean random variable $W_N(t)=\frac{1}{N}\sum_{i=1}^N B_t^i$ as one of its arguments), hence concluding the proof of \eqref{eq:apxf}.
\hfill
$\Box$
\nn

\noindent{\bf Proof of Lemma \ref{lem-diff}.} We shall show that every solution of \eqref{eqM1}-\eqref{eqFrep} is necessarily a solution of \eqref{eqM}-\eqref{eqZ}, where $\chi(s,t)=\int_0^t R(s,u)du$.

First, the same argument as in the beginning of Lemma 5.1 of \cite{BDG2} applied to
\begin{eqnarray*}
h(s,t):=-f'(C(s,s))\chi(s,t)
-\int_0^s\chi(u,t)\nu''(C(s,u))D(s,u) du+
\chi(s,t) \nu'(C(s,s))
\\
-\int_0^{t \wedge s} \nu'(C(s,u)) du - h M(s)\int_0^s \nu''(C(s,u))\chi(u,t) du
\end{eqnarray*}
will show that $t \mapsto \chi(s,t)$ is continuously differentiable on $s\ge t$,
with $\chi(s,t)=\int_0^t R(s,u) du$, where $R(s,s)=1$ for all $s$ and $\chi(s,t)=\chi(s,s)$ for $t > s$, implying that $R(s,t)=0$, for $t > s$.

From \req{eqC1} we have that $C(s,t)-\chi(s,t)$
is differentiable with respect to its
second argument $t$, hence $\partial_2 C(s,t) = D(s,t) + R(s,t) + hM(s)$
Further, $C(s,t)=C(t,s)$ implying that
$\partial_1 C (s,t) = \partial_2 C(t,s) = D(t,s)+R(t,s) + hM(t)$
on $[0,T]^2$. Thus, combining the identity
\begin{eqnarray*}
C(s,t \vee s)\nu'(C(t \vee s,t))-C(s,0)\nu'(C(0,t))&=&
\int_0^{t \vee s} \nu'(C(t,u)) \partial_2 C(s,u) du \\
&+&\int_0^{t \vee s} C(s,u) \nu''(C(t,u)) \partial_2 C(t,u) du \,,
\end{eqnarray*}
with \req{eqD} we have that for all $t,s \in [0,T]^2$,
\begin{eqnarray}
D(s,t) &=& -f'(K(t)) C(t,s)+
\int_0^{t \vee s} \nu'(C(t,u))R(s,u) du + \int_0^{t \vee s} C(s,u)\nu''(C(t,u)) R(t,u) du \,.
\label{eq:tempD}
\end{eqnarray}
Interchanging $t$ and $s$ in \req{eq:tempD} and adding
$R(t,s)=0$ when $s>t$, results for $s>t$ with
\begin{eqnarray*}
\partial_1 C (s,t) &=&
-f'(K(s)) C(s,t)+
\int_0^s\nu'(C(s,u))R(t,u) du +\int_0^s C(t,u)\nu''(C(s,u)) R(s,u) du + hM(t)\,,
\end{eqnarray*}
which is \req{eqC} for $\beta=1$.

Now, from \req{eqM1}, $M(\cdot)$ is differentiable and $M'(t)=h+P(t)$, hence combining the identity
\begin{eqnarray*}
M(t)\nu'(C(t,t))-M(0)\nu'(C(t,0))&=&
\int_0^t \nu'(C(t,u)) M'(u) du \\
&+&\int_0^t M(u) \nu''(C(t,u)) \partial_2 C (t,u) du \,,
\end{eqnarray*}
with \req{eqP} we have that for all $t\in [0,T]$,
\begin{equation}\label{eq:tempP}
P(t)=-f'(K(t)) M(t) + \int_0^{t} M(u)\nu''(C(t,u)) R(t,u) du \,.
\end{equation}
thus showing is \req{eqQ} for $\beta=1$.

Also, since from \req{eqQ1}, $\partial_2 Q(s,t) = H(s,t) + hM(s)$, from the identity
\begin{eqnarray*}
Q(s,t \vee s)\nu'(C(t \vee s,t))-Q(s,0)\nu'(C(0,t))&=&
\int_0^{t \vee s} \nu'(C(t,u)) \partial_2 Q (s,u) du \\
&+&\int_0^{t \vee s} Q(s,u) \nu''(C(t,u)) \partial_2 C (t,u) du \,,
\end{eqnarray*}
with \req{eqFrep} we have that for all $t\in [0,T]$,
\begin{equation}\label{eqFrepTemp1}
X(s,t)=\int_0^{t \vee s} \nu''(C(t,u)) R(t,u) Q(s,u) du\,.
\end{equation}
Similarly,
\begin{equation}\label{eqFrepTemp2}
Y(s,t)=\int_0^{t \vee s} \nu'(Q(t,u)) R(s,u) du\,.
\end{equation}
thus showing is \req{eqQ} for $\beta=1$.

Since $K(s)=C(s,s)$, with $C(s,t)=C(t,s)$
and $\partial_2 C(t,s) = D(t,s)+R(t,s)+hM(t)$, it follows that for all $k>0$,
\begin{eqnarray*}
K(s)-K(s-k) &=& \int_{s-k}^{s}(D(s,u) + R(s,u) + hM(s))du\\
&& + \int_{s-k}^{s}(D(s-k,u) + R(s-k,u) + hM(s-k))du \,.
\end{eqnarray*}
Recall that $R(s,u)=0$ for $u > s$, hence,
dividing by $k$ and taking $k \downarrow 0$, we thus get
by the continuity of $D$ and that of
$R$ for $s \geq t$ that $K(\cdot)$ is differentiable,
% we can also consider [\chi(s+h,s+h)-\chi(s+h,s)]/h
% resulting with h^{-1} \int_s^{s+h} R(s+h,u) du --> 1
% hence same limit for [K(s+h)-K(s)]/h as for [K(s)-K(s-h)]/h.
with $\partial_s K(s)=2 D(s,s)+R(s,s)+2hM(s)=2 D(s,s) + 1 + 2hM(s)$,
resulting by \req{eq:tempD} with \req{eqZ} for $\beta=1$.

Further, it follows from \req{eq:chi} that
$\partial_1 \chi (u,t) = E(u,t)+ 1_{u<t}$.
Hence, combining the identity
\begin{eqnarray*}
\chi(s,t)\nu'(C(s,s))-\chi(0,t)\nu'(C(s,0)) \,=\, \int_0^s \nu'(C(s,u)) \partial_1 \chi (u,t) du
+\int_0^s \chi(u,t) \nu''(C(s,u)) \partial_2 C (s,u) du \,,
\end{eqnarray*}
with \req{eqE} we have that for all $T \geq s \geq t$,
\begin{equation}\label{eq:tempE}
E(s,t)=-f'(K(s)) \chi (s,t)+
\int_0^s \chi (u,t)\nu''(C(s,u)) R(s,u) du
\end{equation}
(recall that $\chi(0,t)=\chi(0,0)=0$). Let
\begin{equation}\label{eq:smE}
g(s,t):=-f'(K(s)) R (s,t)+\int_0^s R(u,t)\nu''(C(s,u)) R(s,u) du \,,
\end{equation}
for $s,t \in [0,T]^2$. Recall that $\chi(s,t)=\int_0^t R(s,v) dv$,
so by Fubini's theorem, \req{eq:tempE} amounts to
$E(s,t)=\int_0^t g(s,v) dv$ for all $s \geq t$. Further, with
$E(s,t)=E(s,s)$ when $t>s$, it follows that
$$
E(s,t)=\int_0^{t \wedge s} g(s,v) dv
$$
for all $s,t \leq T$. Putting this into \req{eq:chi} we have
by yet another application of Fubini's theorem that
$$
\int_0^t R(s,u) du = \chi(s,t)
= t + \int_0^s \int_0^{t \wedge u} g(u,v) dv du
= t + \int_0^t \int_v^s g(u,v) du dv \,,
$$
for any $s \geq t$. Consequently, for every $t \leq s$,
$$
R(s,t) = 1 + \int_t^s g(u,t) du \,,
$$
implying that $\partial_1 R = g$ for a.e. $s>t$,
which in view of \req{eq:smE} gives \req{eqR} for $\beta=1$,
thus completing the proof of the lemma.
$\Box$

\noindent{\bf Proof of Lemma \ref{uniqueness}.} We shall show that the system \eqref{eqM}--\eqref{eqZ} with
initial conditions $C(t,t) = K(t)$, $R(t,t) = 1$, $M(0)=\alpha$ and $Q(0,0) = K(0) = C(0,0) = \vartheta$ admits at most one bounded solution $(M,R,C,Q,K)$ on $[0,T] \times [0,T]^2 \times (\bD \cap [0,T]^2) \times [0,T]^2 \times [0,T]^2$. First notice that if we denote $D(t):=Q(t,t)$, by the symmetry of $Q$, we have $\pars D(t)=2\pars_1 Q(t,t)$. Now consider the difference between the integrated form of \eqref{eqM}--\eqref{eqQ} for two such solutions
$(M,R,C,Q,K,D)$ and $(\bar M,\bar R,\bar C,\bar Q,\bar K,\bar D)$ and define the functions $\Delta V(s,t)=|V(s,t)-\bar V(s,t)|$, when $V$ is one of the functions $C, R$ or $Q$ and $\Delta U(s)=|U(s)-\bar U(s)|$, when $U$ is $M$, $D$ or $K$. Then, since $\nu''$ is uniformly Lipschitz on any compact interval and $C, Q, \bar C, \bar Q$ are continuous, hence bounded on $[0,T]^2$, we have, for $0\le t\le s\le T$,
\begin{align}
\Delta M(t) &\le \kappa_1
\left[\int_0^t \Delta M(v)dv + \int_0^t h(v)dv\right]
\label{tutuM}\\
\Delta R(s,t) &\le \kappa_1
\left[\int_t^s \Delta R(v,t) dv +\int_t^s h(v) dv\right]
\label{tutuR}\\
\Delta C(s,t) &\le \kappa_1 \left[
\int_t^s \Delta C(v,t) dv + \int_t^s h(v) dv + \Delta M(t) + \Delta K(t) + h(t)\right]
\label{tutuC}\\
\Delta Q(s,t) &\le \kappa_1 \left[
\int_t^s \Delta Q(v,t) dv + \int_t^s h(v) dv + \Delta M(t) + \Delta D(t) + h(t)\right]
\label{tutuQ}\\
\Delta K(t) &\le \kappa_1 \left[
\int_0^t \Delta K(v) dv + \int_0^t h(v) dv + \Delta M(t) + h(t)\right]
\label{tutuK}\\
\Delta D(t) &\le \kappa_1 \left[
\int_0^t \Delta D(v) dv + \int_0^t h(v) dv + \Delta M(t) + h(t)\right]
\label{tutuD}
\end{align}
where $h(v) := \int_0^v [\Delta R(v,\theta) + \Delta C(v,\theta) + \Delta Q(v,\theta) + \Delta M(\theta) + \Delta D(\theta) + \Delta K(\theta)] d\theta$ and
$\kappa_1 < \infty$ depends on $T$, $\b$, $\nu(\cdot)$ and
the maximum of  $|M|$, $|R|$, $|C|$, $|Q|$, $|\bar M|$, $|\bar R|$, $|\bar C|$ and $|\bar Q|$ on $[0,T]^2$.
Integrating \eqref{tutuM}-\eqref{tutuD} over $t \in [0,s]$,
since $\Delta R(v,u)=0$ for $u \geq v$, $\Delta C(v,u)=\Delta C(u,v)$ and $\Delta Q(v,u)=\Delta Q(u,v)$, we find that
{\allowdisplaybreaks
\begin{align*}
\int_0^s \Delta M(t) dt &\quad\le\quad \kappa_2 \int_0^s h(v) dv \,, \\
\int_0^s \Delta R(s,t) dt &\quad\le\quad \kappa_2 \int_0^s h(v) dv \,, \\
\int_0^s \Delta C(s,t) dt &\quad\le\quad \kappa_2 \int_0^s h(v) dv \,, \\
\int_0^s \Delta Q(s,t) dt &\quad\le\quad \kappa_2 \int_0^s h(v) dv \,, \\
\int_0^s \Delta K(t) dt &\quad\le\quad \kappa_2 \int_0^s h(v) dv \,, \\
\int_0^s \Delta D(t) dt &\quad\le\quad \kappa_2 \int_0^s h(v) dv \,,
\end{align*}}
for some finite constant $\kappa_2$ (of the same type of
dependence as $\kappa_1$).
Summing the last three inequalities, we see that for all $s \in [0,T]$,
$$
0 \leq h(s) \leq \kappa_3 \int_0^s h(v) dv.
$$
where $\kappa_3=6 \max\{\kappa_1,\kappa_2\}$. Further, $h(0)=0$, so by Gronwall's lemma $h(s)=0$
for all $s \in [0,T]$.  Plugging this result back into
\eqref{tutuM}-\eqref{tutuD} and observing that
$\Delta R(t,t)=\Delta K(0)=\Delta M(0)=\Delta D(0)=0$, $\Delta C(t,t)=\Delta K(t)$ and $\Delta Q(t,t)=\Delta D(t)$, we deduce that
$\Delta R(s,t)=\Delta C(s,t)=\Delta M(t)=\Delta Q(s,t)=\Delta D(t)=\Delta K(s)=0$
for all $0 \le t\le s \le T$, hence, by symmetry, the stated uniqueness.
\hfill
$\Box$

%%%%%%%%%%%%%%%%%%%%%%%%%%%%%%%%%%%%%%%%%%%%%%%%%%%%
%%%%%%%%%%%%%%%%%%%%%%%%%%%%%%%%%%%%%%%%%%%%%%%%%%%%

\section{Limiting Hard Spherical Dynamics}\label{exactdynamics}

%%%%%%%%%%%%%%%%%%%%%%%%%%%%%%%%%%%%%%%%%%%%%%%%%%%%
%%%%%%%%%%%%%%%%%%%%%%%%%%%%%%%%%%%%%%%%%%%%%%%%%%%%

Through this section, we will fix $r>0$ and, for convenience of notation, suppress the $r$ dependence in the subscripts.

The uniform bounds on the moments of $K_N(s)$ used to establish Proposition \ref{prop-self}
(namely equation \eqref{eq:conc2}), will show that $\sup_{t \geq 0} K(t) < \infty$. Further, as $C(s,t)$
is the limit of $C_N (s,t) = \oneN \sum_{i=1}^N x_s^i x_t^i$,
it is a non-negative definite kernel
on $\reals_+ \times \reals_+$ and in particular, $C(s,t)^2 \leq K(s) K(t)$ and $C(t,t)\geq 0$. Also, since
$Q(s,t)$
is the limit of $Q_N (s,t) = \frac{1}{N}\sum_{i=1}^N x_s^{1,i} x_t^{2,i}$, for two iid replicas $\Bx_t^{1}$ and $\Bx_t^{2}$, by the Cauchy-Schwartz inequality and then taking the limit as $N\rightarrow \infty$, we have
$Q(s,t)^2\leq K(s)K(t)$.

%R
To complete the proof of
Theorem \ref{theo-sphere},
we first prove that
%-------------------------------------------------
any solution $(M,R,C,Q,K)$ of \eqref{eqM}--\eqref{eqZ}
consists of positive functions, a key fact in our forthcoming
analysis.
\begin{lem}\label{pos}
For any $f:\R_+ \ra \R$ whose derivative is bounded above
on compact intervals and any $K(0)>0$, $M(0)>0$,  a solution $(M,R,C,Q,K)$
to \eqref{eqM}--\eqref{eqZ}, if it exists, is positive at all times. Furthermore, $\tilde C(s,t):= C(s,t)-M(s)M(t)$ is also non-negative.
\end{lem}
\noindent{\bf Proof of Lemma \ref{pos}.} By definition $K(t)\ge 0$ for all $t\in\R_+$. Define
$$
S_1=\inf\{ u \geq 0: \, C(u,t)\le 0 \mbox{ for some } t \leq u \}\,.
$$
and
$$
S_2=\inf\{ u \geq 0: \, M(u)\le 0\}\,.
$$
and suppose that $S=\min\{S_1,S_2\}<\infty$. By continuity of $(C,K,Q)$, since $K(0)>0$ and $M(0)>0$, also $S_1,S_2>0$, hence $S>0$.
Set $\L(s,t)=\exp(-\int_t^s \mu(u) du) >0$ for
$\mu(u)=f'(K(u))$ which is bounded above on compact intervals,
and $R(s,t)=\L(s,t) H(s,t)$. Then, by \cite{GM}, for $s\ge t$,
\begin{equation}
\label{gmfor}
H(s,t) = 1+\sum_{n\ge 1}\beta^{2n}\sum_{\s\in \NC_n}
\int_{t\le t_1\cdots \le t_{2n}\le s}
\prod_{i\in \cro(\sigma) } \nu''(C(t_i,t_{\s(i)})) \prod_{j=1}^{2n} dt_j
\end{equation}
where  $\NC_n$ denotes the
set of involutions of $\{1,\cdots,2n\}$
without fixed points and without crossings
and   $\cro(\sigma)$
is defined to be the set of indices $1\le i\le 2n$
such that $i<\sigma(i)$.
Consequently,
$$
R(s,t)\ge \L(s,t) > 0 \mbox{ for }t\le s\le S\,,$$
and thus, (\ref{eqC}) implies that
$$
C(s,t)\ge K(t) \L(s,t)>0
\mbox{ for }t\le s\le S \,.
$$
Also, \eqref{eqM} implies that
$$
M(s)\ge M(0) \L(s,0)>0
\mbox{ for }0\le s\le S \,.
$$
Note that in the last two estimates we used the fact that
$\nu'(\cdot)$ and $\nu''(\cdot)$ are non negative on $\R_+$.
Similarly, from the equation \eqref{eqZ} we see that
$\pars [\L(s,0)^{-2} K(s)] \ge \L(s,0)^{-2}$ for all $s\le S$
resulting with
$$
K(s)\ge K(0) \L(s,0)^{-2} + \int_0^s \L(s,v)^{-2} dv  >0
$$
Hence, the continuous functions $R(s,t), C(s,t)$ and  $M(s)$ are
bounded below by a strictly positive constant
for $0\le t\le s\le S$ in contradiction with the definition of $S$.
We thus deduce that $S=\infty$, hence $S_1=S_2=\infty$ and by the preceding argument and the symmetry of $C$, the functions $R(s,t)$, $C(s,t)$ and $M(s)$ are positive.

Similarly, let $S_3=\inf\{ u \geq 0: \, Q(u,t)\le 0 \mbox{ for some } t \leq u \}$ and assume $S_3<\infty$.  Then, from the symmetry of $Q(s,t) = Q(t,s)$, defining $D(t):=Q(t,t)$, we have $\pars D(t) = 2\pars_1 Q(t,t)$, hence by \eqref{eqQ} we have:
$$D(s) \geq D(0)\L^2(s,0)>0\mbox{ for }0\le s\le S_3 \,.$$
and hence, using again \eqref{eqQ}:
$$Q(s,t) \geq Q(t,t)\L(s,t) = D(t)\L(s,t) > 0\mbox{ for }t\le s\le S_3 \,.$$
Hence the continuous function $Q$ is bounded below by a positive constant on $0\le t \le s \le S_3$, contradiction to the definition of $S_3$. Hence $S_3=\infty$ and by the symmetry of $Q$, it is positive on $\R_+^2$. This concludes our proof that $M,R,C,Q,K$ are all positive functions.

Furthermore, from \eqref{eqM} and \eqref{eqC}, we know that $\tilde C(s,t) = C(s,t)-M(s)M(t)$ satisfies:
\begin{eqnarray*}
\pars_1 \tilde C(s,t) &=& - f'(K(s)) \tilde C(s,t) +
\b^2 \int_0^s \tilde C(u,t) R(s,u) \nu''(C(s,u)) du\\
&& + \b^2 \int_0^t \nu'(C(s,u)) R(t,u) du
\end{eqnarray*}
hence
$$
\tilde C(s,t)\ge \tilde C(t,t) \L(s,t)\geq 0
\mbox{ for }t\le s\le S \,
$$
since $\tilde C(t,t)=K(t)-M^2(t)\geq 0$.
\hfill$\Box$

We next show that if $(M_{L,r}, R_{L,r}, C_{L,r}, Q_{L,r}, K_{L,r})$
are solutions of the system \eqref{eqM})-\eqref{eqZ}
with potential $f_{L,r}(\cdot)$ as in \eqref{eq:fdef},
then $K_{L,r}(s) \to r$ as $L \to \infty$, uniformly
over compact intervals. Specifically,

\begin{lem}\label{lem-bdd}
Assuming $K_L(0) = r$, there exist $L_0 > 0$ such that $K_L(s) \ge r - B_0L^{-1}$, for some $B_0 > 0$, 
for all $L>L_0$ and $s \geq 0$.
Further, for any $T$ finite there exists $B(T)<\infty$ (depending on $r$), such that $K_L(s)\leq r+B(T) L^{-1}$ for all $s\le T$ and $L \geq \max\{B(T), L_0\}$.
\end{lem}
\noindent{\bf Proof of Lemma \ref{lem-bdd}.} We first deal with the lower bound on $K_L(\cdot)$.
Fix $L>0$ and let $g_L(x):=1 - 2x f_L'(x)= 1 + 4L x(r-x)- \left(\frac{x}{r}\right)^{2k} - \frac{2 \alpha h x}{r}$. Let $x_L$ be the largest root of $g_L(x)$ smaller than $r$. It is easy to see that $g_L(r) < 0$ and also that $\lim_{L\ra \infty} g_L(r/2) > 0$, so there exist $L_0 > 0$ such that $x_L > r/2$ whenever $L > L_0$. Furthermore, 
$$L(r-x_L) = - \frac{1}{4x_L} + \left(\frac{x_L}{r}\right)^{2k}\frac{1}{4x_L} + \frac{2 \alpha h x_L}{r} \leq B_0$$
for $B_0 = 4r^{-1} + 2 \alpha h$. By Lemma \ref{pos}, we know that the functions $R_L(\cdot,\cdot)$, $C_L(\cdot,\cdot)$ and $M_L(\cdot)$ are non negative, as is $\psi(x)$ for $x \geq 0$,
so from (\ref{eqZ}) we get the lower bound $\pars K_L(s)\ge g_L(K_L(s))$. Since $K_L(0)=r$, it follows that $K_L(s) \geq x_L$, for all $x \geq 0$, so $K_L(s) \geq r - B_0 L^{-1}$, for $L \geq L_0$.

Turning now to the complementary upper bound,
recall that $\psi(x)$ is a polynomial of degree $m-1$, hence
there exists $\kappa<\infty$ such that
$\psi(a b) \leq \kappa (1+a^2)^{m/2} (1+b^2)^{m/2}$ for all $a,b$.
Thus, by (\ref{eq:rbd}), the monotonicity of $\psi(x)$ on $\R_+$
and the non-negative definiteness of
$C_L(s,u)$ we have that for any $s,t,u \geq 0$,
$$
\psi(C_L(s,u))\le \kappa (1+K_L(u))^{\frac{m}{2}}
(1+K_L(s))^{\frac{m}{2}}$$
and
$$\int_0^t R_L(s,u)
du\le \sqrt{t K_L(s)},\qquad M_L(t)\le \sqrt{K_L(t)},
$$
and from \eqref{eqZ} we find that
\begin{equation}\label{eq:Klbd}
\pars K_L(s) \leq g(K_L(s)) + 2 \beta^2 \kappa
\left(1+\sup_{u \leq s} K_L(u)\right)^m \sqrt{K_L(s)} \sqrt{s} + 2h\sqrt{K_L(s)}  \,.
\end{equation}
Setting now $B(T)=\frac{1}{2r}\left(1 + 2\sqrt{r+1} \beta^2 \kappa (r+2)^m \sqrt{T}+2\sqrt{r+1}h\right)$ and
fixing $T<\infty$ and $L \geq \max\{L_0, B(T)\}$, let
$$
\tau := \inf\{ u\ge 0: K_L (u)\ge r+B(T) L^{-1}\} \,.
$$
By the continuity of $K_L (\cdot)$ and the fact that
$K_L(0) = r < r+B(T) L^{-1}$, we have that
$\tau>0$ and further, if $\tau<\infty$ then necessarily
$$
K_L(\tau)=\sup_{u \leq \tau} K_L(u) = r+B(T) L^{-1} \leq r+1 \,.
$$
Recall that $g_L(x) \leq 1 + 4Lx(r-x)$, whereas
from \eqref{eq:Klbd} we see that if $\tau<\infty$ then
\begin{align*}
\pars K_L (\tau) &\,\le\, 1 - 4 K_L(\tau) B(T) +
2\sqrt{r+1} \beta^2 \kappa  (r+2)^m \sqrt{\tau} + 2\sqrt{r+1}h \\
&\,=\, 2r B(\tau) - 4 K_L(\tau) B(T) \leq 2r B(\tau) - 2 r B(T)\,.
\end{align*}
where the last inequality holds since $L \geq L_0$ implies $K_L(s) \geq r/2$, as previously shown. Recall the definition of $\tau<\infty$ implying that
$\pars K_L (\tau) \ge 0$. Hence the above inequality implies $B(\tau)\ge B(T)$, hence $\tau>T$, for our choice of $B=B(T)$. That is,
$K_L(s) \leq r+B(T) L^{-1}$ for all $s \leq T$ and $L \geq \max\{B(T),B_0\}$,
as claimed.
\hfill$\Box$

Let $\mu_L(s)=f'_L(K_L(s))$, $h_L(s)=\pars K_L(s)$.
Fixing hereafter $T<\infty$ (recall $r>0$ is fixed) and denoting $\tilde L = \max\{L_0, B(T)\}$, we next prove the equi-continuity and uniform boundedness of $(M_L, R_L, C_L, Q_L, K_L, \mu_L, h_L)$, en-route to
having limit points for $(M_L, R_L, C_L, Q_L, K_L)$.
\begin{lem}\label{lem-tight}
The continuous functions $M_L(s), K_L(s), \mu_L(s), h_L(s)$ and their derivatives
are bounded uniformly in $L \geq \tilde L$ and $0 \leq s \leq T$. The same is true for
$C_L(s,t), Q_L(s,t)$ in $L \geq \tilde L$ and $0 \leq s,t \leq T$ and also for $R_L(s,t)$ in $L \geq \tilde L$ and $0 \leq t \leq s \leq T$.
\end{lem}
\noindent{\bf Proof of Lemma \ref{lem-tight}.} Recall that by Lemma \ref{lem-bdd}, for any $L \geq \tilde L$,
\begin{equation}\label{eq:kbdd}
\sup_{s\le T}\left|K_L(s)-r\right|\le \frac{\tilde B}{L} \,.
\end{equation}
where $\tilde B = \max\{B(T),B_0\}$. Consequently, the collections
$\{C_L(s,t), 0 \leq s,t \leq T, L \geq \tilde L\}$ and $\{Q_L(s,t), 0 \leq s,t \leq T, L \geq \tilde L\}$ are uniformly bounded (since both $|C_L(s,t)|$ and $|Q_L(s,t)|$ are bounded above by $\sqrt{K_L(s)K_L(t)}$) and also $\{M_L(s), 0 \leq s \leq T, L\geq \tilde L\}$ (since $M_L(s)\leq \sqrt{K_L(s)}$).
By (\ref{eq:kbdd}) and our choice of $f_L(r)$, we have that
$$
|\mu_L(s)| \leq 2L |K_L(s)-r| + \left(\frac{K_L(s)}{r}\right)^{2k-1} \leq 2 \tilde B + \left(\frac{r+1}{r}\right)^{2k-1},
\quad \forall L \geq \tilde L, s \leq T \,.
$$
By \eqref{gmfor}, the collection
$\{H_L(s,t), 0 \leq t \leq s \leq T, L \geq \tilde L\}$ is also uniformly bounded and since
 $R_L(s,t) = H_L(s,t) \exp\left(-\int_t^s\mu_L(u)du\right)$, the collection $\{R_L(s,t), 0 \leq t \leq s \leq T, L \geq \tilde L\}$ is also uniformly bounded.
Further, since by \eqref{eqZ}:
\begin{equation}\label{hlid}
h_L(s)=1-2 K_L(s) \mu_L(s) + 2\beta^2 \int_0^s \psi(C_L(s,u)) R_L(s,u) du + 2hM_L(s) \,,
\end{equation}
it follows from the uniform boundedness of
$K_L$, $M_L$, $\mu_L$, $C_L$ and $R_L$
that $\{h_L(s), s \in [0,T], L \geq \tilde L\}$ is also uniformly
bounded. By the same reasoning, from \eqref{eqM}, \eqref{eqR}, \eqref{eqC} and \eqref{eqQ},
we deduce that $\pars M_L(s)$, $\pars_1 C_L(s,t)$, $\pars_1 R_L(s,t)$, $\pars_1 Q_L(s,t)$ and $\pars D_L(s)$ are bounded uniformly in $L \geq \tilde L$ and $s,t \in [0,T]$.

Next, differentiating the identity \eqref{gmfor} with respect to $t$,
we get for $f=f_L$ that
$$
\pars_2 H_L (s,t) =
\sum_{n\ge 1}\beta^{2n}\sum_{\s\in \NC_n}
\int_{t=t_1 \le t_2\cdots \le t_{2n}\le s}
\prod_{i\in \cro(\sigma) } \nu''(C_L(t_i,t_{\s(i)})) \prod_{j=2}^{2n} dt_j \,,
$$
where $NC_n$ denotes the finite set of non-crossing involutions
of $\{1,\ldots,2n\}$ without fixed points.
With the Catalan number $|NC_n|$ bounded by $4^n$, and
since $C_L(t_i,t_{\s(i)}) \in [0,r+1]$
for $t_i, t_{\s(i)}\le T$, $L \geq \tilde L$, we thus deduce
by the monotonicity of $x \mapsto \nu''(x)$ that
$$
0 \leq \pars_2 H_L(s,t) \leq \sum_{n\ge 1}\frac{\beta^{2n}}{(2n-1)!}
4^n \left(\nu''(r+1)\right)^n (s-t)^{2n-1} \,,
$$
so $\pars_2 H_L(s,t)$ is finite and
bounded uniformly in $L \geq \tilde L$ and $0 \leq t \leq s \leq T$. Since
$$
\pars_2 R_L (s,t) = \mu_L(t) R_L(s,t) +
e^{-\int_t^s \mu_L(u) du} \pars_2 H_L (s,t) \,,
$$
we thus have that $|\pars_2 R_L (s,t)|$ is also bounded uniformly
in $L \geq B(T)$ and $0 \leq t \leq s \leq T$.

Also, due to the symmetry of $C_L$, $\pars_2 C_L(s,t) = \pars_1 C_L(t,s)$, hence $\pars_2 C_L(s,t)$ is also bounded uniformly in $L \geq \tilde L$ and $0 \leq s,t \leq T$. The same argument applied to $Q$, will show that $\pars_2 Q_L(s,t)$ is also bounded uniformly in $L \geq \tilde L$ and $0 \leq s,t \leq T$.

Turning to deal with $\pars h_L(s)$, setting
$g_L(x):=[f_L'(x) x]'-2rL = 4L(x-r)+\frac {k}{r}\left(\frac{x}{r}\right)^{2k-1} + \frac{\alpha h}{r}$,
we deduce from \eqref{eq:kbdd} that
$|g_L(K_L(s))| \leq 4 \tilde B + \frac{k}{r}\left(\frac{r+1}{r}\right)^{2k-1} + \frac{\alpha h}{r}$ for any $s \leq T$
and $L \geq \tilde L$. Differentiating \eqref{hlid} we find that
$\pars h_L(s)= - 4 L r h_L(s) +\kappa_L(s)$ for
$$
\kappa_L(s)= - 2 g_L (K_L(s)) h_L(s)
+ 2\beta^2 \frac{\pars}{\pars s}\left(\int_0^s \psi(C_L(s,u)) R_L(s,u) du\right) + 2h\pars M_L(s)\,,
$$
which is thus bounded uniformly in $L \geq B(T)$ and $s \leq T$
(in view of the uniform boundedness of $h_L$, $C_L$, $R_L$,
$\pars_1 C_L$, $\pars_1 R_L$ and $\pars M_L$). Further,
recall that $K_L(0)=r$, so by \eqref{eqZ} and our choice of $f_L(\cdot)$
we have that $h_L(0)=1-2rf_L'(r)+2h\alpha=0$, resulting with
$$
h_L(s) = \int_0^s e^{-4 L r (s-u)} \kappa_L(u) du \;.
$$
hence for $L \geq \tilde L$,
\begin{equation}\label{eq:hbd}
\sup_{s \leq T} |h_L(s)|\le \frac{\sup \{ |\kappa_L(u)| : L \geq \tilde L, u \leq T\}}{4 Lr} = \frac{A(T)}{4Lr} < \infty \,,
\end{equation}
where $A(T):=\sup \{ |\kappa_L(u)| : L \geq \tilde L, u \leq T\}<\infty$
and the uniform boundedness of $|\pars h_L(s)|$ follows.

Finally, by definition,
$\pars \mu_L(s) = f''_L(K_L(s)) h_L(s)$, yielding for our choice
of $f_L$ that
$$
|\pars \mu_L(s)| \leq \left(2L+\frac{2k-1}{2r^2}\left(\frac{r+1}{r}\right)^{2k-2}\right) |h_L(s)| \,,
\qquad \forall L \geq \tilde L, s \leq T \,,
$$
which by \eqref{eq:hbd} provides the uniform boundedness of $|\pars \mu_L(s)|$.
\hfill$\Box$

\medskip
\noindent{\bf Proof of Theorem \ref{theo-sphere}.}
In Lemma \ref{lem-tight} we have established that the functions
$(M_L(s), R_L(s,t), C_L(s,t), Q_L(s,t))$, $L \geq \tilde L$ are equi-continuous and
uniformly bounded on their respective domains for $0 \leq s,t \leq T$. Further,
$(K_L(s),\mu_L(s),h_L(s))$ are equi-continuous and uniformly bounded
on $s \in [0,T]$. By the Arzela-Ascoli theorem, the collection
$(M_L, R_L, C_L, Q_L, K_L, \mu_L, h_L)$ has a limit point
$(M, R, C, Q, K, \mu, h)$ with respect to uniform convergence on
$[0,T] \ts (\bD \cap [0,T]^2) \ts [0,T]^7$.

By Lemma \ref{lem-bdd} we know that the limit $K(s)=r$
for all $s \leq T$, whereas by \eqref{eq:hbd} we have that
$h(s)=0$ for all $s \leq T$. Consequently, considering
\eqref{hlid} for the subsequence $L_n \to \infty$ for which
$(M_{L_n},R_{L_n},C_{L_n},Q_{L_n},K_{L_n},\mu_{L_n},h_{L_n})$ converges to
$(M,L,R,C,Q,K,\mu,h)$ we find that the latter must satisfy
\eqref{eqZs} for $k=1$. Further, recalling that $R_L(t,t)=1$,
$C_L(t,t)=K_L(t)$, integrating \eqref{eqM}, \eqref{eqR}, \eqref{eqC} and \eqref{eqQ}
we find that
$M_L(s) = M_L(0)+\int_0^s \tilde M_L(\theta) d\theta$ and
$V_L(s,t)= V_L(t,t) + \int_t^s \tilde V_L(\theta,t) d\theta$, for $V$ any of the functions $R$, $C$ or $Q$, where:
{\allowdisplaybreaks
\begin{align*}
\tilde M_L(\theta) &\,=\, - \mu_L(\theta) M_L(\theta) + \b^2 \int_0^\theta
  M_L(u) R_L(\theta,u) \nu''(C_L(\theta,u)) du + h \\
\tilde R_L(\theta,t) &\,=\, - \mu_L(\theta) R_L(\theta,t) + \b^2 \int_t^\theta
 R_L(u,t) R_L(\theta,u) \nu''(C_L(\theta,u)) du , \\
\tilde C_L(\theta,t) &\,=\, - \mu_L(\theta) C_L(\theta,t) + \b^2 \int_0^\theta
  C_L(u,t) R_L(\theta,u) \nu''(C_L(\theta,u)) du \\
&\quad\, + \b^2 \int_0^t \nu'(C_L(\theta,u)) R_L(t,u) du + hM_L(t), \\
\tilde Q_L(\theta,t) &\,=\, - \mu_L(\theta) Q_L(\theta,t) + \b^2 \int_0^\theta
  Q_L(u,t) R_L(\theta,u) \nu''(C_L(\theta,u)) du \\
&\quad\, + \b^2 \int_0^t \nu'(Q_L(\theta,u)) R_L(t,u) du + hM_L(t)
\end{align*}}
Since $\tilde M_{L_n}$, $\tilde R_{L_n}$, $\tilde C_{L_n}$ and $\tilde Q_{L_n}$ converge uniformly on their domains, for $0 \leq s,t \leq T$, to the right-hand-sides of
\eqref{eqMs}, \eqref{eqRs}, \eqref{eqCs} and \eqref{eqQs}, respectively, we deduce that for each limit point $(M,R,C,Q,\mu)$, the functions $M(s)$, $R(s,t)$, $C(s,t)$ and $Q(s,t)$ are differentiable in $s$ in the region that they are defined and all
limit points satisfy the equations \eqref{eqMs}--\eqref{eqZs}.
Further, since $C_L(s,t)$ and $Q_L(s,t)$ are non-negative definite symmetric kernels, the same properties are inherited by their limits. Similarly, since
$R_L(t,t)=1$ and $R_L(s,t)$ satisfy \eqref{eq:rbd},
the same applies for any limit point $R(s,t)$ and also since $C_L(t,t) \to r$, then $C(t,t) = r$.

Using an argument similar to the one in Lemma \ref{uniqueness}, we show that there exist at most one bounded solution $(M, R, C, Q)$ in $\Ca^1[0,T] \ts \Ca^1(\bD \cap [0,T]^2) \ts \Ca^1_s([0,T]^2) \ts \Ca^1_s([0,T]^2)$ to the system \eqref{eqMs}--\eqref{eqZs}, with initial conditions $C(t,t) = Q(0,0) = r$, $R(t,t) = 1$ and $M(0)=\alpha\sqrt{r}$, $\alpha \in [0,1)$ (actually the uniqueness and the result are true for any choice of starting points, however, it will not be relevant for us).
%---------

In conclusion, when $L \to \infty$ the collection $(M_{L,r},R_{L,r},C_{L,r},Q_{L,r},K_{L,r})$
converges towards the unique solution
$(M_r,R_r,C_r,Q_r,K_r \equiv r)$ of
\eqref{eqMs}--\eqref{eqZs}, as claimed.
\hfill $\Box$

%%%%%%%%%%%%%%%%%%%%%%%%%%%%%%%%%%%%%%%%%%%%%%%%%%%%
%%%%%%%%%%%%%%%%%%%%%%%%%%%%%%%%%%%%%%%%%%%%%%%%%%%%

\section{Convergence to the Pure Spin Model}\label{convergencepurespin}

%%%%%%%%%%%%%%%%%%%%%%%%%%%%%%%%%%%%%%%%%%%%%%%%%%%%
%%%%%%%%%%%%%%%%%%%%%%%%%%%%%%%%%%%%%%%%%%%%%%%%%%%%

Let $(M_r, R_r, C_r, Q_r)$ be the solution of \eqref{eqMs}-\eqref{eqZs}, for $h_r = hr^{\frac{m-1}{2}}$ and the initial conditions $R_r(t,t)=1$, $C_r(t,t) = Q_r(0,0) = r$, $M_r(0) = \alpha\sqrt{r}>0$, $\alpha\in(0,1)$. Set:
$$\tilde \mu_r(s) \,=\, \frac{\mu(sr^{1-m/2})}{r^{m/2-1}}\,.$$
and recall the definitions used in Theorem \ref{theo-sk}:
\begin{align*}
\tilde M_r(s) \,=&\, \frac{M_r(sr^{1-m/2})}{\sqrt{r}}, &\tilde R_r(s,t) \,=&\,R_r(sr^{1-m/2},tr^{1-m/2})\\
\tilde C_r(s,t) \,=&\, \frac{C_r(sr^{1-m/2},tr^{1-m/2})}{r}, &\tilde Q_r(s,t) \,=&\, \frac{Q_r(sr^{1-m/2},tr^{1-m/2})}{r}
\end{align*}
The system \eqref{eqMs}-\eqref{eqZs} thus becomes:
\begin{align}
\pars \tilde M_r(s) &\,=\, - \tilde \mu_r(s) \tilde M_r(s) + h + \b^2 \int_0^s \tilde M_r(u) \tilde R_r(s,u) \frac{\nu''(r\tilde C_r(s,u))}{r^{m-2}} du, && s\geq 0
\label{eqMsSK1} \\
\pars_1 \tilde R_r(s,t) &\,=\, - \tilde \mu_r(s) \tilde R_r(s,t) + \b^2 \int_t^s
\tilde R_r(u,t) \tilde R_r(s,u) \frac{\nu''(r\tilde C_r(s,u))}{r^{m-2}} du, && s\geq t\geq 0
\label{eqRsSK1}\\
\pars_1 \tilde C_r(s,t) &\,=\,
 -\tilde \mu_r(s) \tilde C_r(s,t) +
\b^2 \int_0^s \tilde C_r(u,t) \tilde R_r(s,u) \frac{\nu''(r \tilde C_r(s,u))}{r^{m-2}} du
\label{eqCsSK1}\\
&\,\quad+ \b^2 \int_0^t \frac{\nu'(r\tilde C_r(s,u))}{r^{m-1}} \tilde R_r(t,u) du + h\tilde M_r(t), && s\geq t\geq 0
\nonumber\\
\pars_1 \tilde Q_r(s,t) &\,=\,
 -\tilde \mu_r(s) \tilde Q_r(s,t) +
\b^2 \int_0^s \tilde Q_r(u,t) \tilde R_r(s,u) \frac{\nu''(r \tilde C_r(s,u))}{r^{m-2}} du
\label{eqQsSK1} \\
&\,\quad + \b^2 \int_0^t \frac{\nu'(r\tilde Q_r(s,u))}{r^{m-1}} \tilde R_r(t,u) du + h\tilde M_r(t), && s,t \geq 0
\nonumber
%\frac{1}{2}\pars_s \tilde D_b(s)&\!\!\!\! = \!\!\!\!
%& -\tilde \mu_b(s) \tilde D_b(s) +
%\b^2 \int_0^s \tilde Q_b(u,s) \tilde R_b(s,u) \frac{\nu''(b \tilde C_b(s,u))}{b^{m-2}} %du
%+ \b^2 \int_0^s \frac{\nu'(b\tilde Q_b(s,u))}{b^{m-1}} \tilde R_b(s,u) du + h\tilde %M_b(s),\label{eqDsSK1}
\end{align}
where
\begin{equation}\label{eqZsSK1}
\tilde \mu_r(s) \,=\, \frac{1}{2r^{m/2}}+
\b^2 \int_0^s \frac{\psi(r\tilde C_r(s,u))}{r^{m-1}} \tilde R_r(s,u) du +h\tilde M_r(s).
\end{equation}
and $\tilde C_r(t,t) = \tilde R_r (t,t) = \tilde Q_r(0,0) = 1$, $\tilde M_r(0)=\alpha$, $\tilde C_r(t,s)=\tilde C_r(s,t)$ and $\tilde Q_r(t,s) = \tilde Q_r(s,t)$.

Fixing $T<\infty$, the first step of the proof is to establish, in Lemma \ref{lem-tightSK}, that the
function $\tilde M_r$, $\tilde R_r$, $\tilde C_r$, $\tilde Q_r$ and  $\tilde \mu_r$ are equi-continuous and uniformly bounded. Then we will be able to use Arzela-Ascoli theorem to establish the desired limits.

\begin{lem}\label{lem-tightSK}
The continuous functions $\tilde M_r(s), \tilde \mu_r(s), \tilde C_r(s,t)$ and $\tilde Q_r(s,t)$ and their derivatives are uniformly bounded in  $r \geq 1$ and $0 \leq s,t \leq T$. The same is true for $\tilde R_r(s,t)$ in $r \geq 1$ and $0 \leq t \leq s \leq T$.
\end{lem}
\noindent{\bf Proof of Lemma \ref{lem-tightSK}.} Recall Theorem \ref{theo-sphere} implies that $C_r(s,t)$, $Q_r(s,t)$, $M_r^2(s) \in [0,r]$, for all $0 \leq s,t \leq T$. Hence, by construction, $\tilde C_r(s,t)$, $\tilde Q_r(s,t)$ and $\tilde M_r^2(s)$ take values in the interval $[0,1]$, for every $r > 0$, thus showing the uniform boundedness of $\tilde C_r(s,t), \tilde Q_r(s,t)$ and $\tilde M_r(s)$ on $0 \leq s,t \leq T$ and $r \geq 1$.

Also notice that $\tilde R_r(s,t) = \tilde H_r(s,t)\exp(-\int_t^s \tilde \mu_r(u) du)$, for $s\ge t$, where, by \cite{GM}, $\tilde H_r(s,t)$ satisfies:
\begin{equation}
\label{Hr_tilda}
\tilde H_r(s,t) = 1+\sum_{n\ge 1}\beta^{2n}\sum_{\s\in \NC_n}
\int_{t\le t_1\cdots \le t_{2n}\le s}
\prod_{i\in \cro(\sigma) } \frac{\nu''(r\tilde C_r(t_i,t_{\s(i)}))}{r^{m-2}} \prod_{j=1}^{2n} dt_j
\end{equation}
Since $\nu''(x)$ is a polynomial of degree $m-2$, there exist an universal constant $K_1$ (depending on $\nu''$) such that, for any $r\geq 1$, and $x\in[0,1]$, $\frac{\nu''(rx)}{r^{m-2}}<K_1$. Hence the collection $\left\{\tilde H_r(s,t),0 \leq t \leq s \leq T, r \geq 1\right\}$ is also uniformly bounded (since $\tilde C_r(t_i,t_{\sigma(i)})\in[0,1]$).
Since $\tilde \mu_r(s)\geq 0$, for all $r$ and $s$, then $\tilde R_r(s,t)\leq \tilde H_r(s,t)$. Since $\tilde R_r(s,t)\geq 0$, for all $s,t$, the uniform boundedness of $\left\{\tilde R_r(s,t),0 \leq t \leq s \leq T, r \geq 1\right\}$ is established.

Since $\psi(x)$ is a polynomial of degree $m-1$, there exist an universal constant $K_2$ (depending on $\psi$) such that, for any $r \geq 1$, and $x \in [0,1]$, $\frac{\psi(rx)}{r^{m-1}} < K_2$. Since in addition $\tilde \mu_r(s)\geq 0$, \eqref{eqZsSK1} implies that
the family $\left\{\tilde \mu_r(s),0 \leq s \leq T, r \geq 1\right\}$ is uniformly bounded.

Moving over to the partial derivatives, since by \eqref{eqMsSK1}:
\begin{equation*}
\pars \tilde M_r(s) \,=\,
 - \tilde \mu_r(s) \tilde M_r(s) +
\b^2 \int_0^s \tilde M_r(u) \tilde R_r(s,u) \frac{\nu''(r\tilde C_r(s,u))}{r^{m-2}} du
+ h
\end{equation*}
it follows from the uniform boundedness of
$\tilde \mu_r$, $\tilde M_r$, $\tilde R_r$ and $\tilde C_r$
that the family $\left\{\pars \tilde M_r(s),\, 0 \leq s \leq T, r \geq 1\right\}$ is also uniformly bounded. By similar reasoning, using \eqref{eqRsSK1}, \eqref{eqCsSK1} and \eqref{eqQsSK1}, we show that $\pars_1 \tilde R_r(s,t)$, $\pars_1 \tilde C_r(s,t)$ and $\pars_1 \tilde Q_r(s,t)$ are uniformly bounded in $r \geq 1$ and $0 \leq t \leq s \leq T$ (or $s,t\in [0,T]$, whichever is relevant).

Now, differentiating the identity \eqref{Hr_tilda} with respect to $t$, we get
$$\pars_2 \tilde H_r (s,t) \,=\,
\sum_{n\ge 1}\beta^{2n}\sum_{\s\in \NC_n}
\int_{t=t_1 \le t_2\cdots \le t_{2n}\le s}
\prod_{i\in \cro(\sigma) } \frac{\nu''(r\tilde C_r(t_i,t_{\s(i)}))}{r^{m-2}} \prod_{j=2}^{2n} dt_j \,,
$$
where $NC_n$ denotes the finite set of non-crossing involutions
of $\{1,\ldots,2n\}$ without fixed points.
With the Catalan number $|NC_n|$ bounded by $4^n$, and
since $0\leq \frac{\nu''(r\tilde C_r(t_i,t_{\s(i)}))}{r^{m-2}}\leq K_1$
for $0\leq t_i, t_{\s(i)}\le T$ and $r \geq 1$, we thus deduce that
$$
0 \leq \pars_2 \tilde H_r(s,t) \leq \sum_{n\ge 1}\frac{\beta^{2n}}{(2n-1)!}
(4K_1)^n (s-t)^{2n-1} \,,
$$
so $\pars_2 \tilde H_r(s,t)$ is finite and
bounded uniformly when $r \geq 1$ and $0 \leq t \leq s \leq T$. Since
$$
\pars_2 \tilde R_r (s,t) \,=\, \tilde \mu_r(t) \tilde R_r(s,t) +
e^{-\int_t^s \tilde \mu_r(u) du} \pars_2 \tilde H_r (s,t) \,,
$$
we thus have that $|\pars_2 \tilde R_r (s,t)|$ is also bounded uniformly
in $r \geq 1$ and $0 \leq t \leq s \leq T$.
Also, since $\tilde C_r$ is symmetric, $\pars_2 \tilde C_r(s,t) = \pars_1 \tilde C_r(t,s)$, hence $\pars_2 \tilde C_r(s,t)$ is also bounded uniformly in $r \geq 1$ and $s,t\in [0,T]$. Since $\tilde Q_r$ is also symmetric, we derive the same conclusion about $\pars_2 \tilde Q_r(s,t)$.

Finally, by \eqref{eqZsSK1},
\begin{eqnarray*}
\pars \tilde \mu_r(s) &=&
\b^2 \int_0^s \left[\frac{\psi'(r \tilde C_r(s,u))}{r^{m-2}}\pars_1\tilde C_r(s,u)\tilde R_r(s,u) + \frac{\psi(r \tilde C_r(s,u))}{r^{m-1}}\pars_1 \tilde R_r(s,u)\right]du + \frac{\psi(r)}{r^{m-1}} + h \pars\tilde M_r(s).
\end{eqnarray*}
and since $\psi(x)$ is a polynomial of order $m-1$, it follows that $\psi(rx)/r^{m-1}$ and $\psi'(rx)/r^{m-2}$ are uniformly bounded in $r\geq 1$, $x\in [0,1]$. Since $\pars_1 \tilde C_r$, $\pars_1 \tilde R_r$, $\pars \tilde M_r$ and $\tilde R_r$ are uniformly bounded and $\tilde C_r(s,u)\in [0,1]$, it follows that the functions $\pars \tilde \mu_r(s)$ are uniformly bounded on $0\leq s \leq T$ and $r \geq 1$, thus concluding the proof.
\hfill
$\Box$

\medskip
\noindent{\bf Proof of Theorem \ref{theo-sk}.}
In Lemma \ref{lem-tightSK} we have established that the functions
$\tilde M_r(s)$, $\tilde R_r(s,t)$, $\tilde C_r(s,t)$, $\tilde Q_r(s,t)$ and $\tilde \mu_r(s)$ are equi-continuous and uniformly bounded for $r \geq 1$. By the Arzela-Ascoli theorem, the collection
$(\tilde M_r, \tilde R_r, \tilde C_r, \tilde Q_r, \tilde \mu_r)$ has a limit point
$(M, R, C, Q, \mu)$ with respect to uniform convergence on
$\Ca^1[0,T] \ts \Ca^1(\bD \cap [0,T]^2) \ts \Ca^1_s([0,T]^2) \ts \Ca^1_s([0,T]^2) \ts \Ca^1[0,T]$. Let $r_n$ be an increasing sequence going to infinity, such that $(\tilde M_{r_n}, \tilde R_{r_n}, \tilde C_{r_n}, \tilde Q_{r_n}, \tilde \mu_{r_n})$ converges uniformly to $(M, R, C, Q, \mu)$.

Now, since $\tilde C_r(\theta,u) \in [0,1]$, for all $r \geq 1$ and $\theta,u \geq 0$, the same is true for its limit point $C(\theta,u)$. Since $\nu(\cdot)$ is a polynomial of degree $m$ with the dominant coefficient $\frac{a_m^2}{m!}$, $\psi(x)=\nu'(x) + x\nu''(x)$ is a degree $m-1$ polynomial with dominant coefficient $\frac{a_m^2}{(m-1)!} + \frac{a_m^2}{(m-2)!}$. Recalling that $\tilde \psi(x) = \left[\frac{a_m^2}{(m-1)!} + \frac{a_m^2}{(m-2)!}\right] x^{m-1}$, we can easily see that there exist constant $K_3$ (depending only on $\nu(\cdot)$), such that
$$\sup_{0\leq \theta, r\leq T}\left|\frac{\psi(r C(\theta,u))}{r^{m-1}}-\tilde \psi(C(\theta,u))\right|\leq \frac{K_3}{r}$$
Also, since $\tilde C_r, C\in [0,1]$, there exist $K_4$ such that
$$\left|\frac{\psi(r\tilde C_{r}(\theta,u))}{r^{m-1}}-\frac{\psi(r C(\theta,u))}{r^{m-1}}\right|\leq K_4 |\tilde C_{r}(\theta,u) - C(\theta,u)| \leq K_4 \|\tilde C_r- C\|_\infty\,,$$
for every $r$. Altogether, we have shown that:
$$\frac{\psi(r_n\tilde C_{r_n}(\theta,u))}{r_n^{m-1}}\overset{n\rightarrow\infty}{\longrightarrow} \tilde \psi( C(\theta,u))\,,$$
and the convergence is uniform on $[0,T]^2$. Using this result, together with the uniform convergence of $\tilde C_{r_n}, \tilde R_{r_n}$ and $\tilde M_{r_n}$, we conclude that $\tilde \mu_{r_n}(s)$, as it is defined in \eqref{eqZsSK1}, converges to the right hand side of \eqref{eqZs} for $k=0$ and the convergence is uniform on $[0,T]$.

Furthermore, since $\tilde R_r(t,t)=1$,
$\tilde C_r(t,t)=1$, $\tilde M_r(0)=\alpha$ and $\tilde Q_r(0,0)=1$
integrating \eqref{eqMsSK1}, \eqref{eqRsSK1}, \eqref{eqCsSK1} and \eqref{eqQsSK1}
we find that $\tilde M_r(s) = \alpha + \int_0^s \bar M_r(\theta) d\theta$ and $\tilde V_r(s,t)= \tilde V_r(t,t) + \int_t^s \bar V_r(\theta,t) d\theta$, for $V$ any of the functions $R$, $C$ or $Q$, for:
\begin{align*}
\bar M_r(\theta) \quad=&\quad - \tilde \mu_r(\theta) \tilde M_r(\theta) + \b^2 \int_0^\theta
  \tilde M_r(u) \tilde R_r(\theta,u) \frac{\nu''(r\tilde C_r(\theta,u))}{r^{m-2}} du + h \\
\bar R_r(\theta,t) \quad=&\quad - \tilde \mu_r(\theta) \tilde R_r(\theta,t) + \b^2 \int_t^\theta
 \tilde R_r(u,t) \tilde R_r(\theta,u) \frac{\nu''(r\tilde C_r(\theta,u))}{r^{m-2}} du , \\
\bar C_r(\theta,t) \quad=&\quad - \tilde \mu_r(\theta) \tilde C_r(\theta,t) + \b^2 \int_0^\theta
  \tilde C_r(u,t) \tilde R_r(\theta,u) \frac{\nu''(r\tilde C_r(\theta,u))}{r^{m-2}} du\\
  &\quad  + \b^2 \int_0^t \frac{\nu'(r\tilde C_r(\theta,u))}{r^{m-1}} \tilde R_r(t,u) du + h\tilde M_r(t)\\
\bar Q_r(\theta,t) \quad=&\quad - \tilde \mu_r(\theta) \tilde Q_r(\theta,t) + \b^2 \int_0^\theta
  \tilde Q_r(u,t) \tilde R_r(\theta,u) \frac{\nu''(r\tilde C_r(\theta,u))}{r^{m-2}} du\\
&\quad + \b^2 \int_0^t \frac{\nu'(r\tilde Q_r(\theta,u))}{r^{m-1}} \tilde R_r(t,u) du + h\tilde M_r(t)
\end{align*}
Similar arguments as employed earlier will show that:
\begin{align*}
\frac{\nu''(r\tilde C_{r_n}(\theta,u))}{r_n^{m-2}}&\overset{n\rightarrow\infty}{\longrightarrow} \tilde \nu''( C(\theta,u)),&\frac{\nu'(r\tilde C_{r_n}(\theta,u))}{r_n^{m-1}}&\overset{n\rightarrow\infty}{\longrightarrow} \tilde \nu'( C(\theta,u))\,,
\end{align*}
and the same for $\nu''(Q(\theta,u))$ and $\nu'(Q(\theta,u))$, where the convergence is uniform on $[0,T]^2$. Using this result, together with the uniform convergence of the quad-uple $(\tilde M_{r_n}, \tilde R_{r_n}, \tilde C_{r_n}, \tilde \mu_{r_n})$, we conclude that $\bar M_{r_n}(s)$ converges to the right hand side of \eqref{eqMs} and the convergence is uniform on $[0,T]$.

Similarly we show that $\bar R_{r_n}(s,t)$, $\bar C_{r_n}(s)$ and $\bar Q_{r_n}(s,t)$ converge uniformly on $(s,t)\in[0,T]^2$ to the right hand sides of \eqref{eqRs}, \eqref{eqCs} and \eqref{eqQs}, respectively. Thus, we see that for each limit point $(M, R, C, Q, \mu)$, the functions $M(s)$, $R(s,t)$, $C(s,t)$ and $Q(s,t)$ are differentiable in $s$ on $0\leq s,t \leq T$ and all
limit points satisfy the equations \eqref{eqMs}--\eqref{eqZs}.
Since $\tilde R_{r_n}(t,t)=1$ and the functions $\tilde Q_{r_n}$ and $\tilde C_{r_n}$ are non-negative definite symmetric kernels, the same applies for any limit point $R$, $Q$ or $C$.

Finally, using a Gromwell-type argument similar to the one employed in Lemma \ref{uniqueness}, we show that there exist at most one bounded solution $(M, R, C, Q)$ on $\Ca^1[0,T] \ts \Ca^1(\bD \cap [0,T]^2) \ts \Ca^1_s([0,T]^2) \ts \Ca^1_s([0,T]^2)$ to the system \eqref{eqMs}--\eqref{eqZs}, with initial conditions $C(t,t) = R(t,t) = Q(0,0) = 1$ and $M(0)=\alpha \in (0,1)$.

In conclusion, when $r \to \infty$ the collection $(\tilde M_r, \tilde R_r, \tilde C_r, \tilde Q_r, \tilde \mu_r)$
converges towards the unique solution $(M, R, C, Q, \mu)$ of
\eqref{eqMs}--\eqref{eqZs}, as claimed.
\hfill $\Box$

%%%%%%%%%%%%%%%%%%%%%%%%%%%%%%%%%%%%%%%%%%%%%%%%%%%%
%%%%%%%%%%%%%%%%%%%%%%%%%%%%%%%%%%%%%%%%%%%%%%%%%%%%

\section{FDT regime}\label{fdt_regime}

%%%%%%%%%%%%%%%%%%%%%%%%%%%%%%%%%%%%%%%%%%%%%%%%%%%%
%%%%%%%%%%%%%%%%%%%%%%%%%%%%%%%%%%%%%%%%%%%%%%%%%%%%

\subsection{Proof Preliminaries}

%%%%%%%%%%%%%%%%%%%%%%%%%%%%%%%%%%%%%%%%%%%%%%%%%%%%
%%%%%%%%%%%%%%%%%%%%%%%%%%%%%%%%%%%%%%%%%%%%%%%%%%%%

The arguments that are used for the cases $\b$ and $h$ small and, respectively, $\gamma$ small, are very similar, and we will be treating them in parallel. On the high level, we will use a perturbation argument based on the stability of linear and respectively Ricatti differential equations. From now on, we will refer to the case when $\gamma=\frac{\beta}{h}$ is small as \emph{the first case} and when both $\b$ and $h$ are small as \emph{the second case}.

First, notice that, since $r=1$, making the substitution $U_h(s,t)=U(s/h,t/h)$, for $U$ any of $R, C$ or $Q$ and $V_h(s)=V(s/h)$ for $V$ any of $M$ or $D$, the equations \eqref{eqMs}-\eqref{eqZs} are transformed to:

\begin{align}
\pars M_h(s) &\,=\,
 - \mu_h(s) M_h(s) + 1 + \gamma^2 \int_0^s M_h(u) R_h(s,u) \nu''(C_h(s,u)) du, &&s \geq 0
\label{eqMsh}\\
\pars_1 R_h(s,t) &\,=\,
- \mu_h(s) R_h(s,t) + \gamma^2 \int_t^s
R_h(u,t) R_h(s,u) \nu''(C_h(s,u)) du, &&s\geq t\geq 0
\label{eqRsh} \\
\pars_1 C_h(s,t) &\,=\,
-\mu_h(s) C_h(s,t) +
\gamma^2 \int_0^s C_h(u,t) R_h(s,u) \nu''(C_h(s,u)) du &&
\label{eqCsh}\\
&\,\quad+ \gamma^2 \int_0^t \nu'(C_h(s,u)) R_h(t,u) du + M_h(t), && s \geq t \geq 0
 \nonumber\\
\pars_1 Q_h(s,t) &\,=\,
 -\mu_h(s) Q_h(s,t) +
\gamma^2 \int_0^s Q_h(u,t) R_h(s,u) \nu''(C_h(s,u)) du &&
\label{eqQsh} \\
&\,\quad+ \gamma^2 \int_0^t \nu'(Q_h(s,u)) R_h(t,u) du + M_h(t), && s,t\geq 0
\nonumber
\end{align}
with
\begin{equation}\label{eqZsh}
\mu_h(s)= \frac{1}{2h} + M_h(s) + \gamma^2
\int_0^s \psi(C_h(s,u)) R_h(s,u) du.
\end{equation}

From now on, we will be interested in the behavior of the functions $C(s,t)$ and $Q(s,t)$ only for $s \geq t$ (the rest of the plane will be automatically given, by symmetry). We do need, however, to specify initial conditions for $Q(\cdot,\cdot)$. Defining $D(s):=Q(s,s)$, due to the symmetry of $Q$, the function $D$ will satisfy $\pars D(s) = 2\pars_1 Q(s,s)$, hence:
\begin{align}
\label{eqDs}
\frac{\pars D(s)}{2} &\quad=\quad
 - \mu(s) D(s) + hM(s) +
\beta^2\int_0^s Q(u,s) R(s,u) \nu''(C(s,u)) du \\
&\quad\qquad + \beta^2\int_0^s \nu'(Q(s,u)) R(s,u) du
\nonumber
\intertext{hence it's time transform, $D_h(s):=Q_h(s,s)$ solves:}
\frac{\pars D_h(s)}{2} &\quad=\quad
 - \mu_h(s) D_h(s) + M_h(s) +
\gamma^2 \int_0^s Q_h(u,s) R_h(s,u) \nu''(C_h(s,u)) du
\label{eqDsh}\\
&\quad\qquad + \gamma^2\int_0^s \nu'(Q_h(s,u)) R_h(s,u) du,
\nonumber
\end{align}

In the course of the proof, we will establish that, when either $\gamma$ is small or both $\b$ and $h$ are small, the limits:
\begin{eqnarray}
 M^{\fdt} &=& \lim_{t\ra\infty}  M(t) \,, \label{Mfdt-ex} \\
 R^{\fdt}(\tau) &=& \lim_{t\ra\infty}  R(t+\tau,t) \,, \label{Rfdt-ex}\\
 C^{\fdt}(\tau) &=& \lim_{t\ra\infty}  C(t+\tau,t) \,, \label{Cfdt-ex}\\
 Q^{\fdt}(\tau) &=& \lim_{t\ra\infty}  Q(t+\tau,t) \,, \label{Qfdt-ex}
% D^{\fdt} &=& \lim_{t\ra\infty}  D(t) \,, \label{Dfdt-ex}
\end{eqnarray}
are well-defined for $\tau \geq 0$ and, furthermore, that $R^\fdt$ decays to $0$ exponentially fast (i.e. $0\leq R^{\fdt}(\tau)\leq K_1e^{-K_2\tau}$), for some positive constants $K_1$ and $K_2$ depending on $\b$, $h$ and $\alpha = M(0)$.

Notice that if the FDT limits exist for the functions $M_h, R_h, C_h$ and $Q_h$, the same is true for the functions $M, R, C$ and $Q$. We will establish \eqref{Mfdt-ex}-\eqref{Qfdt-ex} for $(M_h,R_h,C_h,Q_h)$, in the first case, and for $(M,R,C,Q)$ in the second. Also, until further notice, we will drop the $h$ subscript in the regime when $\gamma$ is small (i.e. the first case).

Recalling our notation $\bD=\{(s,t): 0 \leq t \leq s \} \subset \R_+ \ts \R_+$, consider the maps $\Psi_i:(M, R, C, Q) \mapsto (\tilde M_i, \tilde R_i, \tilde C_i, \tilde Q_i)$, $i=1,2$, on
\begin{eqnarray*}
\Aa &=&\big\{ (M, R, C, Q)\in \CC^1 (\R_+) \ts \CC^1(\bD) \ts \CC^1_s (\R_+^2) \ts \CC^1_s (\R_+^2)\; | \; M(0)=\alpha\in(0,1], \\
&& R(t,t) = C(t,t) = Q(0,0)=1, \;
C(s,t)=C(t,s), \; Q(s,t)=Q(t,s) \big\} \,,
\end{eqnarray*}
such that for $s \geq 0$,
{\allowdisplaybreaks
\begin{align}
\pars \tilde M_1(s) &\,=\,
 -\left(\frac{1}{2h}+\tilde M_1(s)\right)\tilde M_1(s) + 1
\label{eqM1P} \\
&\,\quad + \gamma^2 \left(\int_0^s M(u) R(s,u) \nu''(C(s,u))du  - M(s)\int_0^s \psi( C(s,u))  R(s,u) du\right)
\nonumber \\
\pars \tilde M_2(s) &\,=\,
 -\mu_2(s)\tilde M_2(s) + h + \beta^2 \int_0^s M(u) R(s,u) \nu''(C(s,u)du \label{eqM2P}
\intertext{and for $s \geq t \geq 0$:}
 \pars_1 \tilde R_i(s,t) &\,=\,
 - \mu_i(s)\tilde R_i(s,t) + \epsilon_i^2 \int_t^s
\tilde R_i(u,t) \tilde R_i(s,u) \nu''( C(s,u)) du,\label{eqRP}\\
\pars_1 \tilde C_i(s,t) &\,=\,
 - \mu_i(s) \tilde C_i(s,t) + k_i \tilde M_i(t) \label{eqCP} \\
&\quad + \epsilon_i^2 \left( \int_0^s  C(u,t)  R(s,u) \nu''( C(s,u)) du
+ \int_0^t \nu'( C(s,u))  R(t,u) du\right) ,\nonumber\\
\pars_1 \tilde Q_i(s,t) &\,=\,
 - \mu_i(s) \tilde Q_i(s,t) + k_i \tilde M_i(t)
\label{eqQP} \\
&\quad + \epsilon_i^2 \left( \int_0^s  Q(u,t)  R(s,u) \nu''( C(s,u)) du
+ \int_0^t \nu'( Q(s,u))  R(t,u) du\right),
\nonumber
\end{align}}
with initial conditions $\tilde R_i(t,t) = \tilde C_i(t,t) = \tilde Q_i(0,0) = 1$,
$\tilde D_i(t):=\tilde Q_i(t,t)$ and symmetry conditions $\tilde C_i(t,s)=\tilde C_i(s,t)$ and $\tilde Q_i(t,s)=\tilde Q_i(s,t)$, where $\tilde D_i$ satisfies:

\begin{align}
\frac{\pars \tilde D_i(s)}{2} &\,=\,
 - \mu_i(s) \tilde D_i(s) + k_i \tilde M_i(t)
 \label{eqDP} \\
&\,\quad + \epsilon_i^2 \left( \int_0^s  Q(u,s)  R(s,u) \nu''( C(s,u)) du
+ \int_0^s \nu'( Q(s,u))  R(s,u) du\right)\nonumber
\end{align}
and
\begin{align}
\mu_1(s) &\quad=\quad \omega_1(s) + \gamma^2 \int_0^s \psi( C(s,u))  R(s,u) du \quad=\quad \frac{1}{2h}  + \tilde M_1(s) + \gamma^2 \int_0^s \psi( C(s,u))  R(s,u) du \label{eqZ1P}\\
\mu_2(s) &\quad=\quad \omega_2(s) + \beta^2 \int_0^s \psi( C(s,u))  R(s,u) du \quad=\quad \frac{1}{2}  + h M(s) + \beta^2\int_0^s \psi( C(s,u))  R(s,u) du \label{eqZ2P}
\end{align}
and $k_1=1$, $k_2=h$, $\epsilon_1=\gamma$, $\epsilon_2=\beta$ and the functions $\omega_1(s)$, $\omega_2(s)$ are defined implicitly above.

Assuming $(M, R, C, Q) \in \Aa$, then both the Ricatti equation, \eqref{eqM1P} and the linear one, \eqref{eqM2P} have unique solutions in $\CC (\R_+)$ for the initial conditions $\tilde M_i(0)=\alpha$. Thus,
$\mu_i(s)$ are continuous and further, by \cite{GM}
there exists a unique non-negative solution $\tilde R_i(s,t)$
of \eqref{eqRP} which is continuous on $\bD$ (see for example
\eqref{gmfor} for existence, uniqueness and non-negativity of the
solution, and the proof of Lemma \ref{lem-tight}
for the differentiability, hence
continuity of $\tilde R_i(s,t)$). With $C$, $R$ and $\tilde M_i$ continuous, clearly there
is also a unique solution $\tilde C_i(s,t)$ to
\eqref{eqCP} which is continuous on $\bD$ and due to the boundary
condition $\tilde C_i(t,t)=1$, its symmetric extension to $\R_+\ts \R_+$ remains
continuous. By the same reasoning, there exist an unique solution $\tilde D_i(s)$ to \eqref{eqDP}, hence also an unique solution $\tilde Q_i(s,t)$ to \eqref{eqQP} defined on $\bD$ with boundary condition $\tilde Q_i(t,t)=\tilde D_i(t)$. Furthermore, by the boundary conditions, its symmetric extension to $\R_+\ts \R_+$ is differentiable, hence continuous. Thus, $\Psi_i$ is well-defined and $\Psi_i(\Aa)\subset \Aa$.

Notice that the solution $(M^h, R^h, C^h, Q^h)$ of \eqref{eqMsh}-\eqref{eqZsh} is a fixed point of the mapping $\Psi_1$ and also that the solution $(M, R, C, Q)$ of \eqref{eqMs}-\eqref{eqZs} is a fixed point of the mapping $\Psi_2$. We will show that, for sufficiently small $\gamma=\frac{\beta}{h}$, any fixed point of $\Psi_1$ is in the space $\Sa(\d,\rho,a,d)$ and also, for sufficiently small $\b$ and $h$, any fixed point of $\Psi_2$ is in the same space, for a suitable choice of constants $\d,\rho,a,d$, independent of $\b$ and $h$. Here:
\begin{align*}
\Sa(\d,\rho,a,d) \,=&\, \{(M, R, C, Q)\in\Ba(\d,\rho,a,d) :\quad
\forall \tau \geq 0 \,,\; \exists R^{\fdt}(\tau) = \lim_{t \to \infty} R(t+\tau,t) \,,\;\\
&\quad \exists C^{\fdt}(-\tau) =
C^{\fdt}(\tau) = \lim_{t \to \infty} C(t+\tau,t)\,,\quad \exists Q^{\fdt}(-\tau) =
Q^{\fdt}(\tau) = \lim_{t \to \infty} Q(t+\tau,t)\,,\;\\
&\quad \exists M^{\fdt} = \lim_{t \to \infty} M(t)\;,\quad \exists Q^{\infty} = \lim_{t \to \infty} Q(t,0)\;
\}\,,
\intertext{and}
\Ba(\d,\rho,a,d) \,=&\, \{(M, R, C, Q)\in \Aa :\quad 0\leq C(s,t),Q(s,t)\le d,
\,\, 0\le R(s,t)\le \rho e^{-\d (s-t)}, \\
&\quad 0 \leq Q(s,s) \leq \frac{d}{2}, \,\, 0\le M(s) \le a, \, \mbox{ for all } s\ge t\}\,.
\intertext{This of course will imply that the FDT limits \eqref{Mfdt-ex}-\eqref{Qfdt-ex} exist and are in the space:}
\Da(\d,\rho,a,d) \,=&\, \{(M, R, C, Q)\,:\quad R,C,Q:\R \rightarrow \R,\,\, M \in \R_+,\\
&\quad C(\tau) = C(-\tau), \quad Q(\tau) = Q(-\tau), \quad 0\leq C(\tau),Q(\tau) \le d,\\
&\quad 0\le R(\tau)\le \rho e^{-\d h \tau},\quad
0\le Q(0) \le \frac{d}{2}, \quad 0\le M \le a, \quad \mbox{ for all } \tau \geq 0\}\,.
\end{align*}

%%%%%%%%%%%%%%%%%%%%%%%%%%%%%%%%%%%%%%%%%%%%%%%%%%%%
%%%%%%%%%%%%%%%%%%%%%%%%%%%%%%%%%%%%%%%%%%%%%%%%%%%%

\subsection{Invariant Spaces}

%%%%%%%%%%%%%%%%%%%%%%%%%%%%%%%%%%%%%%%%%%%%%%%%%%%%
%%%%%%%%%%%%%%%%%%%%%%%%%%%%%%%%%%%%%%%%%%%%%%%%%%%%

We will begin by finding constants $(\d,\rho,a,d)$ such that $\Sa(\d,\rho,a,d)$ is invariant under the mapping $\Psi_i$.

\begin{prop}\label{Psi}There exist $\gamma_1$, $\beta_1$ and $h_1$, depending only on $\alpha$, and a positive, universal constant $c_1$, such that for our choice of constants  $a=\sqrt{\frac{7}{4}}$, $b=\min\left\{\alpha,\frac{1}{2}\right\}$, $\rho=c_1$, $\d=\frac{b}{2}$ and $d = 2\max\left\{1 + \frac{2(a+1)}{b^2}, \frac{a+1}{b} \right\}$, if  $\gamma:=\frac{\b}{h}<\gamma_1$ and $i=1$ or $\beta<\beta_1$, $h<h_1$ and $i=2$, then
\begin{equation}\label{al2}
\Psi_i(\Ba(\d,\rho,a,d)) \subset \Ba(\d,\rho,a,d)\,,
\end{equation}
and
\begin{equation}\label{al3}
\Psi_i(\Sa(\d,\rho,a,d)) \subset \Sa(\d,\rho,a,d).
\end{equation}
Furthermore, under the same conditions, if $(M,R,C,Q)\in \Ba(\d,\rho,a,d)$, then for every $s\geq 0$:
 \begin{equation}
\label{al4}
\mu_i(s) \geq \omega_i(s)\geq b >0
\end{equation}
\end{prop}
\nn
\noindent{\bf Proof of Proposition \ref{Psi}:} We will start by verifying that \eqref{al2} holds.

We will first be dealing with the bounds on $\tilde M_i$. Here, due to the different nature of the equations \eqref{eqM1P} and \eqref{eqM2P} (Ricatti, respectively linear), our analysis will be different. Indeed \eqref{eqM1P} is equivalent to:
\begin{equation*}
\pars \tilde M_1(s) = - \left(\tilde M_1(s)\right)^2 - \frac{ \tilde M_1(s)}{2h}+ 1 + \gamma^2( I_0(s) - I_1(s))\end{equation*}
for
\begin{eqnarray}
\label{I0}
I_0(s) &=& \int_0^s  M(u)  R(s,u) \nu''( C(s,u)du\\
\label{I1}
I_1(s) &=& M(s)\int_0^s \psi(C(s,u))  R(s,u) du
\end{eqnarray}
Since $(M, R, C, Q)\in \Ba(\d,\rho,a,d)$, then we have the bounds:
\begin{equation}
\label{I01bound}
\gamma^2 |I_0(s)-I_1(s)| \leq \gamma^2(|I_0(s)|+|I_1(s)|) \leq \gamma^2\left[\frac{a\nu''(d)\rho}{\d}+\frac{a\psi(d)\rho}{\d} \right]\le \frac{3}{4}
\end{equation}
for $\gamma$ sufficiently small. For $k=1,2$, define $M_{1,k}(\cdot)$, to be the unique solutions to the Ricatti differential equations:
\begin{equation*}
\pars M_{1,k}(s) = - \left(M_{1,k}(s)\right)^2 - \frac{ M_{1,k}(s)}{2h} + 1 + (-1)^k \frac{3}{4},\qquad M_{1,k}(0) = \alpha
\end{equation*}
Since $\pars M_{1,1}(s) \leq \pars M(s) \leq \pars M_{1,2}(s)$, for every $s$ and all three functions start at the same point, we can sandwich $\tilde M_1(\cdot)$ between $M_{1,1}(\cdot)$ and $M_{1,2}(\cdot)$, hence:
\begin{equation}
\label{boundM1}
\underset{s\in [0,\infty)}{\inf}M_{1,1}(s)\leq M_{1,1}(s)\leq \tilde M_1(s) \leq M_{1,2}(s) \leq \underset{s\in [0,\infty)}{\sup}M_{1,2}(s)
\end{equation}
Define the polynomial $P_2(x)=-\left(x^2+\frac{x}{2h}-\frac{7}{4}\right)$. Since its only positive root is $x_2=-\frac{1}{4h}+\sqrt{\frac{1}{(4h)^2}+\frac{7}{4}}<\sqrt{\frac{7}{4}}$ and $M_{1,2}(0)=\alpha\in(0,1)$, a sign analysis of $\pars M_{1,2}$ will show that $M_{1,2}$ must be monotonic on $[0,\infty)$ and $\underset{t\rightarrow\infty}{\lim}M_{1,2}(t) = x_2$. So
\begin{equation*}
\underset{s\in [0,\infty)}{\sup}M_{1,2}(s)\le \max\{\alpha,x_2\}\le \max\left\{\alpha,\sqrt{\frac{7}{4}}\right\}<\sqrt{\frac{7}{4}}
\end{equation*}
which, together with \eqref{boundM1} establishes the upper bound on $\tilde M_1$:
\begin{equation}
\label{bound1M1final}
\tilde M_1(s)\leq \sqrt{\frac{7}{4}}=a
\end{equation}
Define also the polynomial $P_1(x)=-\left(x^2+\frac{x}{2h}-\frac{1}{4}\right)$. Again, its only positive root is $x_1=-\frac{1}{4h}+\sqrt{\frac{1}{(4h)^2}+\frac{1}{4}}>0$ and since  $M_{1,1}(0)=\alpha\in(0,1)$, analyzing the sign of $\pars M_{1,1}$, we conclude that $M_{1,1}$ is monotonic on $[0,\infty)$ and $\underset{t\rightarrow\infty}{\lim}M_{1,1}(t) = x_1$, so:
\begin{equation*}
\underset{\s\in [0,\infty)}{\inf}M_{1,1}(s)\ge \min\{\alpha,x_2\} > 0
\end{equation*}
Combining the above inequality with \eqref{boundM1} and \eqref{bound1M1final} we will finish establishing the desired bounds on $\tilde M_1$:
\begin{equation}
\label{bound2M1final}
0 \leq \tilde M_1(s) \leq a
\end{equation}
The bound on $\omega_1$ will follows suit:
\begin{align}
\label{bound3M1final}
\omega_1(s) \,=&\, \frac{1}{2h}+\tilde M_1(s) \,\ge\, \frac{1}{2h}+\underset{s\in [0,\infty)}{\inf}M_{1,1}(s) \\
\ge&\, \min\left\{\alpha+\frac{1}{2h}, \frac{1}{4h}+\sqrt{\frac{1}{(4h)^2}+\frac{1}{4}}\right\}
\,>\, \min\left\{\alpha,\frac{1}{2}\right\} \,=\, b
\nonumber
\end{align}
Furthermore, since $C$ and $R$ are positive, we are done proving \eqref{al4} for $i=1$.

Now, turning our attention towards $M_2(s)$, first define, for $i=1,2$:
\begin{equation}\label{eqL}
\L_i(s,t)=e^{-\int_t^s \mu_i(u) du} \geq 0\,,
\end{equation}
Solving the linear equation \eqref{eqM2P} (recall $\tilde M_2(0) = \alpha$), we obtain:
\begin{equation}
\label{solutiontM2}
\tilde M_2(s) = \alpha\L_2(s,0) + \beta^2\int_0^s I_0(u)\L_2(s,u)du + h\int_0^s \L_2(s,u)du
\end{equation}
with $I_0$ defined in \eqref{I0}.
Since $\alpha>0$ and $M,R,C$ are positive, then the {\bf RHS} above is positive, hence $\tilde M_2(s)\geq 0$. This implies $\mu_2(s)\geq \omega_2(s)\geq \frac{1}{2}\geq b$, proving \eqref{al4} for $i=2$ and consequently $\L_i(s,t)\leq \exp(-b(s-t))$. Also, since $(M, R, C, Q)\in \Ba(\d,\rho,a,d)$, $I_0(u)$ is positive and bounded above uniformly by $\frac{a \nu''(d)\rho}{\delta}$, hence recalling that $\alpha<1$, we obtain the desired upper bound on $\tilde M_2$:
\begin{equation*}
\tilde M_2(s) \leq 1 + \beta^2\frac{a \nu''(d)\rho}{b \delta} + h\frac{1}{b} \leq \sqrt{\frac{7}{4}} = a
\end{equation*}
holding for $h, \beta$ small enough, as claimed.

Considering next the functions $\tilde R_i$, let $\tilde R_i(s,t)= \L_i(s,t) \tilde H_i(s,t)$, where $\L_i$ is defined as in \eqref{eqL}, with $\tilde H_i(t,t)=1$. Further, from \cite{GM} we have that
for any $(s,t) \in \bD$,
\begin{equation}\label{eqtH}
\tilde H_i(s,t)=1+\sum_{n\ge 1}\epsilon_i^{2n} \sum_{\sigma\in\NC_n}
\int_{t\le t_1\cdots\le t_{2n}\le s}
\prod_{k\in \cro(\sigma) }
\nu''( C(t_k,t_{\sigma_k}))\prod_{j=1}^{2n} dt_j \,.
\end{equation}
Consequently, since $|\NC_n|= (2\pi)^{-1}\int_{-2}^2 x^{2n} \sqrt{4-x^2} dx$
and $C(u,v) \in [0,d]$, by the definition of $\Ba(\d,\rho,a,d)$, we can bound $\tilde H_i$:
\begin{eqnarray}\label{eqtHbd1}
\tilde H_i(s,t)&\le& \sum_{n\ge 0} \left(\epsilon_i^2 \nu''(d)\right)^n
\sum_{\s\in\NC_n}
\int_{t\le t_1\le\cdots\le t_{2n}\le s}  \prod_{j=1}^{2n} dt_j\\
&=& \sum_{n\ge 0} \frac{(\epsilon_i^2 \nu''(d))^n (s-t)^{2n}}{(2n !)}
(2\pi)^{-1}\int_{-2}^{2} x^{2n} \sqrt{4-x^2} dx \nonumber \\
&=& (2\pi)^{-1}\int_{-2}^2 e^{\epsilon_i \sqrt{\nu''(d)} (s-t) x} \sqrt{4-x^2} dx \,.
\nonumber
\end{eqnarray}
It is well known (see for example \cite[(3.8)]{2001}) that
%$$
%\pi^{-1}\int_{-2}^2 e^{\theta x} \sqrt{4-x^2} dx \approx_{\theta \ra \infty}
%(1+|\theta| )^{-\frac{3}{2}} e^{2 |\theta| }$$
%so that we can find
for some universal constant $1 \leq c_1 <\infty$ and all $\theta$,
$$
(2\pi)^{-1}\int_{-2}^2 e^{\theta x} \sqrt{4-x^2} dx\le
c_1(1+|\theta|)^{-3/2}\, e^{2 | \theta |}\,,
$$
from which we thus deduce that:
\begin{equation}\label{eq:tHbd}
\tilde H_i(s,t)\le c_1
\left(1+ \epsilon_i \sqrt{\nu''(d)} (s-t)\right)^{-3/2}\, e^{2\epsilon_i \sqrt{\nu''(d)}(s-t)}\le c_1 e^{2\epsilon_i \sqrt{\nu''(d)}(s-t)}\,.
\end{equation}
Further, since $(M, R, C, Q)\in \Ba(\d,\rho,a,d)$ and $\L_i(s,t)\le e^{-b(s-t)}$, then for $\epsilon_i \le \frac{b}{4\sqrt{\nu''(d)}}$ and for our choice of $\rho=c_1$, and $\d=\frac{b}{2} \leq b-2\epsilon_i\sqrt{\nu''(d)}$, we can establish the desired upper bound on $\tilde R_i$:
\begin{eqnarray}
\label{boundR1final}
\tilde R_i(s,t) \,\le\, c_1 e^{-\left(-b+2\epsilon_i\sqrt{\nu''(d)}\right) (s-t)} \,\le\, \rho e^{-\d (s-t)} \,.
\end{eqnarray}
Finally, since $\L_i > 0$ and $\tilde H_i > 0$ (since $ C \geq 0$), the lower bound on $\tilde R_i$ follows:
\begin{equation}
\label{boundR2final}
\tilde R_i(s,t) \ge 0
\end{equation}

Considering next the function
$\tilde C_i$, recall that $\tilde C_i(t,t)=1$, hence solving the linear equation \eqref{eqCP}, we get, for $(s,t) \in \bD$:
\begin{eqnarray}
\tilde C_i(s,t) &=& \L_i(s,t)+ \epsilon_i^2 \int_t^s \L_i(s,v) I_2(v,t)dt +
\epsilon_i^2 \int_t^s \L_i(s,v) I_3(v,t)dt + k_i\tilde M_i(t)\int_t^s \L_i(s,v)dv
\label{solutiontC}
\end{eqnarray}
where
\begin{eqnarray}
\label{I2}
I_2(v,t) &=& \int_0^v   C(u,t)  R(v,u) \nu''( C(v,u)) du \\
\label{I3}
I_3(v,t) &=& \int_0^t  \nu'(C(v,u))  R(t,u) du
\end{eqnarray}
Since $\L_i, C, R$ and $\tilde M_i$ are positive, then $I_2(v,t), I_3(v,t)\geq 0$ (recall $\nu$ is a polynomial with positive coefficients). Hence the lower bound on $\tilde C_i$ follows easily from \eqref{solutiontC}:
\begin{equation}
\label{boundC1final}
\tilde C_i(s,t)\geq 0
\end{equation}
Now, for the upper bound, since $(M, R, C, Q)\in \Ba(\d,\rho,a,d)$, $I_2$ and $I_3$ are bounded above, uniformly by $\frac{d\nu''(d)\rho}{\delta}$ and $\frac{\nu'(d)\rho}{\delta}$, respectively, hence, \eqref{solutiontC} implies:
\begin{eqnarray}
\label{boundC2final}
\tilde C_i(s,t) &\le& e^{-b(s-t)} + \int_t^s e^{-b(s-v)} dv\left[\epsilon_i^2\left( \frac{d\nu''(d)\rho}{\delta}+ \frac{\nu'(d)\rho}{\delta}\right)+k_ia\right]\\
&\le& 1 + \frac{1}{b}\left[\epsilon_i^2\left( \frac{d\nu''(d)\rho}{\delta}+ \frac{\nu'(d)\rho}{\delta }\right)+ak_i\right]\le 1 + \frac{a+1}{b} < d
\nonumber
\end{eqnarray}
whenever $\epsilon_i^2\left(\frac{d\nu''(d)\rho}{\delta}+\frac{\nu'(d)\rho}{\delta}\right)<1$ and $k_i \leq 1$ (i.e. $\gamma$ is small enough for $i=1$ and $\beta$ is small enough and $h\leq 1$, for $i=2$, respectively).

Now, for $\tilde D_i(s)=\tilde Q_i(s,s)$, recalling that $\tilde D_i(0)=1$, solving \eqref{eqDP} we get:
\begin{eqnarray}
\label{solutiontD}
\tilde D_i(s) &=& \L_i^2(s,0) + 2\epsilon_i^2 \int_0^s \L_i^2(s,v) I_4(v,0)dt +
2\epsilon_i^2 \int_0^s \L_i^2(s,v) I_5(v,0)dt \\
&& + 2 k_i\tilde M_i(0)\int_0^s \L_i^2(s,v)dv
\nonumber
\end{eqnarray}
where
\begin{eqnarray}
\label{I4}
I_4(v,t) &=& \int_0^v   Q(u,t)  R(v,u) \nu''( C(v,u)) du \\
\label{I5}
I_5(v,t) &=& \int_0^t  \nu'(Q(v,u))  R(t,u) du
\end{eqnarray}
Notice that $I_4$ and $I_5$ share the same uniform bounds as $I_2$ and $I_3$, respectively. Recalling that $\L_i(s,t)\leq \exp(-b(s-t))$, we establish the bound:
\begin{equation}
0 \leq \tilde D_i(s) \leq 1 + \frac{2}{b^2}\left[\epsilon_i^2\left( \frac{d\nu''(d)\rho}{\delta}+ \frac{\nu'(d)\rho}{\delta }\right)+ak_i\right]\le 1 + \frac{2(a+1)}{b^2} \leq \frac{d}{2}
\label{boundDfinal}
\end{equation}
for $\gamma$ small for $i=1$ and for $\b, h$ small for $i=2$.

Moving over to $\tilde Q_i(s,t)$, since $\tilde Q_i(s,s) = \tilde D_i(s)$, we can solve the linear equation \eqref{eqQP}:
\begin{align}
\tilde Q_i(s,t) &\,=\, \tilde D_i(t)\L_i(s,t)+ \epsilon_i^2 \int_t^s \L_i(s,v) I_4(v,t)dt +
\epsilon_i^2 \int_t^s \L_i(s,v) I_5(v,t)dt + k_i\tilde M_i(t)\int_t^s \L_i(s,v)dv\,,
\label{solutiontQ}
\end{align}
where $I_4$ and $I_5$ are defined by \eqref{I4} and \eqref{I5}, respectively. Using the same bounds on $\L_i, I_4$ and $I_5$ as above, as well as the controls on $\tilde D_i$ provided by \eqref{boundDfinal}, we show that:
\begin{equation*}
0 \leq \tilde Q_i(s,t) \leq \tilde D_i(t) + \frac{1}{b}\left[\epsilon_i^2\left( \frac{d\nu''(d)\rho}{\delta}+ \frac{\nu'(d)\rho}{\delta }\right)+ak_i\right]\le 1 + \frac{2(a+1)}{b^2} + \frac{a+1}{b} \leq d
\end{equation*}
thus concluding the proof.

So, indeed, for our choices of $a$, $\rho$, $\d$, $d$ and $b$, $(\tilde M_i, \tilde R_i, \tilde C_i, \tilde Q_i) \in \Ba(\d,\rho,a,d)$, for sufficiently small $\gamma=\frac{\b}{h}$ ($i=1$) or sufficiently small $\b$ and $h$ ($i=2$), thus showing \eqref{al2}. Furthermore, $\mu_i(s) \geq \omega_i(s) \geq b$, hence \eqref{al4} is true, under the same regime as above, as claimed.

\bigskip
Our next task is to verify that \eqref{al3}, the second statement of the theorem, holds. Namely,
assuming that $(M, R, C, Q) \in \Sa(\d,\rho,a,d)$ we are to show that the
limits $(\tilde M_i^{\fdt}, \tilde R_i^{\fdt}, \tilde C_i^{\fdt}, \tilde Q_i^{\fdt})$ exist for the solution
$(\tilde M_i, \tilde R_i, \tilde C_i, \tilde Q_i)$ of \eqref{eqRP}--\eqref{eqZ2P}. The main idea used in this section of the proof is to use the exponential decay of $R$ and $\L_i$ to bound all the relevant integrals by ${\bf L}^1$ functions and then apply dominated convergence theorem in order to show the existence of the desired limits. To this end, recall that by \eqref{eqM1P}, \eqref{solutiontM2}, \eqref{solutiontC}, \eqref{solutiontQ}, \eqref{solutiontD}, \eqref{eqL} and \eqref{eqtH}, for any $t \geq 0$ and $\tau \geq v \geq 0$,
\begin{align*}
\pars \tilde M_1(s) \,=&\, - \left(\tilde M_1(s)\right)^2 - \frac{\tilde M_1(s)}{2h}+ 1 +\gamma^2\left(I_0(s)-I_1(s)\right)\\
\tilde M_2(s) \,=&\, \alpha \L_2(s,0) + \beta^2 \int_0^s I_0(u)\L_2(s,u)du + h \int_0^s \L_2(s,u)du\\
\tilde C_i(t+\tau,t) \,=&\, \L_i(t+\tau,t) +
\epsilon_i^2 \int_0^{\tau} \L_i (t+\tau,t+v) I_2(t+v,t) dv\\
&\, + \epsilon_i^2 \int_0^\tau \L_i (t+\tau,t+v) I_3(t+v,t) dv + k_i\tilde M_i(t) \int_0^\tau \L_i (t+\tau,t+v) dv\\
\tilde Q_i(t+\tau,t) \,=&\, \tilde D_i(t)\L_i(t+\tau,t) +
\epsilon_i^2 \int_0^{\tau} \L_i (t+\tau,t+v) I_4(t+v,t) dv\\
&\, + \epsilon_i^2 \int_0^\tau \L_i (t+\tau,t+v) I_5(t+v,t) dv + k_i\tilde M_i(t) \int_0^\tau \L_i (t+\tau,t+v) dv\\
\tilde D_i(t) \,=&\, \L_i^2(t,0) +
2\epsilon_i^2 \int_0^t \L_i^2 (t,v) I_4(v,0) dv
+ 2\epsilon_i^2 \int_0^t \L_i^2 (t,v) I_5(v,0) dv + 2k_i\tilde M_i(0) \int_0^t \L_i^2 (t,v) dv\\
\tilde R_i(t+\tau,t) \,=&\, \L_i(t+\tau,t) \tilde H_i(t+\tau,t) \\
\tilde H_i(t+\tau,t) \,=&\, 1 +
\sum_{n\ge 1}\beta^{2n} \sum_{\s\in\NC_n}
\int_{0\le \theta_1\le \cdots\le \theta_{2n}\le \tau}
\prod_{i\in \cro(\s)} \nu''( C(t+\theta_i,t+\theta_{\sigma(i)}))
\prod_{j=1}^{2n} d\theta_j\\
\L_1(t+\tau,t+v) \,=&\, \exp\left(-\frac{\tau-v}{2h} - I_6(t+\tau,t+v) - \gamma^2 \int_v^\tau I_7(t+u,t) du\right)\\
\L_2(t+\tau,t+v) \,=&\, \exp\left(-\frac{\tau-v}{2} - hI_8(t+\tau,t+v) - \beta^2 \int_v^\tau I_7(t+u,t) du\right)
\end{align*}
where $I_0$ and $I_1$ are given by \eqref{I0} and \eqref{I1}, respectively and:
\begin{align}
%\label{I0t}
%I_0(s) &=& \int_{-s}^0  M(u+s)  R(s,s+u) \nu''( C(s,s+u)du\\
%\label{I1t}
%I_1(s) &=& M(s)\int_{-s}^0 \psi(C(s,s+u))  R(s,s+u) du\\
\label{I2t}
I_2(t+\tau,t) &= \int_{-t}^{\tau}   C(t+u,t)  R(t+\tau,t+u) \nu''(C(t+\tau,t+u)) du \\
\label{I3t}
I_3(t+\tau,t) &= \int_{-t}^0  \nu'(C(t+\tau,t+u))  R(t,t+u) du\\
\label{I4t}
I_4(t+\tau,t) &= \int_{-t}^\tau   Q(t+u,t)  R(t+\tau,t+u) \nu''( C(t+\tau,t+u)) du \\
\label{I5t}
I_5(t+\tau,t) &= \int_{-t}^0  \nu'(Q(t+\tau,t+u))  R(t,t+u) du\\
\label{I6t}
 I_6(t+\tau,t+v) &= \int_{\tau}^{v}  \tilde M_1(t+u) du \\
\label{I7t}
 I_7(t+\tau,t) &= \int_{-t}^\tau  \psi(C(t+\tau,t+u))  R(t+\tau,t+u) du \\
 \label{I8t}
 I_8(t+\tau,t+v) &= \int_{\tau}^{v}  M(u+t) du
\end{align}
We will show that the limits $\hh{I}_k := \underset{s\rightarrow\infty}{\lim}I_k(s)$ exist for $k = 1,2$ and also that $\hh{I}_k(\tau) := \underset{s\rightarrow\infty}{\lim}I_k(t+\tau,t)$ exist, for $k = 3\dots,8$. For $I_0$, begin by dividing the integral into two parts:
\begin{equation}
\label{eqI0}
I_0(s) = \int_0^{s/2}  M(u)  R(s,u) \nu''( C(s,u))du + \int_{-s/2}^0  M(s+u)  R(s,s+u) \nu''( C(s,s+u))du
\end{equation}
Since $\nu''(\cdot)$ is
continuous and $(M, R, C, Q) \in \Sa(\d,\rho,a,d)$,
as $s \to \infty$ the bounded integrand in the second integral above
converges pointwise to the corresponding
expression for $(R^{\fdt}, C^{\fdt},  M^{\fdt})$.
Further, by the exponential tails of $R$
the afore-mentioned integrands
are uniformly in $s$ bounded by $f(\theta):=a\rho \nu''(d) e^{\d \theta}$, which is integrable on $(-\infty,0]$.
Thus, by dominated convergence theorem, we deduce that
\begin{eqnarray*}
\underset{s\rightarrow\infty}{\lim} \int_{-s/2}^0  M(s+u)  R(s,s+u) \nu''( C(s,s+u))du = \int_0^{\infty}  M^{\fdt}  R^{\fdt}(u) \nu''(C^{\fdt}(u))du
\end{eqnarray*}
The first integral in \eqref{eqI0} is bounded above by $\rho \d^{-1} \nu''(s)(e^{-\d  s/2}-e^{-\d s})$ that converges to $0$ as $s\rightarrow\infty$, hence:
\begin{equation}
\label{eqI0l}
\hh{I}_0:=\underset{s\rightarrow\infty}{\lim}I_0(s)= M^{\fdt}\int_0^{\infty}   R^{\fdt}(u) \nu''(C^{\fdt}(u))du
\end{equation}
Applying a similar argument to $I_1$, we conclude that:
\begin{equation}
\label{eqI1l}
\hh{I}_1:=\underset{s\rightarrow\infty}{\lim}I_1(s)= M^{\fdt}\int_0^{\infty}   R^{\fdt}(u) \psi(C^{\fdt}(u))du
\end{equation}
Now, due to the above limits, for any $0<\epsilon<\frac{1}{8\gamma_1^2}$ there exist $s_{\epsilon}>0$ such that if $s>s_{\epsilon}$, $\left|[I_0(s)-I_1(s)]-[\hh{I}_0-\hh{I}_1]\right|<\epsilon$. Recalling the Ricatti equation \eqref{eqM1P} that characterizes $\tilde M_1$, we can sandwich $\tilde M_1$ between the functions $M_{1,3}$ and $M_{1,4}$ that are defined for $s \geq s_\epsilon$ as the unique solutions of the differential equations:
\begin{equation*}
\pars M_{1,k}(s) = - \left(M_{1,k}(s)\right)^2 - \frac{M_{1,k}(s)}{2h} + 1 + \gamma^2 \left((\hh{I}_0-\hh{I}_1)+(-1)^k\epsilon\right)\,,
\end{equation*}
while for $s \leq s_\epsilon$, $M_{1,3}(s) = M_{1,4}(s) = \tilde M_1(s)$. Using the joint bound on $I_0$ and $I_1$ provided by \eqref{I01bound} and observing that our choice of $\epsilon$ guarantees $\gamma^2\epsilon < \frac{1}{8}$, we can conclude that the polynomials $P_k(X)= -X^2 - \frac{X}{2h} + 1 + \gamma^2\left((\hh{I}_0-\hh{I}_1)+(-1)^k\epsilon\right)$, for $k=3,4$, have exactly one positive root and one negative root. Furthermore, denoting with $x_k(\epsilon)$ the afore-mentioned positive roots, it is easy to see that:
\begin{equation*}
\underset{t\rightarrow\infty}{\lim}M_{1,k}(t)=x_k(\epsilon)=-\frac{1}{4h}+\sqrt{\frac{1}{(4h)^2}+1+\gamma^2 \left(\hh{I}_0-\hh{I}_1+(-1)^k\epsilon\right)}
\end{equation*}
Recalling that $\tilde M_1$ is bounded above by $M_{1,4}$ and below by $M_{1,3}$, we obtain:
\begin{equation*}
x_3(\epsilon)\leq \underset{t\rightarrow\infty}{\liminf}\;\tilde M_1(s) \leq \underset{t\rightarrow\infty}{\limsup} \; \tilde M_2(s) \leq x_4(\epsilon)
\end{equation*}
Since $\underset{\epsilon\rightarrow 0}{\lim} \;x_3(\epsilon)=\underset{\epsilon\rightarrow 0}{\lim} \;x_4(\epsilon)=-\frac{1}{4h}+\sqrt{\frac{1}{(4h)^2}+1+\gamma^2\left(\hh{I}_1-\hh{I}_2\right)}$ we can conclude that:
\begin{equation}
\tilde M_1^{\fdt}:=\underset{t\rightarrow\infty}{\lim}\;\tilde M_1(s)=-\frac{1}{4h}+\sqrt{\frac{1}{(4h)^2}+1+\gamma^2\left(\hh{I}_0-\hh{I}_1\right)}
\label{eqM1fdt}
\end{equation}
Consequently, applying again dominated converge theorem, this time to \eqref{I6t}, we show:
\begin{equation}
\hh{I}_6(\tau, v) := \lim_{t\ra\infty} I_6(t+\tau,t+v) = (\tau-v)\tilde M_1^{\fdt}
\label{eqI6l}
\end{equation}
Also, since $M(s)$ converges as $s\rightarrow \infty$:
\begin{equation}
\hh{I}_8(\tau, v) := \lim_{t\ra\infty} I_8(t+\tau,t+v) = (\tau-v) M^{\fdt}
\label{eqI8l}
\end{equation}

Since $\psi(\cdot)$, $\nu''(\cdot)$ and $\nu'(\cdot)$ are
continuous and $(M, R, C, Q) \in \Sa(\d,\rho,a,d)$,
as $t \to \infty$ the bounded integrands in \eqref{I2t}, \eqref{I3t}, \eqref{I4t}, \eqref{I5t} and \eqref{I7t}  converge pointwise to the corresponding
expression for $(M^{\fdt}, R^{\fdt}, C^{\fdt}, Q^{\fdt})$.
Further, by the exponential tail of $R$,
the integrals over $[-t,-m]$ in afore-mentioned formulas,
are bounded uniformly in $t$ by $\rho \d^{-1} \psi(d) e^{-\d m}$.
Thus, applying dominated convergence theorem for the integrals
over $[-m,v]$, then taking $m \to \infty$, we deduce that
for each fixed $v \geq 0$,
\begin{eqnarray}
\label{eqI2l}
\hh{I}_2(\tau) &:=& \lim_{t\ra\infty} I_2(t+\tau,t) = \int_{0}^{\infty}   C^{\fdt}(\tau-\theta)  R^{\fdt}(\theta) \nu''(C^{\fdt}(\theta)) d\theta\,, \\
\label{eqI3l}
\hh{I}_3(\tau) &:=& \lim_{t\ra\infty} I_3(t+\tau,t) = \int_{\tau}^\infty  \nu'(C^{\fdt}(\theta))  R^{\fdt}(\theta-\tau) d\theta\,,\\
\label{eqI4l}
\hh{I}_4(\tau) &:=& \lim_{t\ra\infty}I_4(t+\tau,t) = \int_{0}^{\infty}   Q^{\fdt}(\tau-\theta)  R^{\fdt}(\theta) \nu''(C^{\fdt}(\theta)) d\theta \,,\\
\label{eqI5l}
\hh{I}_5(\tau) &:=& \lim_{t\ra\infty} I_5(t+\tau,t) = \int_{\tau}^\infty  \nu'(Q^{\fdt}(\theta))  R^{\fdt}(\theta-\tau) d\theta\,,\\
\label{eqI7l}
\hh{I}_7 &:=& \lim_{t\ra\infty} I_7(t+\tau,t) = \int_{0}^\infty  \psi(C^{\fdt}(\theta))  R^{\fdt}(\theta) d\theta\,,
\end{eqnarray}
hence also:
\begin{eqnarray}
\hh{\L}_1(\tau-v) &:=& \lim_{t \to \infty}
\L_1(t+\tau,t+v) =
\exp\left(-(\tau-v)\left(\frac{1}{2h}+\tilde M_1^\fdt+\gamma^2\hh{I}_7\right)\right)
 \label{eqL1l} \\
 &=& \exp (-(\tau-v)\hh\omega_1)\,,
\nonumber \\
%\exp\left(-\frac{\tau-v}{2h}-\hh{I}_6(\tau-v)-\gamma^2 (\tau-v)\hh{I}_7\right)\\
\hh{\L}_2(\tau-v) &:=& \lim_{t \to \infty}
\L_2(t+\tau,t+v) = \exp\left(-(\tau-v)\left(\frac{1}{2}+hM^\fdt+\beta^2\hh{I}_7\right)\right) \label{eqL2l} \\
&=& \exp (-(\tau-v) \hh\omega_2)\,.
\nonumber
%\exp\left(-\frac{\tau-v}{2}-h\hh{I}_8(\tau-v)-\beta^2 (\tau-v)\hh{I}_7\right)\\
\end{eqnarray}
with $\hh \omega_i = -\frac{\log \L_i(t)}{t} \geq b > 0$.

Moving over to $\tilde M_2$, we first split each integral from the right hand side of \eqref{solutiontM2} into $[0,s/2]$ and $[s/2,s]$. Since the integral over $[0,s/2]$ is bounded below by $0$ and above by $[\exp(-bs/2)-\exp(-bs)]a\rho\nu''(d)\d^{-1}$, it converges to $0$ as $s \ra \infty$. The integrand over $[s/2,s]$ is dominated by the integrable function $\exp(-bs)a\rho\nu''(d)\d^{-1}$ hence we can and will apply dominated  convergence theorem, concluding:
\begin{equation}
\label{eqM2l}
\tilde M_2^\fdt : = \lim_{s \ra \infty} \tilde M_2(s) = \frac{\beta^2 \hh I_0 + h}{\hh \omega_2}
\end{equation}
A similar argument will show that
\begin{equation}
\label{eqbarI4}
\bar I_4 : = \lim_{s \ra \infty} I_4(s,0) = Q^\infty \int_0^\infty R^\fdt(u)\nu''(C^\fdt(u))du
\end{equation}
Since trivially $\bar I_5 := \lim_{t \ra \infty}I_5(t,0) = 0$, similar arguments applied to the integrals in \eqref{solutiontD} and \eqref{solutiontC} will show:
\begin{eqnarray}
\label{eqDWl}
\tilde D_i^\fdt &:=& \lim_{t \ra \infty}\tilde D_i(t) = \frac{\epsilon_i^2 \bar I_4 + k_i \alpha}{\hh \omega_i} = \tilde Q_i^\infty := \lim_{t \ra \infty}\tilde Q_i(t,0)
%\label{eqWl}
%\tilde W_i^\fdt &:=& \lim_{t \ra \infty}\tilde Q_i(t,0) = \frac{\epsilon_i^2 \bar I_4 + %k_i \tilde M_i^\fdt}{\hh \omega_i}
\end{eqnarray}

By the preceding discussion we also know that for all $v,t \geq 0$ and $i\in \{2,3,4,5,7\}$,
$0 \leq I_i(t+v,t) \leq \rho \psi(d) \d^{-1}$ and the same bound holds for $I_0(t)$, uniformly in $t$. Since
$0\leq \L_i(t+\tau,t+v) \leq \exp(-b(\tau-v))$, we can bound all the integrands in the right hand sides of \eqref{solutiontC} and \eqref{solutiontQ} by the integrable function $\rho \psi(d) \d^{-1} \exp(-bx)$ and then apply dominated convergence theorem, concluding:
\begin{eqnarray}
\label{eqCfl}
\tilde C_i^{\fdt} (\tau) &:=& \lim_{t \to \infty}
\tilde C_i(t+\tau,t) = \hh{\L}_i(\tau) +
\epsilon_i^2 \int_0^{\tau} \hh{\L}(\tau-v) \hh{I}_2(v) dv
\\
\nonumber
&&\quad + \epsilon_i^2 \int_0^\tau \hh{\L}_i(\tau-v) \hh{I}_3(v) dv
+ k_i\tilde M_i^{\fdt}\int_0^\tau \hh{\L}_i(v) dv\,.\\
\label{eqQfl}
\tilde Q_i^{\fdt} (\tau) &:=& \lim_{t \to \infty}
\tilde Q_i(t+\tau,t) = \tilde D_i^{\fdt}\hh{\L}_i(\tau) +
\epsilon_i^2 \int_0^{\tau} \hh{\L}(\tau-v) \hh{I}_4(v) dv
\\
\nonumber
&&\quad + \epsilon_i^2 \int_0^\tau \hh{\L}_i(\tau-v) \hh{I}_5(v) dv
+ k_i\tilde M_i^{\fdt}\int_0^\tau \hh{\L}_i(v) dv\,.
\end{eqnarray}

We also have that for any $n\in \Z_+$, all $\s\in\NC_n$ and
each fixed $\theta_1,\ldots,\theta_{2n} \geq 0$,
$$
\lim_{t\ra\infty}
\prod_{i\in\cro(\sigma)}\nu''( C (t+\theta_i, t+\theta_{\sigma(i)})) =
\prod_{i \in\cro(\sigma)}\nu''( C^{\fdt}(\theta_i- \theta_{\sigma(i)}))\,,
$$
By dominated convergence, the
corresponding integrals over
$0 \leq \theta_1 \leq \cdots \leq \theta_{2n} \leq \tau$ converge.
Further, the
non-negative series \eqref{eqtH} is dominated in $t$ by a summable
series (see \eqref{eqtHbd1}), so by dominated convergence,
\begin{align}
\label{eqHfl}
\tilde H_i^{\fdt} (\tau) &:=
\lim_{t \to \infty} \tilde H_i(t+\tau,t) \\
&= 1 + \sum_{n\ge 1}\epsilon_i^{2n} \sum_{\s\in\NC_n}
\int_{0\le \theta_1\le \cdots\le \theta_{2n}\le \tau}
\prod_{i\in \cro(\s)} \nu''( C^{\fdt}(\theta_i-\theta_{\sigma(i)}))
\prod_{j=1}^{2n} d\theta_j \,.
\nonumber
\end{align}
It thus follows that
\begin{equation}
\label{eqRfl}
\tilde R_i^{\fdt} (\tau) := \lim_{t \to \infty}
\tilde R_i(t+\tau,t)= \hh{\L}_i(\tau) \tilde H_i^{\fdt} (\tau) \,,
\end{equation}
exists for each $\tau \geq 0$, which establishes our claim \eqref{al3}
(we have already shown that $\tilde M_i^{\fdt}$, $\tilde C_i^\fdt(\tau)$,  $\tilde Q_i^\fdt(\tau)$ and $\tilde Q_i^{\infty}$ exists).
\hfill$\Box$

%%%%%%%%%%%%%%%%%%%%%%%%%%%%%%%%%%%%%%%%%%%%%%%%%%%%
%%%%%%%%%%%%%%%%%%%%%%%%%%%%%%%%%%%%%%%%%%%%%%%%%%%%

\subsection{Contraction Mapping}

%%%%%%%%%%%%%%%%%%%%%%%%%%%%%%%%%%%%%%%%%%%%%%%%%%%%
%%%%%%%%%%%%%%%%%%%%%%%%%%%%%%%%%%%%%%%%%%%%%%%%%%%%

The next step in our proof is to establish that the mappings $\Psi_i$ are contractions on $\Sa(\d,\rho,a,d)$. Thus we will be able to conclude that their unique fixed point, that coincides with the solution of our system, will be stationary in the limit, hence the FDT limits \eqref{Mfdt-ex}-\eqref{Qfdt-ex} are well-defined.

\begin{prop}\label{FDT1}
For $\d, \rho, a, b, d, \gamma_1, h_1, \beta_1$ of Proposition \ref{Psi}, there exist $0<\gamma_2\leq\gamma_1$, $0<\beta_2\leq\beta_1$ and $0<h_2\leq h_1$, such that the mappings $\Psi_i$ are contractions on $\Sa(\d,\rho,a,d)$, equipped with the norm
\begin{align}
\| (M,R,C,Q)\|&:= \sup_{s\in \R_+}|M(s)| + \sup_{s,t \in \R_+} |Q(s,t)| + \sup_{s,t \in \R_+} |C(s,t)| + \sup_{(s,t) \in \bD} |R(s,t)e^{\xi (s-t)}|\,,
\label{eq:unorm}
\end{align}
whenever $\gamma\in[0,\gamma_2]$ (for $i=1$) or $\beta\in[0,\beta_2]$ and $h\in[0,h_2]$ (for $i=2$), for $\xi=\frac{b}{3}>0$. Also
the solution $(M, R, C, Q)$ of \eqref{eqMs}-\eqref{eqZs} is also
the unique fixed point of $\Psi_1$ in $\Sa(\d h,\rho,a,d)$ and of $\Psi_2$ in
$\Sa(\d,\rho,a,d)$. Consequently,
the functions $M^{\fdt}, R^{\fdt}, C^{\fdt}$ and $Q^\fdt$ of \eqref{Mfdt-ex}-\eqref{Qfdt-ex}
are then the unique solution in $\Da(\d h,\rho,a,d)$, respectively $\Da(\d,\rho,a,d)$ of the FDT equations
\begin{align}
\label{eqMFDT}
0 &\quad=\quad - \mu  M
+ h + \beta^2 M\int_0^\infty  R (\theta) \nu''( C (\theta)) d\theta,\\
 R'(\tau) &\quad=\quad
 - \mu  R(\tau) + \beta^2 \int_0^\tau  R(\tau-\theta)
 R(\theta) \nu''( C(\theta))d\theta ,\label{eqRFDT}\\
 C'(\tau) &\quad=\quad
 - \mu  C(\tau)
+\beta^2 \int_0^\infty  C(\tau-\theta)  R (\theta) \nu''(C(\theta)) d\theta + \beta^2 \int_\tau^\infty \nu'( C(\theta))  R(\theta-\tau)d\theta + hM,
\label{eqCFDT}\\
 Q'(\tau) &\quad=\quad
 - \mu  Q(\tau)
+ \beta^2 \int_0^\infty  Q(\tau-\theta)  R (\theta) \nu''(C(\theta)) d\theta + \beta^2 \int_\tau^\infty \nu'( Q(\theta))  R(\theta-\tau)d\theta + hM,
\label{eqQFDT}
\intertext{where}
\mu &\quad=\quad \frac{1}{2} + \beta^2
\int_0^\infty \psi( C(\theta))  R(\theta) d\theta +  hM \,, \label{eqZFDT}
\end{align}
with initial conditions $D(0)=R(0)=1$ and $Q'(0)=0$.
\end{prop}

\noindent {\bf Proof of Proposition \ref{FDT1}:}
Keeping $\d,\rho,a,b$ and $d$ as in Proposition \ref{Psi}, we will show that
$\Psi_i$ is a contraction on $\Sa(\d,\rho,a,d)$ equipped with the
uniform norm $\|((M,R,C,Q)\|$ of \eqref{eq:unorm}, for
any $\gamma$ small enough ($i=1$) or $\beta, h$ small enough ($i=2$). We will first recall
that in Proposition \ref{Psi}, we have shown that if $(M,R,C,Q)\in \Sa(\d,\rho,a,d)$,
then $\omega_i(s)\geq b$, for all $s\geq 0$, a critical fact that we will use in our upcoming proof.
For simplicity of notation, we will denote by $E(s,t)=R(s,t)e^{\xi (s-t)}$

%Then the $\Psi$ can be
%seen as mapping $( Q,  D,  A)$ into $(\tilde Q, \tilde D, \tilde A)$.

Consider a pair of elements in $\Sa(\d,\rho,a,d)$, $(M_k, R_k, C_k, Q_k)$ for $k=1,2$ and consider their images through $\Psi_i$, namely $(\tilde M_{i,k}, \tilde R_{i,k}, \tilde C_{i,k}, \tilde Q_{i,k}) = \Psi_i(M_k, R_k, C_k, Q_k)$ for $i = 1,2$. We will also use the already established notation $D_k(s):=Q_k(s,s)$. We will denote hereafter in short $\D f(s,t)=f_1(s,t)- f_2(s,t)$
and $\bar \D f(s)=\sup_{0\le u\le v\le s} |\D f(v,u)|$ when $f$ is
one of the functions of interest to us, such as $Q$, $C$, $R$, $E$,
$\L$ or $H$. A similar notation will be used for functions $f$ of only one variable, for example $M$ or $D$, namely $\D f(s)=f_1(s)- f_2(s)$  and $\bar \D f(s)=\sup_{0\le u\le s} |\D f(u)|$

Denoting by $\vartheta_1 = \gamma^2$ and $\vartheta_2 = \beta^2 + h$, we shall show that for $i=1,2$, there exist finite positive constants $L_{M,i}, L_{E,i}, L_{C,i}$ and $L_{Q,i}$ depending on $\d, \rho, a, b$ and $d$, such that for any finite $s \geq 0$,
\begin{eqnarray}
\bar \D \tilde M_i (s) &\leq& \vartheta_i L_{M,i} [ \bar \D  M(s) + \bar \D  E(s) + \bar \D  C(s) + \bar \D  Q(s)] \,,
\label{eq:lipM}\\
\bar \D \tilde E_i (s) &\leq& \vartheta_i L_{E,i} [ \bar \D  M(s) + \bar \D  E(s) + \bar \D  C(s) + \bar \D  Q(s)] \,,
\label{eq:lipE}\\
\bar \D \tilde C_i (s) &\leq& \vartheta_i L_{C,i} [ \bar \D  M(s) + \bar \D  E(s) + \bar \D  C(s) + \bar \D  Q(s)] \,, \label{eq:lipC} \\
\bar \D \tilde Q_i (s) &\leq& \vartheta_i L_{Q,i} [ \bar \D  M(s) + \bar \D  E(s) + \bar \D  C(s) + \bar \D  Q(s)]
\label{eq:lipQ}
\end{eqnarray}
whenever $\gamma \in [0,\gamma_1]$ and $i=1$ or $h\in[0,h_1]$, $\beta \in [0,\beta_1]$ and $i=2$. Here $\gamma_1$, $h_1$, $\beta_1$, $a$, $d$, $\rho$, $\d$ and $b$ are the ones of Proposition \ref{Psi}.

So, if $\vartheta_i$ is small enough (i.e. $\vartheta_i \leq \min\left\{1/(5L_M), 1/(5L_E), 1/(5L_C),  1/(5L_Q)\right\}$, $\vartheta_1 \leq \gamma_1^2$ and $\vartheta_2 \leq \b_1^2+h_1$), then from \eqref{eq:lipM}-\eqref{eq:lipQ} we deduce that
\begin{align*}
\| (\Delta \tilde M_i, \Delta \tilde R_i, \Delta \tilde C_i, \Delta \tilde Q_i)\| &\,=\, \sup_{s \geq 0} \bar \D \tilde M_i(s) + \sup_{s \geq 0} \bar \D \tilde E_i(s) + \sup_{s \geq 0} \bar \D \tilde C_i(s) + \sup_{s \geq 0} \bar \D \tilde Q_i(s)\\
&\,\leq\, \frac{4}{5} \left[\sup_{s \geq 0} \bar \D  M(s) + \sup_{s \geq 0} \bar \D  E(s) + \sup_{s \geq 0} \bar \D  C(s) + \sup_{s \geq 0} \bar \D  Q(s)\right]\\
&\,=\, \frac{4}{5} \| (\Delta M, \Delta R, \Delta C, \Delta Q)\| \,.
\end{align*}
In conclusion, the mapping
$\Psi_i$ is then a contraction on $\Ba(\d,\rho,a,d)$, since
\begin{equation}\label{eq:Psicont}
\| \Psi_i(M_1, R_1, C_1, Q_1) - \Psi_i(M_2, R_2, C_2, Q_2) \|
\leq \frac{4}{5} \| (M_1, R_1, C_1, Q_1) - (M_2, R_2, C_2, Q_2) \| \,,
\end{equation}
whenever $(M_k, R_k, C_k, Q_k) \in \Ba(\d,\rho,a,d)$, for $k=1,2$.

From now until the end of the proof, for simplifying the notations, we will denote:
$${\bf \bar \Delta}(s)\, :=\, \bar \D  M(s) + \bar \D  E(s) + \bar \D  C(s) + \bar \D  Q(s)$$

Before we start, recall that $I_{0,k}$ is defined by \eqref{I0} for $(M_k, C_k, R_k, Q_k)$ and $I_{1,k}$ is defined by \eqref{I1}. Notice that for every $i \in \{0,1\}$ and $k \in \{1,2\}$, that $I_{i,k}$ is of the form $\int_0^s R_k(s,u) T_{k;i}(u;s,t)du$, where $T_{k;i}(u;s,t)$ are polynomial function depending only on $C_k(\theta_1,\theta_2)$, $M_k(\theta_1)$ and $Q_k(\theta_1,\theta_2)$, for $\theta_1, \theta_2 \in \{s,t,u\}$. By the definition of $\Sa(\d,\rho,a,d)$ the family $\left\{\int_0^s R_k(s,u)du\right\}_{s \geq 0}$ is uniformly bounded above by $\rho\d^{-1}$, hence:
\begin{align}
\label{generalIbound1}
0 &\leq I_{i,k}(s) \leq K_I, && i = 0,1
\intertext{for $K_I = \frac{\rho}{\d} \phi(d) \max\{a,d\}$. Similar arguments will show also that:}
\label{generalIbound2}
0 &\leq I_{i,k}(s,t) \leq K_I && i = 2,3,4,5,7
\end{align}
where $I_{2,k}$, $I_{3,k}$, $I_{4,k}$, $I_{5,k}$ and $I_{7,k}$ are defined by \eqref{I2}, \eqref{I3}, \eqref{I4}, \eqref{I5} and \eqref{I7t}, respectively, for $(M, R, C, Q) = (M_k, R_k, C_k, Q_k)$. Now consider the difference between $I_{0,1}$ and $I_{0,2}$. Since $\int_0^t |\D R(t,u)|du \leq \frac{\bar \D E(t)}{\xi}$(by the definition of $E_k$), the difference between $I_{0,1}$ and $I_{0,2}$ can be controlled, yielding:
\begin{equation*}
\bar \D I_{0}(s) \leq L_{I_0}[ \bar \D  M(s) + \bar \D  E(s) + \bar \D  C(s) + \bar \D  Q(s)] = L_{I_0} {\bf \bar \Delta}(s)
\end{equation*}
for $L_{I_0} = \max\left\{\frac{\rho \nu''(d)}{\d}, \frac{a\nu''(d)}{\xi}, \frac{a\rho \nu'''(d)}{\d}\right\}$. In a similar manner, we obtain analogous bounds for $\bar \D I_i(s)$, for $i=1,2,3,4,5,7$, for positive and finite constants $L_{I_i}$, depending only on $a$, $d$, $\rho$, $\d$ and $\xi$. Hence, defining $L_I:= \max_{i\in \{0,1,2,3,4,5,7\}} L_{I_i}$, we establish an uniform Lipschitz control on $I_i$:
\begin{align}
\label{eqIlip}
\bar \D I_{i}(s) &\leq L_I {\bf \bar \Delta}(s), && i=0,1,2,3,4,5,7
\end{align}

\noindent $\bu$ {\it The Lipschitz bound \eqref{eq:lipM} on $\tilde M_1$.}
Recall that $\tilde M_{1,k}$ satisfies:
\begin{equation}
\label{eqM1lip}
\pars \tilde M_{1,k}(s) = - \left(\tilde M_{1,k}(s)\right)^2 - \frac{\tilde M_{1,k}(s)}{2h}+ 1 + \gamma^2 (I_{0,k}(s)-I_{1,k}(s))
\end{equation}
%where
%\begin{equation*}
%I_{0,k}(s)-I_{1,k}(s)=\int_0^s  M_k(u)  R_k(s,u) \nu''(C_k(s,u))du -  M_k(s)\int_0^s \psi(C_k(s,u))  R_k(s,u) du
%\end{equation*}
Let $\Theta(s,t)=\exp\left(-\int_s^t\left(\tilde M_{1,1}(\theta) + \tilde M_{1,2}(\theta) + \frac{1}{2h}\right)d\theta\right)$. In Proposition \ref{Psi} we have shown that if $(M_k, R_k, C_k, Q_k) \in \Ba(\d,\rho,a,d)$ then both $\tilde M_{1,k}(s) \geq 0$ and $\omega_{1,k}(s) = \tilde M_{1,k}(s) + \frac{1}{2h} \geq b$ are true, hence $0 \leq \Theta(s,t)\le \exp(-(t-s)b)$. Now, considering the difference between the realizations of \eqref{eqM1lip} for $k=1$ and $k=2$, respectively, we get:
\begin{equation*}
\pars \Delta \tilde M_1(s) = - \Delta\Tilde M_1(s) \left(\tilde M_{1,1}(\theta) + \tilde M_{1,2}(\theta) + \frac{1}{2h}\right) + \gamma^2(\Delta I_0(s) - \Delta I_1(s))
\end{equation*}
and since $\Delta \tilde M_1(0) = 0$ we get:
\begin{equation*}
\Delta \tilde M_1(s)=\gamma^2 \int_0^s (\Delta I_{0,k}(u) - \Delta I_{1,k}(u)) \Theta(u,s)du
\end{equation*}
hence:
\begin{align}
\bar \Delta \tilde M_1(s) &\le \gamma^2 [\bar \Delta I_0(s) + \bar \Delta I_1(s)] \int_0^s e^{-(s-u)b}du
\le L_{M,1}\gamma^2 {\bf \bar \Delta}(s) \label{eqlipM11}
\end{align}
with $L_{M,1}=\frac{2L_I}{b}$, where in the last inequality we have used the Lipschitz bound on $I_i$'s established in \eqref{eqIlip}.

%\eqref{eq:lipM} for $i=1$ follows, from \eqref{eqlipM11} and the above, for our %choice of $K_{M,1}=K_1K_2$.

%Also, since both $C_1$ and $C_2$ are $[0,d]$-valued symmetric functions and both $\nu''(\cdot)$ and $\psi(\cdot)$ are
%non-negative and monotone non-decreasing on $\R_+$, it follows that $|\Delta \nu''(C(s,u))|\le |\Delta C(s,u)|\nu'''(d)$ and $|\Delta \psi(C(s,u))|\le |\Delta  C(s,u)|\psi'(d)$, respectively. Hence:
%\begin{eqnarray*}
%|\Delta I_0(s)| + |\Delta I_1(s)| &\le& \int_0^s |\Delta\left( M(u)  R(s,u) \nu''( C(s,u))\right)|du + \int_0^s |\Delta\left(M(s)\psi(C(s,u))  R(s,u)\right)| du\\
%&\le& \bar \Delta  M(s)\int_0^s \nu''(d)\rho e^{-(s-u)\delta}du +
%\bar \Delta  E(s)\int_0^s a\nu''(d)\rho e^{-(s-u)\xi}du +
%\bar \Delta  C(s)\int_0^s a\nu'''(d)\rho e^{-(s-u)h\delta}du\\
%&+& \bar \Delta  M(s)\int_0^s \psi(d)\rho e^{-(s-u)\delta}du +
%\bar \Delta  E(s)\int_0^s a\psi(d)\rho e^{-(s-u)\xi}du +
%\bar \Delta  C(s)\int_0^s a\psi'(d)\rho e^{-(s-u)\delta}du\\
%&\leq& K_2\left[\bar \Delta  M(s)+\bar \Delta  E(s)+\bar \Delta  C(s)+\bar \Delta  %Q(s)\right]
%\end{eqnarray*}
%where $K_2=\max\left\{\frac{\rho(\nu''(d)+\psi(d))}{\delta},
%\frac{a\rho(\nu''(d)+\psi(d))}{\xi},\frac{a\rho(\nu'''(d)+\psi'(d))}{\delta}\right\}$. %Hence \eqref{eq:lipM} for $i=1$ follows, from \eqref{eqlipM11} and the above, for our %choice of $K_{M,1}=K_1K_2$.

\noindent $\bu$ {\it The Lipschitz bound \eqref{eq:lipM} on $\tilde M_2$.}
We will first establish the Lipschitz bounds on $\mu_i$ and $\L_i$, $i=1,2$, that will be needed later. Namely, for $i=1$, from \eqref{eqZ1P}:
\begin{align*}
|\D \mu_1(v)| &\leq |\D \tilde M_1(v)| + \gamma^2 |\D I_7(v,0)| \leq (L_I + K_{M,1}) \gamma^2 {\bf \bar \Delta}(s)
\end{align*}
where in the last inequality we have used the bounds in \eqref{eqIlip} and \eqref{eq:lipM} for $i=1$.
Since $|e^{-x}-e^{-y}| \leq |x-y|$ for all $x,y \geq 0$ and $\mu_{k,i}(s) \ge b$, $i,k=1,2$, denoting $K_4:=L_I + K_{M,1}$, we get that
\begin{align}
|\D \L_1(s,t)| &\le e^{-(s-t)b}\int_t^s |\D \omega_1(v)| dv \le \left[K_4 e^{-b(s-t)}(s-t)\right]\gamma^2 {\bf \bar \Delta}(s)
\label{eqL1lip}
\end{align}
Similarly, for $i=2$, we get from \eqref{eqZ2P}:
\begin{align*}
|\D \mu_2(v)| &\le h |\D M(v)| + \beta^2|\D I_7(v,0)| \le (L_I+1)(h+\beta^2) {\bf \bar \Delta}(s)
%&\le& h \bar \D M(v) + \beta^2 \bar \Delta C(v) \int_0^v \psi'(d)\rho e^{-\delta (v-u)}du + \beta^2 \bar \Delta E(v) \int_0^v \psi(d) e^{-\xi (v-u)}du\\
\end{align*}
%for $K_5=\max\left\{\frac{\rho\psi'(d)}{\delta},\frac{\psi(d)}{\xi}, 1 \right\}$,
and a similar argument as above, for $K_5:=L_I+1$, will establish:
\begin{align}
|\D \L_2(s,t)| \le \left[K_5 e^{-b(s-t)}(s-t)\right](\beta^2 + h) {\bf \bar \Delta}(s)
\label{eqL2lip}
\end{align}
Hence, from \eqref{eqL1lip} and \eqref{eqL2lip} we establish the Lipschitz bound for $\L_i$:
\begin{eqnarray}
\bar \D \L_i(s) &\le& K_6\vartheta_i {\bf \bar \Delta}(s)
\label{eqLlip}
\end{eqnarray}
with $K_6 :=\max\{K_4,K_5\}\sup_{\theta \geq 0} \left(\theta e^{-b\theta}\right)$ and also:
\begin{eqnarray}
\int_0^s|\D \L_i(s,u)|du &\le& K_7 \vartheta_i {\bf \bar \Delta}(s)
\label{eqLintlip}
\end{eqnarray}
where $K_7:=\max\{K_4,K_5\}\sup_{\theta \geq 0} \left(e^{-b\theta}\theta^2\right)$.

We can now establish the Lipschitz bound \eqref{eq:lipM} for $\tilde M_2$. Recalling that $\tilde M_{2,k}$ satisfies \eqref{solutiontM2}, we get:
\begin{eqnarray*}
|\D \tilde M_2(s)| &\leq & \alpha |\D \L_{2}(s,0)| + \beta^2 \int_0^s
|\D (I_0(u)\L_2(s,u))|du + h \int_0^s |\D \L_2(s,u)| du\\
&\le&  \alpha |\bar \D \L_2(s)| + \frac{\beta^2}{b}\bar \D I_0 + (\beta^2 K_I + h)\int_0^s |\D \L_2(s,u)|du\\
&\leq& K_8(\beta^2 + h) {\bf \bar \Delta}(s) \,=\, K_8 \vartheta_2 {\bf \bar \Delta}(s)
\end{eqnarray*}
with $K_8 := \alpha K_6 + \frac{L_I}{b} + K_7(\beta_1^2 K_I + h_1)$, where in the last line of the derivation above we have used the bounds in \eqref{generalIbound1}, \eqref{eqIlip}, \eqref{eqLlip} and \eqref{eqLintlip}.

\nn
$\bu$
{\it The Lipschitz bound \eqref{eq:lipE} on $\tilde E$.}
\nn
We rely on the formulas \eqref{eqtH}
and $\tilde R_{i,k}(s,t)=\tilde H_{i,k}(s,t) \L_{i,k}(s,t)$.
Indeed, since $ C_1$ and $ C_2$ are $[0,d]$-valued
symmetric functions, $t_i \in [0,s]$ and
both $\nu''(\cdot)$ and $\nu'''(\cdot)$ are
non-negative and monotone non-decreasing, it follows that
for any $n$, $t_{2n} \leq s$ and $\sigma \in\NC_n$,
\begin{equation*}
\left| \prod_{i\in \cro(\sigma) } \nu''(C_1(t_i,t_{\sigma_i}))
-
 \prod_{i\in \cro(\sigma) } \nu''(C_2(t_i,t_{\sigma_i})) \right|
\leq n \nu''(d)^{n-1} \nu'''(d) \bar \D  C(s) \,.
\end{equation*}
Thus we easily deduce from \eqref{eqtH} that
\begin{eqnarray}
|\D\tilde H_i(s,t)|
&\le&4\epsilon_i^{2} \nu'''(d)(s-t)^2
\sum_{n\ge 1}n(2n!)^{-1} [2 r (s-t)]^{2(n-1)}
\bar \D  C(s)
\label{DH}\\
&\le&  \epsilon_i^2 K_9 (s-t)^2 e^{2 \epsilon_i \sqrt{\nu''(d)} (s-t)} \bar \D  C(s) \,.
\nonumber
\end{eqnarray}
for $K_9=2\nu'''(d)$. Recalling that $\epsilon_i^2 \leq \vartheta_i$ and since
$\tilde E_{i,k} (s,t)=\tilde R_{i,k}(s,t)e^{\xi (s-t)}=\tilde H_{i,k}(s,t) \L_{i,k}(s,t)e^{\xi (s-t)}$ we now obtain from
\eqref{eq:tHbd}, \eqref{DH}, \eqref{eqL1lip} and \eqref{eqL2lip} that:
\begin{align*}
\D \tilde E_i(s,t) &\le e^{\xi (s-t)}\left[\L_{i,1}(s,t) \D \tilde H_i(s,t)+ \tilde
H_{i,2}(s,t) \D \L_i(s,t)\right] \\
&\le \vartheta_i e^{\left(-b+\xi +2\epsilon_i \sqrt{\nu''(d)}\right)(s-t)}[K_9  (s-t)^2 + c_1 (K_4+K_5)(s-t)] {\bf \bar \Delta}(s)\\
&\le \vartheta_i e^{-(b/3)(s-t)}[K_9  (s-t)^2 + c_1 (K_4+K_5)(s-t)] {\bf \bar \Delta}(s) \\
&\le \vartheta_i L_{E,i} {\bf \bar \Delta}(s)
\end{align*}
for $\epsilon_i<\frac{b}{6\sqrt{\nu''(d)}}$ and for the finite positive constant
\begin{equation*}
L_{E,i}:= \sup_{\theta \geq 0} \,
e^{-b\theta/3}\left[K_9 \theta^2 + c_1(K_4+K_5)\theta \right] \,.
\end{equation*}

\smallskip\nn
$\bu${\it The Lipschitz bounds \eqref{eq:lipC} and \eqref{eq:lipQ} on $\tilde C$ and $\tilde Q$, respectively.}
Recalling the solution \eqref{solutiontC} of $C_{i,k}$, we have:
\begin{align*}
\Delta \tilde C_{i}(s,t) &= \Delta \L_{i} (s,t) + \epsilon_i^2 \int_t^s \D (\L_{i}(s,v) I_{7}(v,t)) dv\\
&\quad + \epsilon_i^2 \int_t^s \D (\L_{i}(s,v) I_{8}(v,t)) dv + k_i \int_t^s \D (\tilde M_{i}(t)\L_{i}(s,v)) dv
\end{align*}
Using the Lipschitz bounds in \eqref{eqIlip}, \eqref{eqLlip} and \eqref{eqLintlip}, the first two integrals above are each bounded by:
$$(\vartheta_i K_7 K_I + L_I) {\bf \bar \Delta}(s)$$
while by \eqref{eq:lipM}, the last one is bounded by:
$$\vartheta_i\left(\frac{L_{M,i}}{b} + a K_6\right) {\bf \bar \Delta}(s)$$
Wrapping all together, we get:
\begin{eqnarray*}
|\bar \D \tilde C_{i}(s)| &\leq& \vartheta_i L_{C,i} {\bf \bar \Delta}(s)
\end{eqnarray*}
for $L_{C,i} = (K_5+K_6) + 2((\beta_1+\gamma_1)K_7 K_I + L_I) + \left(\frac{L_{M,i}}{b} + a K_6\right)(h_1+1)$ and consequently, \eqref{eq:lipC} holds.

Similarly, by the solution \eqref{solutiontD} of $D_{i,k}(s):=Q_{i,k}(s,s)$, we have:
\begin{align*}
\Delta \tilde D_{i}(s) &= \Delta (\L_{i}^2 (s,0)) + 2\epsilon_i^2 \int_t^s \D (\L_{i}^2(s,v) I_{4}(v,0)) dv +
2\epsilon_i^2 \int_0^s \D (\L_{i}^2(s,v) I_{5}(v,0)) dv\\
&\quad + 2\alpha k_i \int_0^s \D (\L_{i}^2(s,v)) dv
\end{align*}
Since $\L_i(s,t)\in [0,1]$, then $|\D L_i^2(s,t)| \leq 2 |\D L_i(s,t)|$, hence similarly as above, we get:
\begin{eqnarray*}
|\bar \D \tilde D_{i}(s)| &\leq& L_{D,i}\vartheta_i {\bf \bar \Delta}(s)
\end{eqnarray*}
for $L_{D,i} = 4L_{C_i}$. Moving over to $\tilde Q_{i,k}$, since:
\begin{align*}
\Delta \tilde Q_{i}(s,t) &= \Delta (\tilde D_i(t)\L_{i} (s,t)) + \epsilon_i^2 \int_t^s \D (\L_{i}(s,v) I_{4}(v,t)) dv\\
&\quad + \epsilon_i^2 \int_0^s \D (\L_{i}(s,v) I_{5}(v,t)) dv + k_i \int_t^s \D (\tilde M_i(t)\L_{i}(s,v)) dv
\end{align*}
using the Lipschitz bound on $\tilde D_i$ and similar reasonings as above, we get:
\begin{eqnarray*}
|\bar \D \tilde Q_{i}(s)| &\leq& L_{Q,i}\vartheta_i {\bf \bar \Delta}(s)
\end{eqnarray*}
for $L_{Q,i} = L_{D,i} + \max\{d,1\}L_{C,i}$, thus concluding the argument that $\Psi_i$ is a contraction.

\medskip
Now suppose that, for a choice of parameters $\beta$ and $h$, the constants $d, \rho, a, b$ and $d$ are such that $\Psi_i$ is a contraction on $\Ba(\d,\rho,a,d)$,
hence also on its non-empty subset $\Sa(\d,\rho,a,d)$.
%Regarding the mapping $\Psi$ as a map from $( Q,  D,  A)$ to $(\tilde Q, \tilde D, \tilde A)$,
Proposition \ref{Psi} shows that both $\Ba(\d,\rho,a,d)$ and $\Sa(\d,\rho,a,d)$ are invariant under $\Psi_i$. We start at some $S_{i,0}=(M_0, R_0, C_0, Q_0) \in \Sa(\d,\rho,a,d)$ and construct recursively
the sequences $S_{i,k}=\Psi_i(S_{i,k-1})$ for $k = 1,2,\dots$,
in $\Sa(\d,\rho,a,d)$. For  $i=1,2$, since $\Psi_i$ is a contraction,
clearly $\{S_{i,k}\}_{k \in \Z_+}$ is a Cauchy sequence for
the uniform norm $\|\cdot\|$ of \eqref{eq:unorm}. Hence,
$S_{i,k} \to S_{i,\infty} = (M_{i,\infty}, R_{i,\infty}, C_{i,\infty}, Q_{i,\infty})$ in the Banach space
$(\CC(\R_+) \ts \CC(\bD) \ts \CC_s(\R_+^2) \ts \CC_s(\R_+^2),\|\cdot\|)$.
Note that $\Ba(\d,\rho,a,d)$ is a closed
subset of this Banach space, so
$S_{i,\infty} \in \Ba(\d,\rho,a,d)$. Further, fixing $\tau \geq 0$, since $S_{i,k} \in \Sa(\d,\rho,a,d)$ we have that
\begin{align*}
& \lim_{T \to \infty} \sup_{t,t' \geq T}
|C_{i,\infty}(t+\tau,t)-C_{i,\infty}(t'+\tau,t')|\\
& \quad\leq 2 \|C_{i,\infty}-C_{i,k}\|_{\infty} +
\lim_{T \to \infty} \sup_{t,t' \geq T}
|C_{i,k}(t+\tau,t)-C_{i,k}(t'+\tau,t')| = 2 \|S_{i,\infty}-S_{i,k}\| \,.
\end{align*}
Taking $k \to \infty$ we deduce
that, for any $\tau\geq 0$, $t \mapsto C_{i,\infty}(t+\tau,t)$ is a Cauchy function from
$\R_+$ to $[0,d]$, hence
$C_{i,\infty}(t+\tau,t)$ converges as $t \to \infty$. A similar bounding procedure as above will show that the same is true for $E_{i,\infty}$ and $Q_{i,\infty}$ and will also show that $M_{i,\infty}(t)$ converges as $t \ra \infty$. Now, since, by definition, $R_{i,\infty}(s,t) = E_{i,\infty}(s,t)e^{-\xi(s-t)}$, then $R_{i,\infty}$ will inherit the limiting property from $E_{i,\infty}$. Hence $S_{i,\infty} \in \Sa(\d,\rho,a,d)$ and
further $S_{i,\infty}$ is the unique fixed point of the contraction
$\Psi_i$ on the metric space $(\Sa(\d,\rho,a,d),\|\cdot\|)$.

By our construction of  $\Psi_i$, it follows that $(M_{1,\infty}, R_{1,\infty}, C_{1,\infty}, Q_{1,\infty})$
 satisfies \eqref{eqMsh}-\eqref{eqZsh} and also that $(M_{2,\infty}, R_{2,\infty}, C_{2,\infty}, Q_{2,\infty})$ satisfies \eqref{eqMs}-\eqref{eqZs}. Recalling that any solution of \eqref{eqMsh}-\eqref{eqZsh} is a solution of \eqref{eqMs}-\eqref{eqZs} that has been time-scaled by a factor of $h$, we can conclude that the unique solution of \eqref{eqMs}-\eqref{eqZs} is in $\Sa(\d h,\rho,a,d)$, for $\gamma \in [0,\gamma_2]$ and in $\Sa(\d,\rho,a,d)$, respectively, for $\beta \in [0,\beta_2]$ and $h \in [0, h_2]$. As noted before, this shows that the FDT limits $M^\fdt$, $R^{\fdt}(\tau)$, $C^{\fdt}(\tau)$ and $Q^\fdt(\tau)$ exist, for the unique solution of \eqref{eqMs}-\eqref{eqZs} and furthermore, $(M^{\fdt}, R^{\fdt}, C^{\fdt}, Q^\fdt) \in \Da(\d h,\rho,a,d)$ if $\gamma \leq \gamma_2$ and $(M^{\fdt}, R^{\fdt}, C^{\fdt}, Q^\fdt) \in \Da(\d ,\rho,a,d)$ if $\beta\leq \beta_2$ and $h\leq h_2$.

In order to conclude the proof, we will show that $M^\fdt, R^{\fdt}(\cdot), C^{\fdt}(\cdot)$ and $Q^{\fdt}(\cdot)$ are the unique solution in $\Da(\d h,\rho,a,d)$, respectively $\Da(\d,\rho,a,d)$, of \eqref{eqMFDT}-\eqref{eqZFDT}. While proving Proposition \ref{Psi} we found that on
$\Sa(\d,\rho,a,d)$, the mapping $\Psi_i$ induces a mapping
$\Psi_i^{\fdt}:(M^\fdt, R^{\fdt}, C^{\fdt}, Q^{\fdt}) \to (\tilde M_i^{\fdt}, \tilde R_i^{\fdt}, \tilde C_i^{\fdt}, \tilde Q_i^{\fdt})$
such that
{\allowdisplaybreaks
\begin{align*}
0 &\quad=\quad -\left(\tilde M_1^{\fdt}\right)^2 - \frac{\tilde M_1^{\fdt}}{2h} + 1 + \gamma^2\left(\hh I_0 - \hh I_1\right)\\
0 &\quad=\quad -\hh \omega_2 \tilde M_2^\fdt + h + \beta^2 \hh I_0 \\
\tilde R_i^{\fdt} (\tau) &\quad=\quad \hh{\L}(\tau)
\sum_{n\ge 0}\epsilon_i^{2n} \sum_{\s\in\NC_n}
\int_{0\le \theta_1\le \cdots\le \theta_{2n}\le \tau}
\prod_{i\in \cro(\s)} \nu''(C^{\fdt}(\theta_i-\theta_{\sigma(i)}))
\prod_{j=1}^{2n} d\theta_j \,,\\
\tilde C_i^{\fdt} (\tau) &\quad=\quad \hh{\L}_i(\tau) +
\epsilon_i^2 \int_0^{\tau} \hh{\L}_i(\tau-v) \hh{I}_2(v) dv
+ \epsilon_i^2 \int_0^\tau \hh{\L}_i(\tau-v) \hh{I}_3(v) dv + k_i \tilde M_i^{\fdt} \int_0^\tau \hh{\L}_i(v) dv\,,\\
\tilde Q_i^{\fdt} (\tau) &\quad=\quad \tilde D_i^\fdt \hh{\L}_i(\tau) +
\epsilon_i^2 \int_0^{\tau} \hh{\L}_i(\tau-v) \hh{I}_4(v) dv
+ \epsilon_i^2 \int_0^\tau \hh{\L}_i(\tau-v) \hh{I}_5(v) dv +
k_i \tilde M_i^{\fdt} \int_0^\tau \hh{\L}_i(v) dv\,,\\
0 &\quad=\quad -\hh \omega_i \tilde D_i^\fdt +
\epsilon_i^2 \hh{I}_4(0) + \epsilon_i^2 \hh{I}_5(0) +
k_i \tilde M_i^{\fdt}\,,
\end{align*}}
where $\hh{I}_0, \hh{I}_1, \hh{I}_2, \hh{I}_3, \hh{I}_4$ and $\hh{I}_5$ are given by \eqref{eqI0l}, \eqref{eqI1l}, \eqref{eqI2l}, \eqref{eqI3l}, \eqref{eqI4l} and \eqref{eqI5l}, respectively. In particular,
$\tilde C^{\fdt}$, $\tilde R^{\fdt}$ and $\tilde Q^{\fdt}$ are differentiable on $\R_+$, and, for $\tau \geq 0$,
{\allowdisplaybreaks
\begin{align}
\label{eqfPsiM1}
0 \quad=&\quad -\left(\tilde M_1^{\fdt}\right)^2 - \frac{\tilde M_1^{\fdt}}{2h} + 1 + \gamma^2 M^{\fdt}\int_0^{\infty}   R^{\fdt}(\theta) \nu''( C^{\fdt}(\theta))d\theta\\
\nonumber
&\quad - \gamma^2 M^{\fdt}\int_0^{\infty}   R^{\fdt}(\theta) \psi( C^{\fdt}(\theta))d\theta\\
0 \quad=&\quad -\hh \omega_2 \tilde M_2^{\fdt} +  \beta^2M^{\fdt}\int_0^{\infty}   R^{\fdt}(\theta) \nu''( C^{\fdt}(\theta))d\theta + h\\
\label{eqfPsiR}
\pars \tilde R_i^{\fdt} (\tau) \quad=&\quad -\hh \omega_i \tilde R_i^{\fdt} (\tau)
+ \epsilon_i^2 \int_0^\tau \tilde R_i^{\fdt} (\tau-\theta) \tilde R_i^{\fdt} (\theta)
\nu''(C^{\fdt}(\theta)) d\theta \,,\\
%-\hh \omega_i \tilde C_i^\fdt (\tau) + \epsilon_i^2 \hh{I}_1(\tau)
%+ \epsilon_i^2 \hh{I}_2(\tau) + k_i \tilde M_i^{\fdt} \\
\pars \tilde C_i^{\fdt} (\tau) \quad=&\quad
 -\hh \omega_i \tilde C_i^{\fdt} (\tau) + \epsilon_i^2 \int_0^\infty
C^{\fdt} (\tau-\theta)  R^{\fdt} (\theta) \nu''( C^{\fdt} (\theta)) d\theta
\label{eqfPsiC}\\
&\quad + \epsilon_i^2 \int_\tau^\infty \nu'( C^{\fdt} (\theta))  R^{\fdt} (\theta-\tau)d\theta + k_i \tilde M_i^{\fdt}\,
\nonumber \\
\pars \tilde Q_i^{\fdt} (\tau) \quad=&\quad -\hh \omega_i \tilde Q_i^{\fdt} (\tau) + \epsilon_i^2 \int_0^\infty
Q^{\fdt} (\tau-\theta)  R^{\fdt} (\theta) \nu''( C^{\fdt} (\theta)) d\theta
\label{eqfPsiQ}\\
&\quad +\epsilon_i^2 \int_\tau^\infty \nu'( Q^{\fdt} (\theta))  R^{\fdt} (\theta-\tau)d\theta + k_i \tilde M_i^{\fdt}\,
\nonumber
\end{align}
with $\tilde R_i^{\fdt}(0)=1$, $\tilde C_i^{\fdt}(0)=1$, $\pars \tilde Q_i^{\fdt}(0) = 0$ and
\begin{eqnarray}
\label{eqfPsiZ1}
\hh \omega_1 &=&  \frac{1}{2h} + \gamma^2 \int_0^\infty \psi(C^\fdt(\theta))  R^\fdt(\theta) d\theta +  \tilde M_1^{\fdt}\\
\label{eqfPsiZ2}
\hh \omega_2 &=&  \frac{1}{2} + \beta^2 \int_0^\infty \psi(C^\fdt(\theta))  R^\fdt(\theta) d\theta + hM^{\fdt}
\end{eqnarray}}
where in the derivation of \eqref{eqfPsiR} we have used the results in \cite{GM}.

Recall that if the functions $M, R, C$ and $Q$ solve \eqref{eqMs}-\eqref{eqZs}, then the functions $M_h, R_h, C_h$ and $Q_h$ are the unique solution of \eqref{eqMsh}-\eqref{eqZsh}, hence the unique fixed point of $\Psi_1$. Then, by \eqref{eqfPsiM1}--\eqref{eqfPsiZ2} the corresponding quad-uple $(M_h^\fdt, R_h^\fdt, C_h^\fdt, Q_h^\fdt)$ is a fixed points of $\Psi_1^{\fdt}$. Then $M^\fdt:=M_h^\fdt$, $R^\fdt(\tau):=R_h^\fdt(h\tau)$, $C^\fdt(\tau):=C_h^\fdt(h\tau)$ and $Q^\fdt(\tau):=Q_h^\fdt(h\tau)$ satisfy the FDT equations \eqref{eqMFDT}-\eqref{eqZFDT}. Noticing that the quad-uple $(M^\fdt, R^\fdt, C^\fdt, Q^\fdt)$ that we have just defined coincide with the FDT limits of the original $(M, R, C, Q)$, we have established that, for $\gamma \in [0,\gamma_2]$, $(M^\fdt, R^\fdt, C^\fdt, Q^\fdt)$ satisfy \eqref{eqMFDT}-\eqref{eqZFDT}.

Also, if $(M, R, C, Q)$ is the unique solution of \eqref{eqMs}-\eqref{eqZs}, it is the unique fixed point of $\Psi_2$, hence $(M^\fdt, R^\fdt, C^\fdt, Q^\fdt)$ is a fixed point of $\Psi_2^{\fdt}$, hence it satisfies  \eqref{eqMFDT}-\eqref{eqZFDT}.

Now, denoting by $E^{\fdt}(\tau)=e^{\xi \tau} R^{\fdt}(\tau)$, by the same arguments as in the Lipschitz estimates \eqref{eq:lipM}-\eqref{eq:lipQ}
of Proposition \ref{FDT1}, we show that:
\begin{eqnarray*}
\bar \Delta \tilde M_i^{\fdt} &\leq& \vartheta_i L_{M,i} [ \bar \D M^{\fdt} + \bar \D E^{\fdt} (\infty) + \bar \D C^{\fdt}(\infty) + \bar \D Q^{\fdt}(\infty)],\\
\bar \Delta \tilde E_i^{\fdt} (\tau) &\leq& \vartheta_i L_{E,i} [ \bar \D M^{\fdt} + \bar \D E^{\fdt} (\tau) + \bar \D C^{\fdt}(\tau) + \bar \D Q^{\fdt}(\tau)],\\
\bar \Delta \tilde C_i^{\fdt} (\tau) &\leq& \vartheta_i L_{C,i} [ \bar \D M^{\fdt} + \bar \D E^{\fdt} (\tau) + \bar \D C^{\fdt}(\tau) + \bar \D Q^{\fdt}(\tau)],\\
\bar \Delta \tilde Q_i^{\fdt} (\tau) &\leq& \vartheta_i L_{Q,i} [ \bar \D M^{\fdt} + \bar \D E^{\fdt} (\tau) + \bar \D C^{\fdt}(\tau) + \bar \D Q^{\fdt}(\tau)]
\end{eqnarray*}
for all $\tau<\infty$, where $\bar \D f(s)=\sup_{0\leq u\le s} |f_1(u)-f_2(u)|$ when $f$ is one of the function of interest $E$, $C$ or $Q$, and $\bar \D M = |M_1-M_2|$, thus showing that the mappings $\Psi_i^\fdt$ are also contractions, they have unique fixed points in $\Da(\d h,\rho,a,d)$ and $\Da(\d,\rho,a,d)$, respectively. So \eqref{eqRFDT}-\eqref{eqZFDT} have an unique solution in $\Da(h\d,\rho,a,d)$, for $\gamma\in[0,\gamma_2]$ and in $\Da(\d,\rho,a,d)$, for $\beta\in[0,\b_2]$ and $h\in[0,h_2]$, as claimed.
\hfill $\Box$

%%%%%%%%%%%%%%%%%%%%%%%%%%%%%%%%%%%%%%%%%%%%%%%%%%%%
%%%%%%%%%%%%%%%%%%%%%%%%%%%%%%%%%%%%%%%%%%%%%%%%%%%%

\subsection{Exponential Decay of the Covariance}

%%%%%%%%%%%%%%%%%%%%%%%%%%%%%%%%%%%%%%%%%%%%%%%%%%%%
%%%%%%%%%%%%%%%%%%%%%%%%%%%%%%%%%%%%%%%%%%%%%%%%%%%%

One consequence of Proposition \ref{FDT1} is that if either $\gamma$ is small or both $\b$ and $h$ are small, the response function is positive and decays to $0$ exponentially fast. In the next proposition we will establish an analogous result for the covariance. Namely, we show:

\begin{prop}\label{expdecay}
For $\gamma_2, \beta_2, h_2 > 0$ of Proposition \ref{FDT1}, if $\gamma\in[0,\gamma_2]$ or $\beta\in[0,\beta_2]$ and $h\in [0,h_2]$ there exist $M = M(\beta,h,\alpha)>0$ and $\eta = \eta(\beta,h,\alpha)$ such that for every $s\geq t\geq 0$:
\begin{eqnarray}
\label{expdecayC}
|C(s,t)-Q(s,t)| &\leq& M e^{-(s-t)\eta}
\end{eqnarray}
\end{prop}
\nn
\noindent{\bf Proof of Proposition \ref{expdecay}:} Let $COV(s,t):=C(s,t)-Q(s,t)$ and respectively $COV_h(s,t):=C_h(s,t)-Q_h(s,t)$, with $U_h(s,t):=U(s/h,t/h)$, whenever $U$ is one of $C$ or $Q$.
Subtracting \eqref{eqQs} from \eqref{eqCs}, we get:
\begin{align}
\pars_1 COV(s,t) &\,=\,
 -\mu(s) COV (s,t) +
\b^2 \int_0^s COV (u,t) R(s,u) \nu''(C(s,u)) du &&\label{eqDeltas} \\
&\qquad+ \b^2 \int_0^t COV (s,u) P(C(s,u),Q(s,u)) R(t,u) du,&& s \geq t \geq 0
\nonumber
\end{align}
for the multivariate polynomial $P(X,Y) = \frac{\nu'(X)-\nu'(Y)}{X-Y}$, where $\mu$ is defined by \eqref{eqZs}, hence
\begin{equation}\label{eqC2}
COV(s,t)= \L (s,t) + \beta^2 \int_t^s \L(s,v) I_9(v,t) dv +
\b^2 \int_t^s \L (s,v) I_{10}(v,t) dv
\end{equation}
with $\L(s,v)=\exp(-\int_v^s \mu(u) du)$,
\begin{eqnarray}
I_9(v,t)&=&\int_0^v COV (u,t) R(v,u) \nu''(C(v,u)) du \,, \label{eqI9} \\
I_{10}(v,t)&=& \int_0^t COV (v,u) P(C(v,u),Q(v,u)) R(t,u) du \,.  \label{eqI10}
\end{eqnarray}
By Proposition \ref{FDT1} we know that, whenever $\beta<\beta_2$ and $h<h_2$, $R(s,t)\leq \rho e^{-(s-t)\d}$ and $\mu(s)\geq b$ implying $\L(s,v) \le e^{-b(s-v)}$. Also, Theorem \ref{theo-sphere} shows $C(s,t), Q(s,t) \in [0,1]$, implying $P(C(s,t),Q(s,t))\leq \nu''(1)$, since $\nu(\cdot)$ is a polynomial with positive coefficients. So, we get:
{\allowdisplaybreaks
\begin{align*}
|I_9(v,t)| &\quad\le\quad \nu''(1) \int_0^v |COV(u,t)| \rho e^{-\d (v-u)}du \le \nu''(1)\rho e^{-\d (v-t)},
\\
|I_{10}(v,t)| &\quad\le\quad \nu''(1)  \rho \d^{-1} \sup_{u\le t} |COV(u,v)|.
\end{align*}}
and hence, with the symmetric function $\D(t,s):=\sup_{u\le t,v \le s} \, |COV(u,v)|$
we deduce from (\ref{eqC2}) that for $s \ge t\ge 0$,
\begin{eqnarray*}
\D(t,s)&\le&  e^{-b(s-t)}
+\beta^2 \nu''(1) \rho \int_t^s e^{-b(s-v)}[
\int_0^v e^{-\d (v-u)}
du + \d^{-1}\D(t,v)] dvdu\\
&\le& e^{-b(s-t)}
+\beta^2 \rho \nu''(1)  \int_t^s e^{-b(s-v)}
\int_0^t  e^{-\d (v-u)} du dv\\
&&
+\beta^2 \rho \nu''(1)\int_t^s \D(t,v)[\d^{-1}
e^{-b(s-v)}+\int_t^v e^{-b(s-v)-\d (v-u)}
du] dv
\end{eqnarray*}
Since for any $\d \in (0, b/2)$ and $s \geq t$,
\begin{equation}\label{eq:integ}
\int_t^s e^{-b(s-v)-\d(v-t)} dv \le 2 b^{-1} e^{-\d (s-t)}
\end{equation}
and with $\d \in (0,b)$ we thus obtain for $s \geq t$ the bound
\begin{eqnarray*}
\D(t,s)&\le& M_\beta e^{-{\d_\beta} (s-t)}
+A_\beta \int_t^s \D(t,v) e^{-{\d_\beta} (s-v)} dv \,,
\end{eqnarray*}
with $M = 1 + 2\beta^2 \rho \nu''(1)(b\d)^{-1}$ and
$A = \beta^2\rho\nu''(1)\d^{-1}(1 + 2b^{-1})$. Therefore, fixing $t \geq 0$,
the function $h_t(s)=e^{{\d}(s-t)} \D(t,s)$ satisfies
$$h_t(s)\le M+ A\int_t^s h_t(v)dv, \quad s\ge t, $$
and so by  Gronwall's lemma $h_t(s)\le  M e^{A(s-t)}$.
We therefore conclude that for any $s \geq t$,
$$|C(s,t)-Q(s,t)|\le  M e^{-(\d-A)( s-t)} \,,
$$
which proves the lemma in this case, since for $\beta \to 0$
we have that $A = A(\beta) \to 0$ (and so
$\eta = \d - A>0$ for any $\beta>0$ small enough).

Similarly, from \eqref{eqQsh} from \eqref{eqCsh}, we get:
\begin{align}
\pars_1 COV_h(s,t) &\,=\,
 -\mu_h(s) COV_h (s,t) +
\gamma^2 \int_0^s COV_h (u,t) R_h(s,u) \nu''(C_h(s,u)) du
\label{eqDeltash}\\
&\qquad+ \gamma^2 \int_0^t COV_h (s,u) P(C_h(s,u),Q_h(s,u)) R_h(t,u) du, \qquad s \geq t \geq 0
\nonumber
\end{align}
where $\mu_h$ is defined by \eqref{eqZsh}. Recalling that if $\gamma \leq \gamma_2$, $\mu_h(s) \geq b$, the same argument as before, with $\gamma$ in the place of $\beta$, will show that $\Delta_h (s,t) := \Delta (s/h,t/h) \leq M e^{-(\delta-A)(s-t)}$, that is equivalent to:
$$|C(s,t)-Q(s,t)|\le  M e^{-h(\d-A)(s-t)}$$
hence concluding out proof.
\hfill$\Box$

%%%%%%%%%%%%%%%%%%%%%%%%%%%%%%%%%%%%%%%%%%%%%%%%%%%%
%%%%%%%%%%%%%%%%%%%%%%%%%%%%%%%%%%%%%%%%%%%%%%%%%%%%

\subsection{Simplifying the FDT System}

%%%%%%%%%%%%%%%%%%%%%%%%%%%%%%%%%%%%%%%%%%%%%%%%%%%%
%%%%%%%%%%%%%%%%%%%%%%%%%%%%%%%%%%%%%%%%%%%%%%%%%%%%

The final step of the proof is to relate the solutions of the limiting equations \eqref{eqMFDT}-\eqref{eqZFDT} to
the FDT equations \eqref{FDTC} and \eqref{FDTQ}, hence concluding the proof of Theorem \ref{FDT}.

\begin{prop}\label{FDT2} There exist $\gamma_3, \b_3, h_3 > 0$ such that whenever $\gamma\in[0,\gamma_3]$ or $\b \in [0,\b_3]$ and $h\in [0,h_3]$, the equations \eqref{FDTC} and \eqref{FDTQ} have unique solutions $C(\cdot)$ and $Q$. Furthermore, the quadruple $(M, C, R, Q)$, where $R(\tau): = -2\pars C(\tau)$ and $Q(\tau) := Q$ solves the system \eqref{eqMFDT}-\eqref{eqZFDT} with initial conditions $C(0)=R(0)=1$, $Q'(0)=0$. Furthermore, $R(\tau)$ is positive and decays exponentially fast to $0$ and $C(\tau)$ is positive and bounded, converging to $Q$ as $\tau \ra \infty$.
\end{prop}
\nn
\noindent{\bf Proof of Proposition \ref{FDT2}:} Consider the function $f(x)=4(x-1)^2[\b^2\nu'(x) + h^2]-x$. Since for any $h>0$, $f(1-(2h)^{-1})>0$ and $f(1)<0$ and also $f(0)>0$, there exist at least a solution to $f(x)=0$ in $[(1-(2h)^{-1})\wedge 0,1]$. By definition, any of these solutions satisfies \eqref{FDTQ}. Fix $Q$ to be one of them.

Let $C$ be the unique $[0,1]$-valued solution of \eqref{FDTC} for $\phi(x)=1/2 - 2\b^2Q\nu'(Q)+2h^2(1-Q)+2\b^2\nu'(x)$ (see Proposition 1.4 of \cite{DGM} for existence and uniqueness of the solution). Also, since $Q \in [(1-(2h)^{-1})\wedge 0,1]$, it is easy to see that for small enough $\gamma$, the following bound holds:
$$2\b^2(\nu'(1)-\nu'(Q)) \geq \b \gamma \nu''(1) \geq 2\sqrt{\b^2 \nu'(1)}$$
and if $\b$ is small enough, then:
$$\frac{1}{2}\geq 2\sqrt{\b^2 \nu'(1)}$$
thus concluding that in both scenarios, $\phi(1)>2\sqrt{b\phi'(1)}$, hence, according to the above-mentioned result, $C'$ decays exponentially to $0$ with some positive exponent (it is easy to see that $\phi$ is convex, so the conditions in the quoted proposition are satisfied).

Moreover, by the same result, $C$ converges as $t \ra \infty$ to
$$C_\infty := \sup \left\{x\in[0,1]\;:\;\phi(x)(1-x)\geq \frac{1}{2}\right\}$$
Now, from the definition of $Q$, it is easy to see that $\phi(Q)(1-Q)=1/2$ and since $Q\in[0,1]$, $C_\infty \geq Q$. Also, for $\gamma$ sufficiently small, for $x\in [Q,1]$,
$$2\beta^2\left(\frac{\nu'(x)-\nu'(Q)}{x-Q}\right) \leq 2\gamma^2 h^2 \nu''(1) < 4h^2 \leq \frac{1}{(1-Q)(1-x)}$$
hence $\phi(x)(1-x) < 1/2$ for $x\in [Q,1]$, implying $C_\infty = Q$. Similarly, for $\b$ small,
$$2\beta^2\left(\frac{\nu'(x)-\nu'(Q)}{x-Q}{(1-Q)(1-x)}\right) \leq 2\beta^2 \nu''(1) < 1$$
so $\phi(x)(1-x) < 1/2$ for $x\in [Q,1]$, hence $C_\infty = Q$.

Now, denoting by $R(\tau):=-2\pars C(\tau)$, and $Q(\tau) \equiv Q$, since $Q = \lim_{t \ra \infty} C(\tau)$, some simple algebra will show that $(M, R, C, Q)$ satisfy \eqref{eqMFDT}-\eqref{eqZFDT} with initial conditions $C(0) = 1$, $R(0) = 1$, $Q'(0) = 0$, if and only if:
\begin{eqnarray}
\label{FDTM1}
0 &=& -\mu M + h + 2\beta^2M \left(\nu'(1)-\nu'(Q)\right)\\
\label{FDTQ1}
0 &=& -\mu Q +2\beta^2 \left(Q\nu'(1)-2Q\nu'(Q)+\nu'(Q)\right) + hM
\end{eqnarray}
with
\begin{equation}
\nonumber
\mu = \frac{1}{2} + 2\beta^2(\nu'(1)-Q\nu'(Q)) + hM
\end{equation}
It's easy to check that $M:=2h(1-Q)$ and $Q$ are a solution to \eqref{FDTM1}-\eqref{FDTQ1}, hence $(M,R,C,Q)$
satisfy \eqref{eqMFDT}-\eqref{eqZFDT}. Furthermore, $M,Q \in [0,1]$, as needed.

Now, for every root of \eqref{FDTQ}, we can use the same procedure as above to construct a quad-uple $(M,R,C,Q)$, that solves the system. Since $Q\in [(1-(2h)^{-1})\wedge 0,1]$, the same arguments as above will conclude that $(M,R,C,Q)$ are positive, $C(\cdot)$ is bounded and $R(\cdot)$ decays to $0$ exponentially fast.
Since according to Proposition \ref{FDT1}, the system \eqref{eqMFDT}-\eqref{eqZFDT} has an unique solution with these properties, the injectivity of the mapping $Q \mapsto (M,R,C,Q)$ shows that \eqref{FDTQ} has a unique root in $[(1-(2h)^{-1})\wedge 0,1]$, thus concluding the proof.
\hfill
$\Box$

\medskip
Now we have all the ingredients we need to finalize the proof of our theorem:

\noindent{\bf Proof of Theorem \ref{FDT}:} Fix $\gamma_0 = \min\{\gamma_i : i=1,2,3\}$, $\b_0 = \min\{\b_i : i=1,2,3\}$ and $h_0 = \min\{h_i : i=1,2,3\}$, for $\gamma_1, \beta_1, h_1$ of Proposition \ref{Psi}, $\gamma_2, \beta_2, h_2$ of Proposition \ref{FDT1} and $\gamma_3, \beta_3, h_3$ of Proposition \ref{expdecay}. Then, according to Proposition \ref{FDT1}, the FDT limits \eqref{Mfdt-ex}-\eqref{Qfdt-ex} exist and are the unique solution of \eqref{eqMFDT}-\eqref{eqZFDT} with initial conditions $C(0) = R(0) = 1$, $Q'(0) = 0$, in the space of positive functions such that $C(\cdot), Q(\cdot)$ are bounded above and $R(\cdot)$ decays exponentially to $0$.

By Proposition \ref{FDT2}, for the same possible values of the parameters $\b$ and $h$, $C(\tau)$, $R(\tau):=-2\pars C(\tau)$, $Q(\tau) := Q$ and $M:=2h(1-h)$ are a solution of \eqref{eqMFDT}-\eqref{eqZFDT} and furthermore, $R$ decays exponentially fast to $0$ and $0\leq M, Q(\tau), C(\tau) \leq 1$, so, by the afore-mentioned uniqueness result, they are indeed the solution of \eqref{eqMFDT}-\eqref{eqZFDT}, thus concluding the proof.
\hfill
$\Box$

\end{document}